\documentclass[a4paper]{article}

\usepackage[pages=all, color=black, position={current page.south}, placement=bottom, scale=1, opacity=1, vshift=5mm]{background}
\SetBgContents{
	\textit{Preprint submitted to Computational Geosciences}
}      

\usepackage[margin=1in]{geometry} 

\usepackage{amsmath}
\usepackage{amsthm}
\usepackage{amssymb}
\usepackage{booktabs}

\usepackage{graphicx} 
\usepackage{float}
\usepackage{algorithm}  
\usepackage{algpseudocode}
\usepackage{color}
\usepackage{setspace}

\usepackage{hyperref}
\usepackage{amsmath}
\usepackage{algorithm}
\usepackage{algpseudocode}
\usepackage{emptypage} 
\usepackage{multicol} 
\usepackage{graphicx}
\usepackage{caption} 
\usepackage{float}
\usepackage{nomencl}
\usepackage{amsmath}
\usepackage{amsthm}
\usepackage{amssymb}
\usepackage{amsfonts}
\usepackage{bm}
\usepackage{tabularx}
\usepackage{longtable}
\usepackage{algorithm}
\usepackage{algpseudocode}
\usepackage{subcaption}
\usepackage{multirow} 
\usepackage{xcolor}  

\usepackage[utf8]{inputenc}
\usepackage{hyperref}
\hypersetup{
	unicode,
	pdfauthor={Guido Di Federico, Wenchao Teng, Louis J. Durlofsky},
	pdftitle={Data assimilation for subsurface flow using latent diffusion model parameterization: performance of ensemble-Kalman and Monte Carlo techniques},
	pdfsubject={Data assimilation for subsurface flow using latent diffusion model parameterization: performance of ensemble-Kalman and Monte Carlo techniques},
	pdfkeywords={Geological parameterization, diffusion models, latent diffusion, data assimilation, history matching, deep learning},
	pdfproducer={LaTeX},
	pdfcreator={pdflatex}
}

\usepackage[authoryear]{natbib}

\theoremstyle{plain}

\theoremstyle{definition}

\usepackage{graphicx, color}
\graphicspath{{fig/}}

\usepackage{algorithm, algpseudocode} 
\usepackage{mathrsfs} 

\usepackage{lipsum}

\title{Data assimilation for subsurface flow using latent diffusion model parameterization: performance of ensemble-Kalman and Monte Carlo techniques}
\author{Guido Di Federico$^1$ \and Wenchao Teng$^1$ \and Louis J. Durlofsky$^1$}

\date{
	\textit{\small$^1$Department of Energy Science \& Engineering, Stanford University, Stanford, CA, 94305, USA}\\ \small{\texttt{\{gdifede, wenchaot, lou\}@stanford.edu}}\\
}

\begin{document}
	\maketitle
	\hrule
\begin{abstract}
Data assimilation (DA) in subsurface flow entails calibrating model parameters to match observed data, typically available at wells, while preserving geological realism. Latent diffusion models (LDMs) provide efficient mappings from high-dimensional geomodel space to a low-dimensional latent variable, reducing the dimensionality of the inverse problem while maintaining plausibility in posterior models. However, the high nonlinearity in the LDM mapping may degrade the performance of Kalman-gain-based ensemble updates. We present a systematic comparison of DA algorithms applied to large-scale 3D channelized geomodels with hierarchical geological uncertainty. We compare model-space and latent-space DA using the ensemble smoother with multiple data assimilation (ESMDA), and demonstrate a key trade-off: model-space updates achieve significant uncertainty reduction but produce geologically unrealistic posterior models, while latent-space updates preserve realism but exhibit limited uncertainty reduction. Motivated by this, we explore formal Markov chain Monte Carlo (MCMC) and Sequential Monte Carlo (SMC) sampling algorithms in the 3D-LDM latent space. To accommodate their high computational demands, we develop a fast surrogate flow model that approximates well-rate responses. MCMC and SMC are evaluated against ESMDA across three synthetic test cases, with DA performed in LDM latent space. All models maintain geological realism due to the LDM parameterization. MCMC and SMC are consistent with one another and achieve lower data mismatch and more uncertainty reduction than latent-space ESMDA. Our overall results demonstrate that ensemble-Kalman methods may provide overestimated posterior uncertainty with highly nonlinear parameterizations, while formal Monte Carlo sampling, enabled by fast surrogate models, can provide a more reliable alternative.
\end{abstract}

Keywords: Data assimilation, latent diffusion models, geological parameterization, Markov~chain Monte~Carlo, Sequential Monte Carlo, ensemble-Kalman methods

\maketitle

\section{Introduction}
\label{sec:intro}

Data assimilation (DA) involves calibrating model parameters to minimize the mismatch between observed data and simulation predictions. In subsurface flow applications, this process is commonly referred to as history matching. Because of the high degree of prior uncertainty, and the fact that pressure, saturation, and flow-rate measurements are sparse and often noisy, history matching can provide uncertainty reduction but not the ``true'' geomodel. The model parameters of interest may include permeability, porosity, and rock-type (facies) distributions, along with flow functions such as relative permeabilities. Geomodel properties are assigned, block-by-block, to the discretized domain on which the governing flow equations are solved. This may result in, e.g., ${\mathcal O}(10^6)$ parameters to estimate during DA. Directly varying grid-based model parameters may not preserve the complex spatial correlations exhibited by realistic systems, such as channelized or deltaic fan systems characterized by multipoint statistics. In these cases, a mapping between the geomodel and a low-dimensional latent variable can be introduced. DA is then performed in a statistically well-behaved latent space, which involves many fewer parameters than the grid-based model. Performing DA through the manipulation of latent-space variables also acts to preserve geological realism in the posterior models, since the mapping is specifically designed to assure this.

In this work, we perform systematic DA for parameterized geological models of (non-Gaussian) fluvial systems. We consider large 3D geomodels, containing channel, levee and mud facies, that are parameterized using a latent diffusion model (LDM) representation. The flow problem entails immiscible, two-phase flow. While ensemble-Kalman-based algorithms are commonly applied due to their computational efficiency (e.g., 4-20 iterations with $\sim$200-1000 ensemble members, resulting in ${\mathcal O}(10^3-10^4)$ flow simulations), we show that they may exhibit inaccuracy when combined with parameterized geomodels. We therefore present a DA workflow that involves LDM parameterization combined with formal Monte Carlo-based posterior sampling. To enable the large number of forward evaluations required in this framework, we introduce a fast surrogate model for flow simulation. This workflow is applied for a range of cases, allowing us to highlight the differences among methods in terms of uncertainty reduction, data mismatch, and posterior geomodel realism.

Earlier parameterization methods for latent-space DA include wavelet~\citep{Jafarpour_2010} and discrete cosine transform~\citep{Jafarpour,zhao_dct_channels_esmda} methods, as well as principal component analysis (PCA)-based methods~\citep{Vo2015, chen_pca_hm}. Truncated pluri-Gaussian simulation~\citep{liu_enkf_facies, chang_enkf_param_facies} and level-set representations~\citep{chang_enkf_param_facies} are parameterization techniques developed specifically for discrete facies geomodels. More recently, a number of deep learning (DL)-based parameterization methods have been proposed. These DL methods are referred to as generative, in that they can learn complex spatial patterns and generate unseen geological realizations from a latent variable input. In complex geological settings, like fluvial channel or deltaic fan systems, these methods typically outperform the earlier methods, though they do require $\mathcal{O}(10^{3}-10^{4})$ prior realizations for training. 

DL-based methods include those that use convolutional neural networks, CNNs~\citep{LIU2021104676,meng_esmda_conference}, variational autoencoders, VAEs~\citep{canchumuni_esmda_vae, Canchumuni_2019}, and generative adversarial networks, GANs~\citep{Laloy_2018, BAO2020125443, Canchumuni_2020, SONG2023129493}. Diffusion models (DMs) address some of the limitations of these earlier DL methods, which include blurred VAE outputs, unstable GAN training, and the need for a significant amount of hard data in CNN-PCA methods. DMs are able to generate high-quality realizations~\citep{dhariwal2021diffusion}, though they do require increased computational time for generation relative to VAEs and GANs. DMs have been used for both geomodel generation~\citep{LEE2025105750,Ovanger2025} and for DA tasks~\citep{DIFEDERICO2025105755, lin_diffusion_inversion, fu_diffusion_pca}. More broadly, DL-based parameterizations have also been extended to treat multiple geological scenarios, first with 2D or relatively small 3D models~\citep{multiscenario_MohdRazak2020, Song2021_multiscenario, multi_disentangled}, and more recently in larger 3D domains~\citep{DiFederico2025,song_gan_multiscenario}. In addition, conditional generation on both static (e.g., facies indicators, geological trends) and dynamic data (e.g., flow rate, pressure, saturation) has been achieved with DMs, either through supervised conditional training~\citep{Feng2024, zhan,LEE2025105750} or posterior-guided generation~\citep{wang_diffusion_1, MIELE2026106076}. 

Ensemble-Kalman-based methods, such as the ensemble smoother with multiple data assimilation (ESMDA,~\citet{EMERICK20133}) or the ensemble Kalman filter (EnKF,~\citet{enkf_evensen}), would appear to be natural choices for latent-space DA due to their computational efficiency and Gaussian assumptions on prior model parameters. As such, they have been applied with a variety of geological parameterizations. Their performance can degrade, however, when they are used with DL-based parameterizations. This is because these parameterizations introduce an additional, potentially highly nonlinear, mapping between DA variables and model parameters. This increased nonlinearity can amplify entanglement between latent variables and geomodel features, as noted by~\citet{multi_disentangled} for a VAE and~\citet{rongier2025geologicalinferenceprocessbaseddeep} for a GAN. Consistent with this, various investigators have observed issues with posterior sampling accuracy when combining deep learning parameterizations with both ensemble-Kalman and gradient-based methods. The problems identified include inconsistency with the correct Bayesian posterior~\citep{meng_esmda_conference}, limited uncertainty reduction relative to observed data~\citep{DIFEDERICO2025105755, fan2026deep_fracture}, and trapping in local minima in latent space~\citep{laloy_gradient}. Although partial mitigation strategies have been developed~\citep{lopez_nonlinear_generator, fully_conv_vae, multi_disentangled, Merzoug2025}, the formal assessment of posterior sampling accuracy with DL-based parameterizations remains limited in the literature. One reason for this may be that ensemble-Kalman methods classically suffer from uncertainty underestimation due to ensemble collapse. The overestimation of posterior uncertainty, which can occur with LDMs or other DL-based parameterizations, may therefore be taken to suggest adequate performance.

Rigorous posterior sampling methods, such as rejection sampling (RS) or MCMC, target the exact Bayesian posterior without Gaussian or model linearity assumptions. However, they may require ${\mathcal O}(10^6)$ or more iterations. The high computational cost of forward simulations and the number of parameters to calibrate motivate the use of surrogate flow models. Trained on a dataset of full-physics simulation runs, surrogate models approximate reservoir behavior for new geomodels at greatly reduced computational cost. Earlier studies primarily focused on reduced-order models~\citep{SAYARPOUR201196, HE201354} and physics-constrained neural networks~\citep{WANG2022111419, tartakovsky_pinn}. More recently, data-driven neural network surrogates have gained significant attention. The literature on subsurface flow surrogates for history matching is extensive, with architectures ranging from recurrent (RNN) and convolutional neural networks~\citep{TANG2020109456, mo_dcae, han_surrogate_mcmc} to Fourier neural operators (FNO,~\citet{wen_ufno_surr}), diffusion models~\citep{wang_diffusion_2}, and transformers~\citep{han2025recurrenttransformerunetsurrogate}. We reiterate that the full-physics simulator is still required to generate training data (and potentially for post-history-matching validation), while the surrogate is used for function evaluations within the DA workflow.

In this work, we address the key issues raised in the above discussion. We first provide a systematic assessment of data assimilation with ESMDA for both full (model-space) and LDM-parameterized geomodels. For model-space ESMDA, we demonstrate uncertainty reduction consistent with the observed data, but degraded geological realism in the posterior models. Using latent-space DA, we observe the opposite behavior, that is, geologically realistic posterior models but limited uncertainty reduction relative to observed data. \textcolor{black}{This behavior has been observed to some extent in previous studies (as discussed above), but it has not been evaluated in great detail.}

\textcolor{black}{Motivated by this discrepancy, we develop a fast and accurate flow surrogate to enable data assimilation based on formal MCMC and Sequential Monte Carlo (SMC) algorithms operating on latent-space (LDM) variables. This flow surrogate is a new, fit-for-purpose procedure specially developed to provide the flow-rate quantities required for DA. 
Monte Carlo algorithms have previously been used with a range of parameterizations for subsurface flow inversion. In particular, MCMC was applied with VAE~\citep{LALOY2017387} and GAN~\citep{Laloy_2018} latent spaces, while SMC has been combined with linear~\citep{beskos2015smc} or geometric channel parameterizations~\citep{ruchi2019transform}, where it was shown to outperform ensemble-Kalman inversion.
However, to the best of our knowledge, neither MCMC nor SMC has been applied with LDM parameterizations for subsurface flow inversion in high-dimensional 3D problems. We show that these approaches retain geological realism while providing appropriate (relative to observed data) and consistent estimates of posterior uncertainty. This is demonstrated across multiple synthetic true models and evaluation metrics. The systematic and detailed algorithmic comparisons, performed over multiple DA methods, are another novel feature of this work. 
Taken in total, our findings highlight the potential limitations of using DL-based parameterizations with standard ensemble-Kalman methods, and the improvements that can be achieved with formal sampling enabled through use of flow surrogates.}

This paper proceeds as follows. In Section~\ref{sec:geomodels_setup}, we describe the geomodels, including the uncertain scenario parameters, and the flow simulation problem. In Section~\ref{sec:3d_ldm_esmda} we review the 3D-LDM model and explain the data assimilation setup. Results for both model-space and latent-space ESMDA, for a synthetic true model, are provided in Section~\ref{sec:esmda_results}. In Section~\ref{sec:surr_and_mc} we present our new surrogate flow model, and we briefly describe the MCMC and SMC algorithms. Comparative results across all methods, for three different true models using LDM-based parameterizations, are provided in~Section~\ref{sec:hm_results}. In~Section~\ref{sec:conclusions} we summarize this work and provide directions for future research. 

\section{Geomodels and flow simulation setup}
\label{sec:geomodels_setup}

In this section, we describe the geomodels considered in this work.
We then present the flow problem and simulation specifications. The geomodels and simulation setup are the same as those described in~\citet{DiFederico2025}.

The geomodels are intended to represent channelized systems containing channel, levee, and mud facies. These facies are identified by cell values of 1, 0.5, and 0, respectively. Realizations are generated through object-based modeling (OBM) using Petrel~\citep{Schlumberger}. Each realization, denoted as $\mathbf{m}_0 \in \mathbb{R}^{N_c}$, contains $N_c = N_x \times N_y \times N_z$ = $128 \times 128 \times 32$ = 524,288~grid blocks. The geomodels are conditioned to either channel or mud facies at $N_w = 9$ wells ($N_{\text{inj}}=3$ injectors and $N_{\text{prod}}=6$ producers) along the full vertical thickness, resulting in $N_h = 288$ conditioning locations. 

The channelized models are characterized in terms of three uncertain geological scenario parameters: mud fraction $f_m \in [0.72, 0.87]$, average channel orientation $\theta_{\rm ch} \in [30^\circ, 60^\circ]$, and average channel width $w_{\rm ch} \in [4, 6]$~grid blocks. In generating realizations within Petrel, these parameters are sampled from uniform distributions over their specified ranges, with a tolerance of $\pm 10\%$ in the OBM procedure. \textcolor{black}{ The channel depth, wavelength, and amplitude parameters do not change across realizations. These are prescribed to have values of 19, 25, and 15~grid blocks, respectively, with tolerances of $\pm 15\%$ in the OBM procedure. The ranges for $f_m$, $\theta_{\rm ch}$ and $w_{\rm ch}$, which are similar to those used by~\citet{Song_Mukerji_Hou_2021, Song2021_multiscenario}, were chosen to reflect the level of uncertainty that might be expected in realistic settings, where some amount of prior information on depositional setting and channel geometry is available. The bounded uniform distributions reflect weakly informative priors. If simulation results over these ranges are in clear disagreement with observed data, modification of the prior would be required before proceeding to DA. Our assumption here is that this prior falsification step has already been performed.} 

The resulting geomodels exhibit a hierarchical uncertainty structure, as they are characterized by uncertain higher-level scenario parameters, with each set of scenario parameters corresponding to an (essentially) infinite number of geobody realizations. The channel bodies have a truly 3D structure -- they are not simply extruded extensions of planar features. More specifically, individual channels exhibit decreasing width as a function of depth. Channels may be confined to specific depth intervals, with some present only in the upper portion and some in the lower portion of the model.
 
\begin{figure}[htbp]
    \centering
    \captionsetup[subfigure]{font=footnotesize}
    \hfill
    \begin{subfigure}[b]{0.32\textwidth}
        \centering
        \includegraphics[width=\textwidth, trim=50 50 50 50, clip]{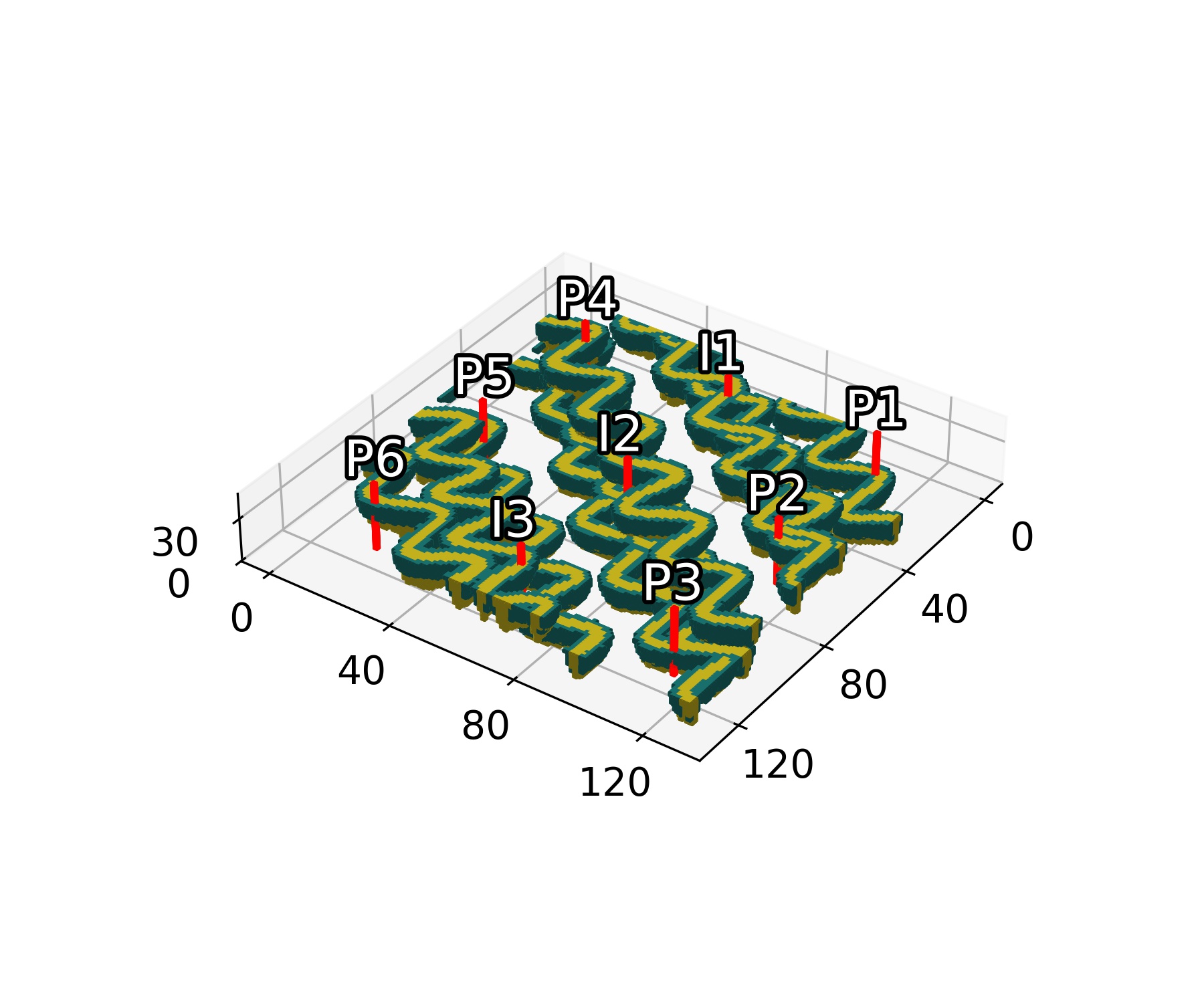}
        \caption{$f_m{=}0.79$, $\theta_{\rm ch}{=}38^\circ$, $w_{\rm ch}{=}5.3$}
    \end{subfigure}
    \hfill
    \begin{subfigure}[b]{0.32\textwidth}
        \centering
        \includegraphics[width=\textwidth, trim=50 50 50 50, clip]{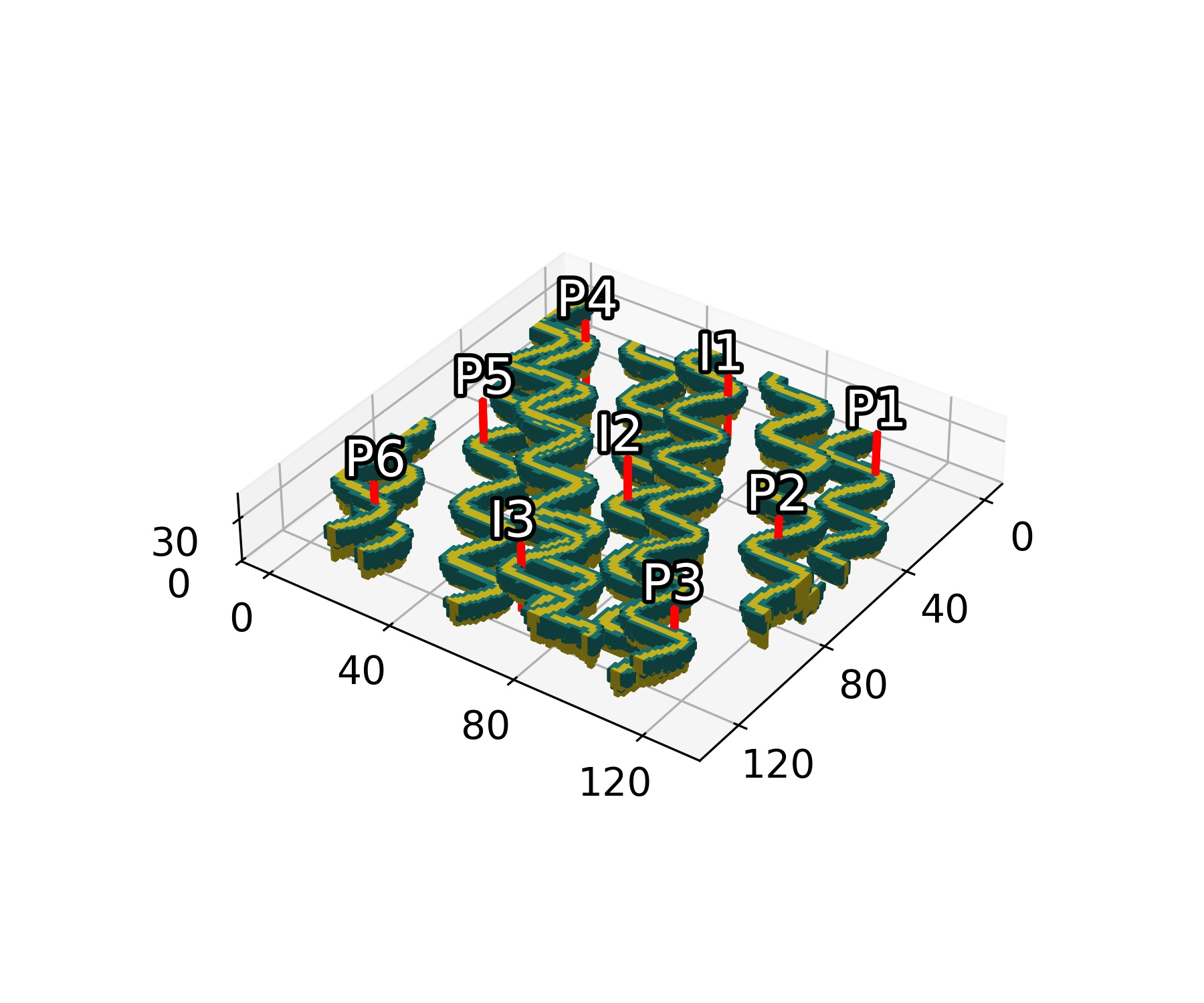}
        \caption{$f_m{=}0.84$, $\theta_{\rm ch}{=}55^\circ$, $w_{\rm ch}{=}4.2$}
    \end{subfigure}
    \hfill
    \begin{subfigure}[b]{0.32\textwidth}
        \centering
        \includegraphics[width=\textwidth, trim=50 50 50 50, clip]{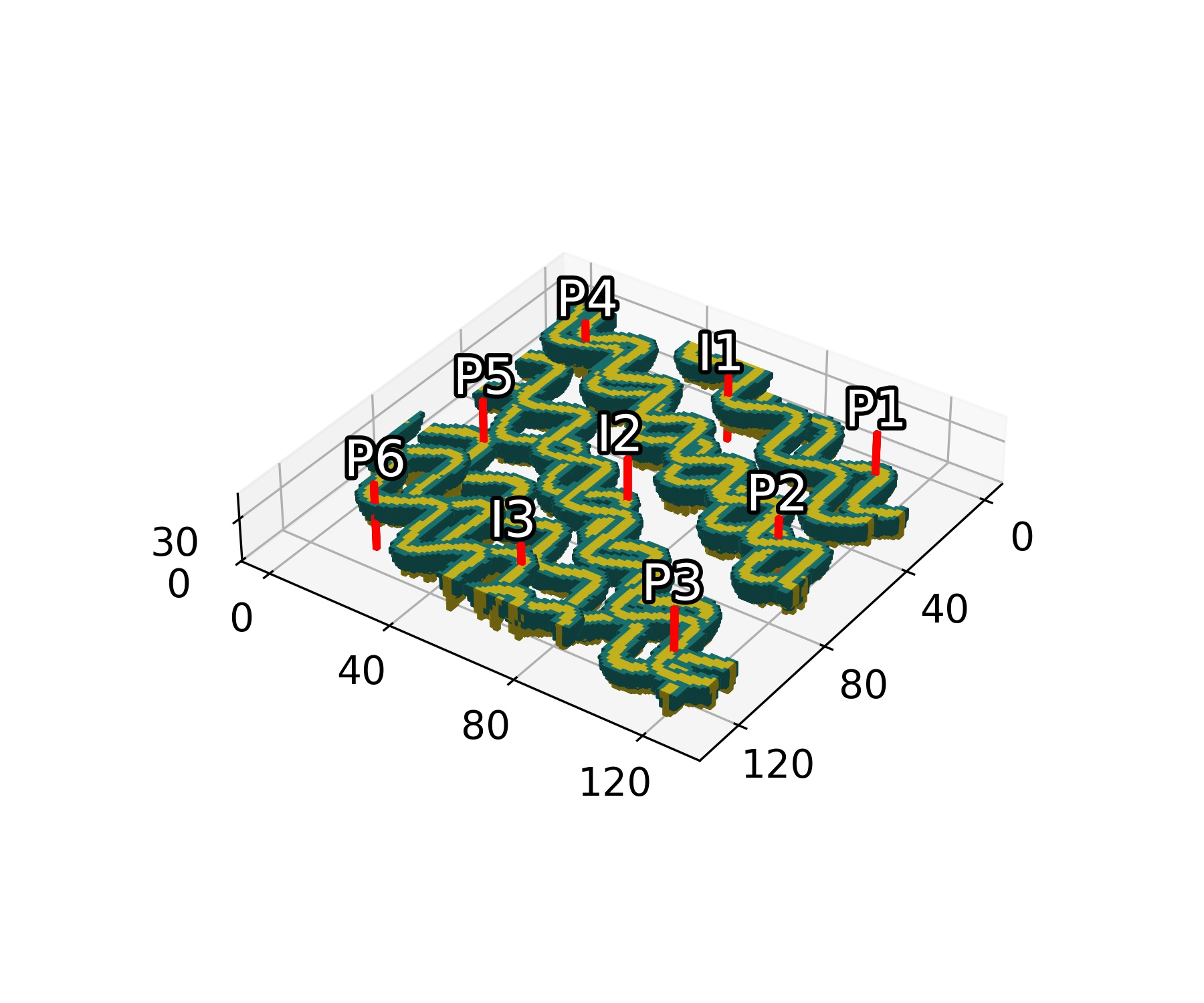}
        \caption{$f_m{=}0.81$, $\theta_{\rm ch}{=}36^\circ$, $w_{\rm ch}{=}4.6$}
    \end{subfigure}
\vfill
    \begin{subfigure}[b]{0.32\textwidth}
        \centering
        \includegraphics[width=\textwidth, trim=50 50 50 50, clip]{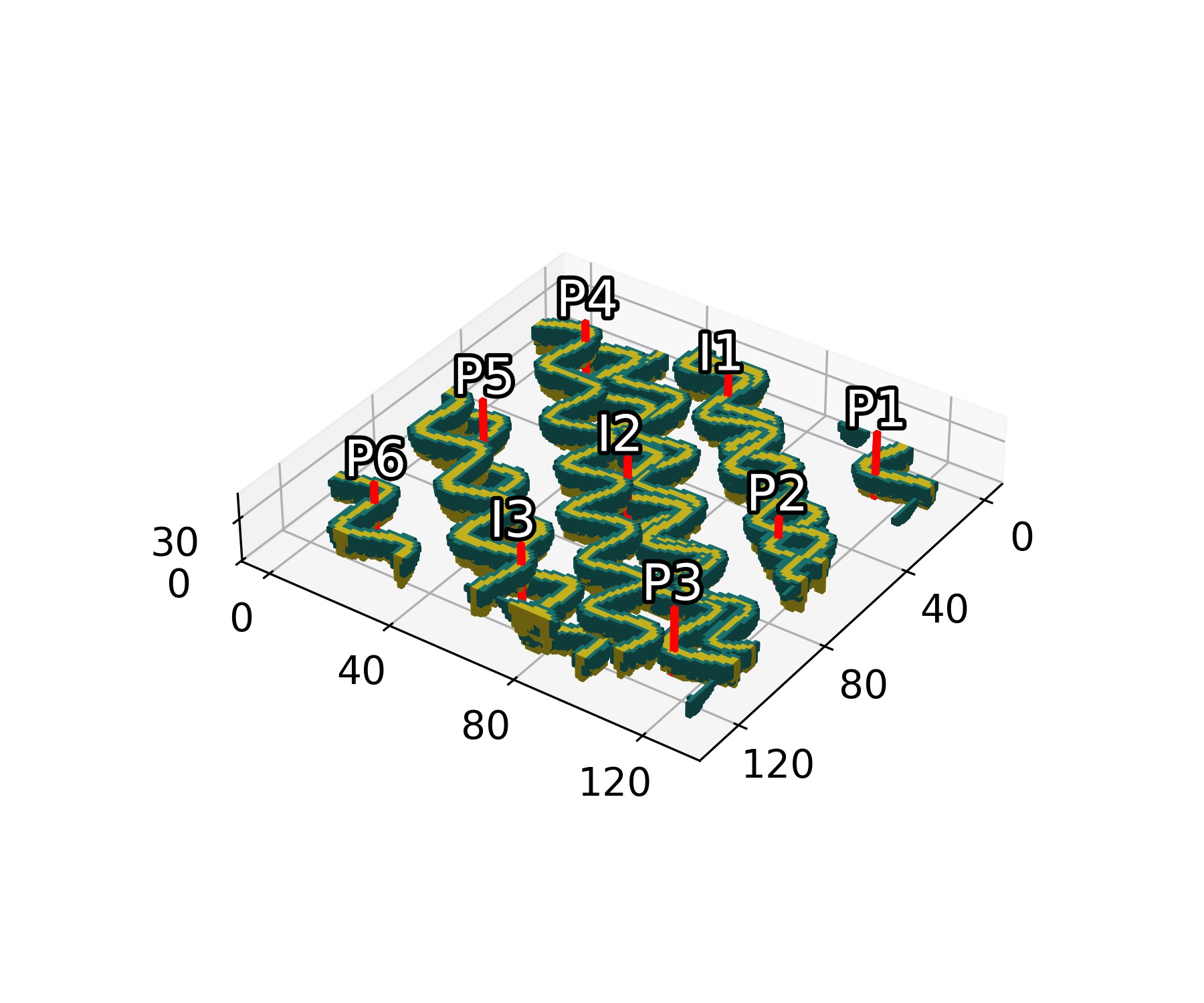}
        \caption{$f_m{=}0.85$, $\theta_{\rm ch}{=}45^\circ$, $w_{\rm ch}{=}4.0$}
    \end{subfigure}
    \hfill
    \begin{subfigure}[b]{0.32\textwidth}
        \centering
        \includegraphics[width=\textwidth, trim=50 50 50 50, clip]{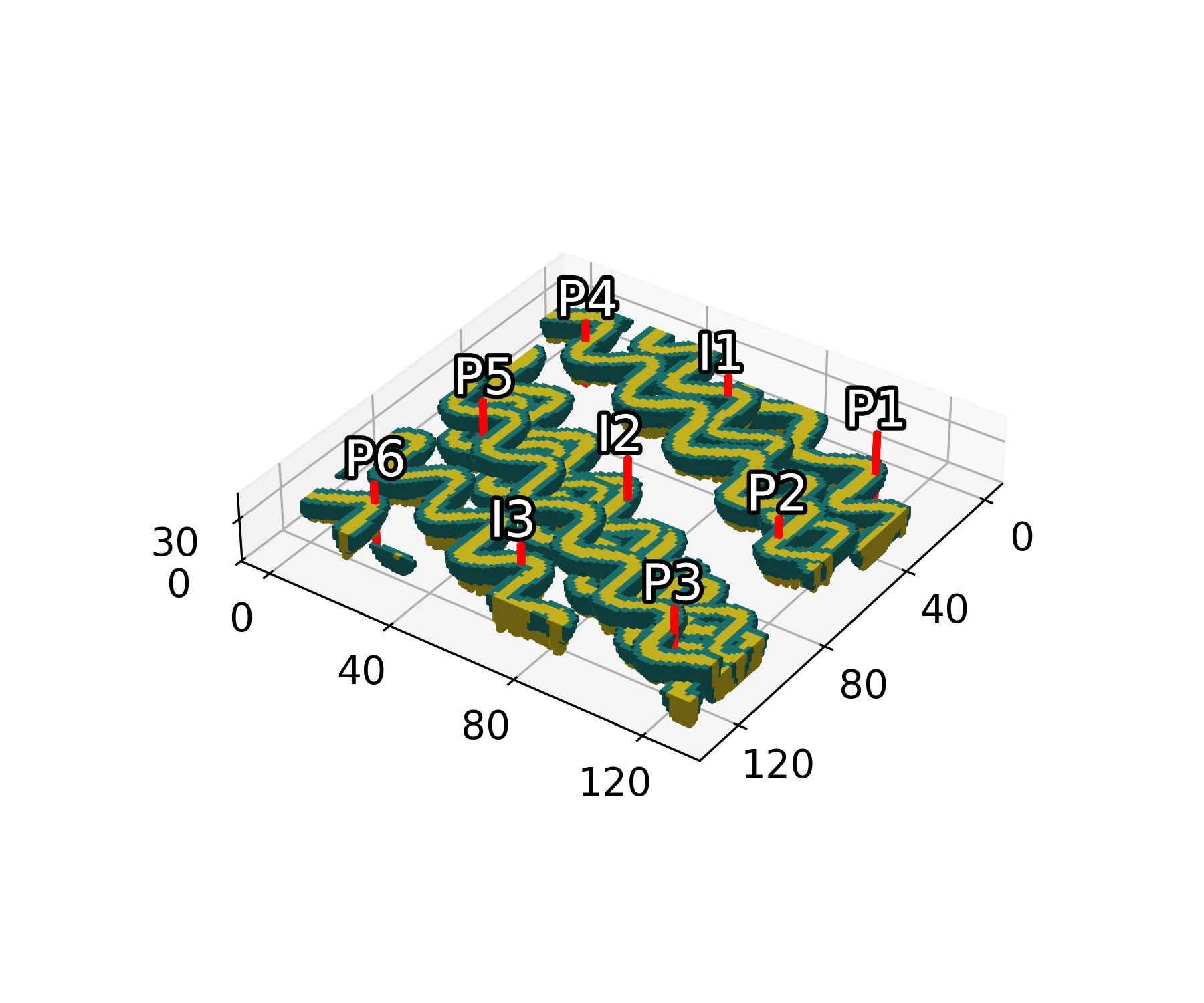}
        \caption{$f_m{=}0.78$, $\theta_{\rm ch}{=}33^\circ$, $w_{\rm ch}{=}5.5$}
    \end{subfigure}
    \hfill
    \begin{subfigure}[b]{0.32\textwidth}
        \centering
        \includegraphics[width=\textwidth, trim=50 50 50 50, clip]{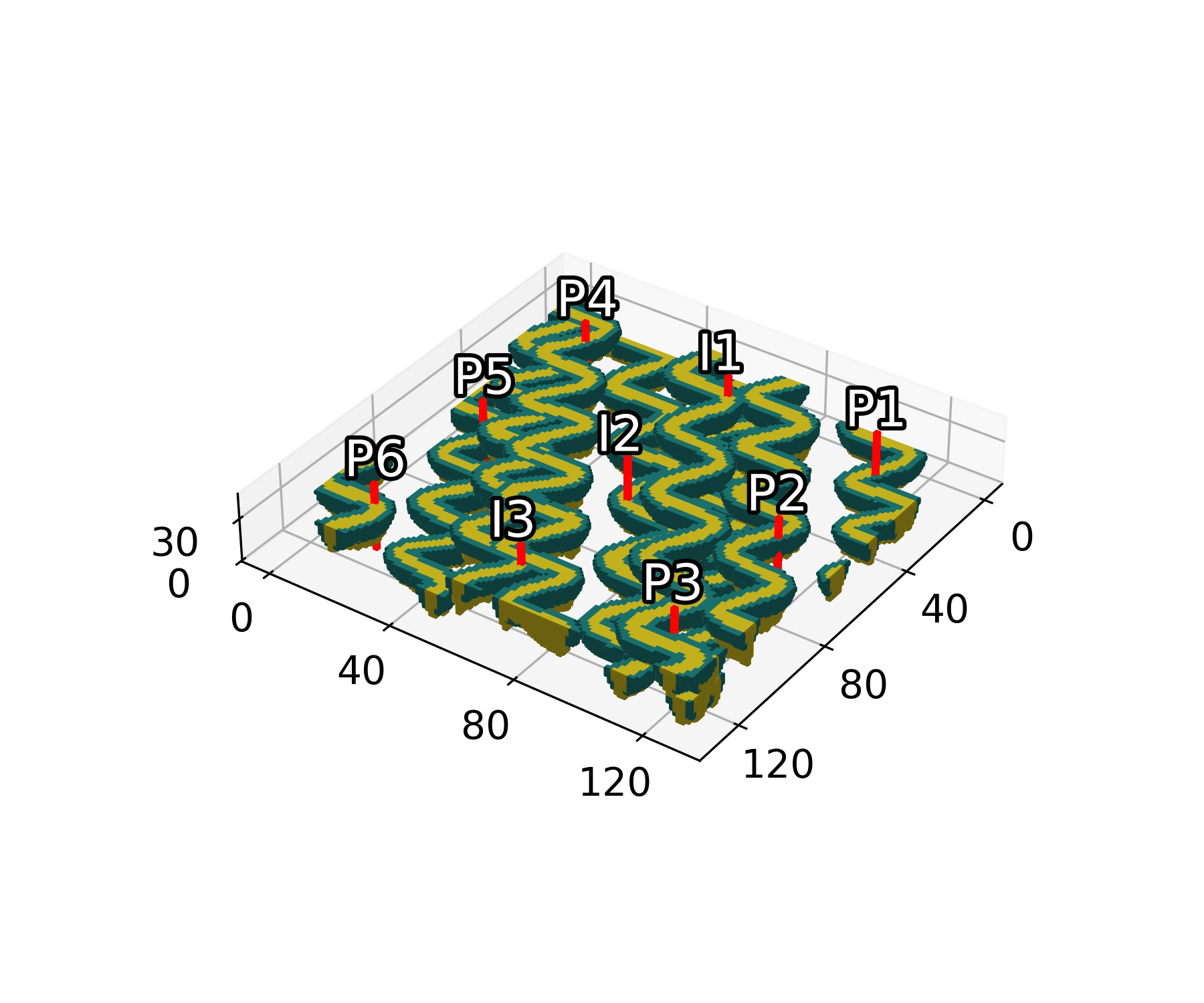}
        \caption{$f_m{=}0.75$, $\theta_{\rm ch}{=}58^\circ$, $w_{\rm ch}{=}5.9$}
    \end{subfigure}
    \caption{Example Petrel realizations and associated scenario parameters ($f_m$ is mud fraction, $\theta_{\rm ch}$ is channel orientation, $w_{\rm ch}$ is channel width in grid cells). Injection (I1--I3) and production (P1--P6) wells, which honor hard data, are also shown. Realizations in (d), (e), (f) are used later as true models (in Cases~1--3, respectively) for DA.}
    \label{fig:examples_petrel}
\end{figure}

Sample realizations and their associated scenario parameters are shown in Figure~\ref{fig:examples_petrel}. In these images, the mud (background) facies has been omitted for better visualization. Differences in channel orientation ($\theta_{\rm ch}$) are quite noticeable from realization to realization. The variability in $f_m$ and $w_{\rm ch}$, while impactful for flow, is more difficult to discern visually. The models on the bottom row will be used as the synthetic ``true'' models for DA (later denoted as Cases~1--3). These models are not included in the training set (the LDM architecture and training are described in Section~\ref{sec:3d_ldm_esmda}) and are characterized by a diverse set of scenario parameters. 

The simulations involve two-phase immiscible flow \textcolor{black}{(with capillary pressure neglected)} driven by three injection wells and six production wells in a line-drive configuration. These wells, denoted I1--I3 and P1--P6, are shown in Figure~\ref{fig:examples_petrel}. This setup may represent a waterflood for oil production or groundwater remediation involving an immiscible contaminant. We will refer to the two phases as oil and water. Grid blocks are of physical dimensions 65~ft~$\times$~65~ft~$\times$~6.5~ft (19.8~m~$\times$~19.8~m~$\times$~1.98~m).  Porosity ($\phi$) and permeability ($k$) vary by facies, though they are taken to be constant within each rock type. We specify $k_x = k_y$ and $k_z = 0.2k_x$ for all facies, where $k_x$, $k_y$ and $k_z$ denote directional permeabilities. The specific ($\phi,~k_x$) values are (0.2, 1000~mD) for channel, (0.15, 200~mD) for levee, and (0.05, 25~mD) for mud. 

The initial reservoir pressure, at a depth of 4160~ft (1268~m), is 4500~psi (31.03~MPa). \textcolor{black}{No-flow conditions are specified at all boundaries. Properties at $14.7\,\text{psi}$ (0.10135~MPa) are as follows. The oil phase is modeled as a ``dead'' oil with a formation volume factor of 1.0, constant compressibility of $1.7 \times 10^{-5}\,\text{psi}^{-1}$, and viscosity of 5~cp. Water also has a formation volume factor of 1.0. Its compressibility and viscosity are $5.0 \times 10^{-7}\,\text{psi}^{-1}$ and 1~cp. Surface densities are $40.0\,\text{lbm/ft}^3$ for oil and $62.24\,\text{lbm/ft}^3$ for water. Rock compressibility is neglected. Oil--water relative permeabilities ($k_{rw}$ and $k_{ro}$) are given by $k_{rw} = (S_w)^{1.5}$ and $k_{ro} = (1 - S_w)^{1.5}$, where $S_w$ denotes water saturation (connate water and residual oil saturation are both set to zero).} 


\textcolor{black}{Wells are perforated only in channel facies. Based on the geomodel conditioning, this means wells I1, I3, P2, P4 and P6 are perforated in the upper portion of the model (layers 1--16), and wells I2, P1, P3 and P5 are perforated in the lower portion of the model (layers 16--32). All wells are bottom-hole pressure (BHP) controlled, with injectors operating at 4700~psi (32.41~MPa) and producers at 4300~psi (29.65~MPa). The simulation time frame is 1500~days, and the maximum time step in any run is 50~days (there are $N_t = 30$ recorded time steps in each run).}

All simulations are performed using tNavigator~\citep{rock_flow_dynamics}. Each run requires approximately 10~minutes on 16~cores for an AMD EPYC CPU. For additional details on the geomodeling and flow simulation setup, please refer to~\citet{DiFederico2025}. \textcolor{black}{See the ``Data and code availability'' section at the end of this paper to access the tNavigator simulation and fluid property files.}

\section{LDM parameterization and data assimilation setup}
\label{sec:3d_ldm_esmda}

In this section we describe the parameterization procedure used in this work along with the basic setup for the DA problem.

\subsection{3D-LDM parameterization} 
The LDM geological parameterization used here was first proposed for 2D systems~\citep{DIFEDERICO2025105755} and was extended to 3D systems with variable geological scenario parameters by~\citet{DiFederico2025}. The method is based on the original LDM formulation by~\citet{rombach2022highresolution}. These authors extended standard generative diffusion models~\citep{ho2020denoising}, which operate directly in the pixel (or voxel) space, to include a low-dimensional latent encoding. \textcolor{black}{The dataset used in this study consists of 3000 realizations of the fluvial channel system described in Section~\ref{sec:geomodels_setup}. We use an 80-10-10\% split for training, validation, and testing, respectively.} These realizations represent our prior knowledge of the distribution of subsurface properties before DA is applied.

Diffusion models are based on two processes, each performed over $T$ discrete steps -- a forward noising stage, in which Gaussian perturbations are progressively added to a data sample, and a learned denoising stage (typically by a U-net,~\citep{unet}), through which noise is progressively removed. In this way, new samples from the dataset distribution can be generated by denoising input Gaussian noise. In the 3D-LDM, both stages operate in a compressed latent space. A VAE~\citep{kingma2022autoencodingvariationalbayes} is used to perform the mapping between the original data and its encoded representation.

The VAE comprises an encoder $\mathcal{E}: \mathbb{R}^{N_x \times N_y \times N_z} \rightarrow \mathbb{R}^{n_x \times n_y \times n_z}$ and a decoder $\mathcal{D}: \mathbb{R}^{n_x \times n_y \times n_z} \rightarrow \mathbb{R}^{N_x \times N_y \times N_z}$ where $n_d = N_d/f$ for $d \in \{x,y,z\}$, and $f$ is the downsampling ratio. The overall compression is thus $f^3$. Note that in the above both the geomodel and LDM latent space are 3D fields. Because they are treated as flattened vectors within the history matching workflow, we will express them as $\mathbf{m}_0 \in \mathbb{R}^{N_c}$ and $\boldsymbol{\xi}_0 \in \mathbb{R}^{n_c}$, with $n_c = n_x \times n_y \times n_z$. The encoder outputs the parameters of a Gaussian distribution, from which the latent variable is sampled via reparameterization as $\boldsymbol{\xi}_0 = \mathcal{E}_\mu(\mathbf{m}_0) + \mathcal{E}_\sigma(\mathbf{m}_0) \odot \boldsymbol{\epsilon}$, where $\boldsymbol{\epsilon} \sim \mathcal{N}(\mathbf{0}, \mathbf{I}_{n_c})$. For simplicity, we denote this operation as $\boldsymbol{\xi}_0 = \mathcal{E}(\mathbf{m}_0)$. 

The VAE is trained by minimizing the loss
\begin{equation}
    L_{\text{VAE}}(\theta) = L_{\text{recon}} + \lambda_{\text{KL}} L_{\text{KL}} + \lambda_{\text{h}} L_{\text{h}} + \lambda_{\text{perc}} L_{\text{perc}}.
    \label{eqn:loss_vae}
\end{equation}
In the above, $L_{\text{recon}} = \|\mathbf{m}_0 - \mathcal{D}(\mathcal{E}(\mathbf{m}_0))\|_2^2$ is the pixel-wise reconstruction loss between the original geomodel and the decoded latent representation. The loss term $L_{\text{KL}} = D_{\text{KL}}(\mathcal{N}(\mathcal{E}_\mu(\mathbf{m}_0),\mathcal{E}_\sigma^2(\mathbf{m}_0)) || \mathcal{N}(\mathbf{0}, \mathbf{I}_{n_c}))$ is the Kullback-Leibler divergence, which regularizes the latent space towards a multi-Gaussian distribution. The term $L_{\text{h}} = \frac{1}{N_{\text{h}}} ||H(\mathbf{m}_0 - \mathcal{D}(\mathcal{E}(\mathbf{m}_0)) ||_2^2$ enforces conditioning at hard-data locations, selected from the full geomodel by the binary selection matrix $H$. Finally, $L_{\text{perc}}$ denotes a perceptual loss that compares deep feature representations of input and reconstructed geomodels, as extracted by a pretrained network (see~\citep{DiFederico2025} for additional details). In our implementation, the loss weights are set to $\lambda_{\text{KL}} = 10^{-6}$, $\lambda_{\text{h}} = 10^{-2}$, and $\lambda_{\text{perc}} = 10^{-3}$. A downsampling ratio of $f = 8$ (total compression factor $f^3 = 512$) is used, such that $n_c=1024$.

Once the VAE is trained and its weights are frozen, a U-net is trained to perform denoising in the latent space. In the forward (noising) stage, Gaussian noise is added to a ``clean'' latent variable $\boldsymbol{\xi}_0 = \mathcal{E}(\mathbf{m}_0)$, up to a randomly selected step $t$, to obtain the noisy latent variable $ \boldsymbol{\xi}_t$ as
\begin{equation}
    \boldsymbol{\xi}_t = \sqrt{\bar{\alpha}_t}\,\boldsymbol{\xi}_0 + \sqrt{1 - \bar{\alpha}_t}\,\boldsymbol{\epsilon}_t, \quad \boldsymbol{\epsilon}_t \sim \mathcal{N}(\mathbf{0}, \mathbf{I}_{n_c}).
    \label{eqn:noising}
\end{equation}
The coefficients $\alpha_s$ for $\bar{\alpha}_t = \prod_{s=1}^{t} \alpha_s$ are defined by a scheduler function of $t$, which typically decreases linearly from $\alpha_1$ to $\alpha_T$. Eq.~\ref{eqn:noising} and the $\bar{\alpha}_t$ parameters act in a manner that, at step $T$, the latent variable is indistinguishable from pure Gaussian noise. In the reverse (denoising) stage, the U-net learns to predict the added noise $\boldsymbol{\epsilon}_t$ using Eq.~\ref{eqn:noising} and the noisy latent variable $\boldsymbol{\xi}_t$. It is trained by minimizing
\begin{equation}
    L_{\text{U-net}}(\theta) = \|\boldsymbol{\epsilon}_t - \boldsymbol{\epsilon}_\theta(\boldsymbol{\xi}_t, t)\|_2^2,
    \label{eqn:loss_unet}
\end{equation}
where $\boldsymbol{\epsilon}_\theta(\boldsymbol{\xi}_t, t)$ is the output of the U-net. 

We adopt a linear scheduler, with $T = 20$ steps and coefficients $\alpha_1 = 0.9985$ and $\alpha_T = 0.9805$. Full architectural and training details are provided in~\citet{DiFederico2025}. 

Since $\boldsymbol{\xi}_T$ is approximately Gaussian, a new realization can be generated by sampling $\boldsymbol{\xi}_T \sim \mathcal{N}(\mathbf{0}, \mathbf{I}_{n_c})$ and reversing the forward noising process in Eq.~\ref{eqn:noising}. Using the predicted noise $\boldsymbol{\epsilon}_\theta(\boldsymbol{\xi}_t, t)$ from the U-net, $\boldsymbol{\xi}_T$ is iteratively denoised as
\begin{equation}
    \boldsymbol{\xi}_{t-1} = \sqrt{\bar{\alpha}_{t-1}} \left(\frac{\boldsymbol{\xi}_t - \sqrt{1 - \bar{\alpha}_t}\, \boldsymbol{\epsilon}_\theta(\boldsymbol{\xi}_t, t)}{\sqrt{\bar{\alpha}_t}}\right) + \sqrt{1 - \bar{\alpha}_{t-1}}\, \boldsymbol{\epsilon}_\theta(\boldsymbol{\xi}_t, t), \quad t = T, \ldots, 1,
    \label{eqn:denoising}
\end{equation}
until a fully denoised latent variable $\boldsymbol{\xi}_0$ is obtained. Eq.~\ref{eqn:denoising} refers specifically to the denoising diffusion implicit model (DDIM), proposed by~\citet{song2021denoising}. The final geomodel realization is recovered by decoding through the VAE, $\mathbf{m}_0^{\text{LDM}} = \mathcal{D}(\boldsymbol{\xi}_0)$. We note that the overall mapping $\mathbf{m}_0^{\text{LDM}}(\boldsymbol{\xi}_{T})$ is highly nonlinear but it is deterministic, which is essential for our DA procedures. 

\begin{figure}[htbp]
    \centering
    \captionsetup[subfigure]{font=footnotesize}
    \begin{subfigure}[b]{0.32\textwidth}
        \centering
        \includegraphics[width=\textwidth, trim=50 50 50 50, clip]{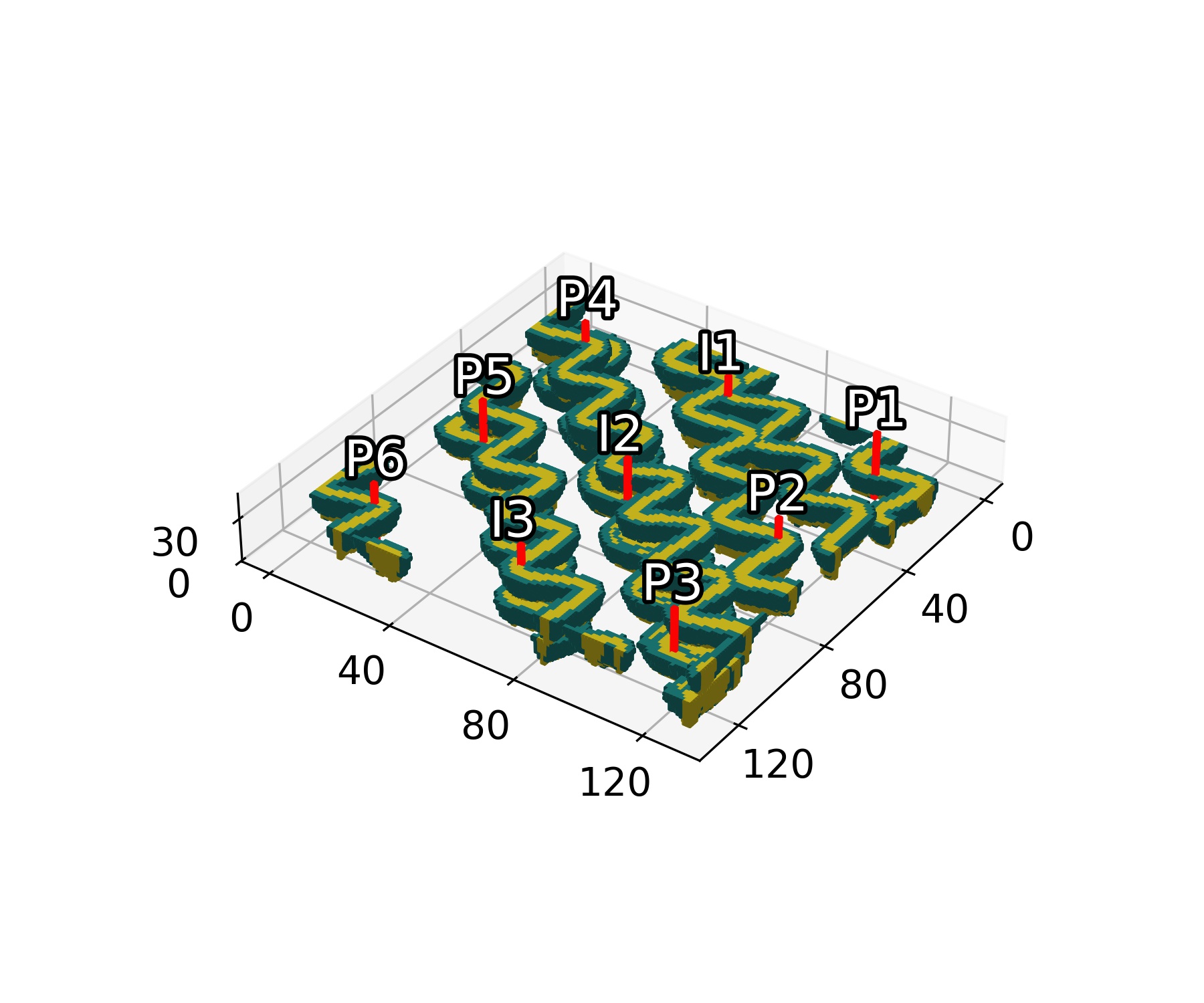}
        \caption{$f_m{=}0.77$, $\theta_{\rm ch}{=}42^\circ$, $w_{\rm ch}{=}5.5$}
    \end{subfigure}
    \hfill
    \begin{subfigure}[b]{0.32\textwidth}
        \centering
        \includegraphics[width=\textwidth, trim=50 50 50 50, clip]{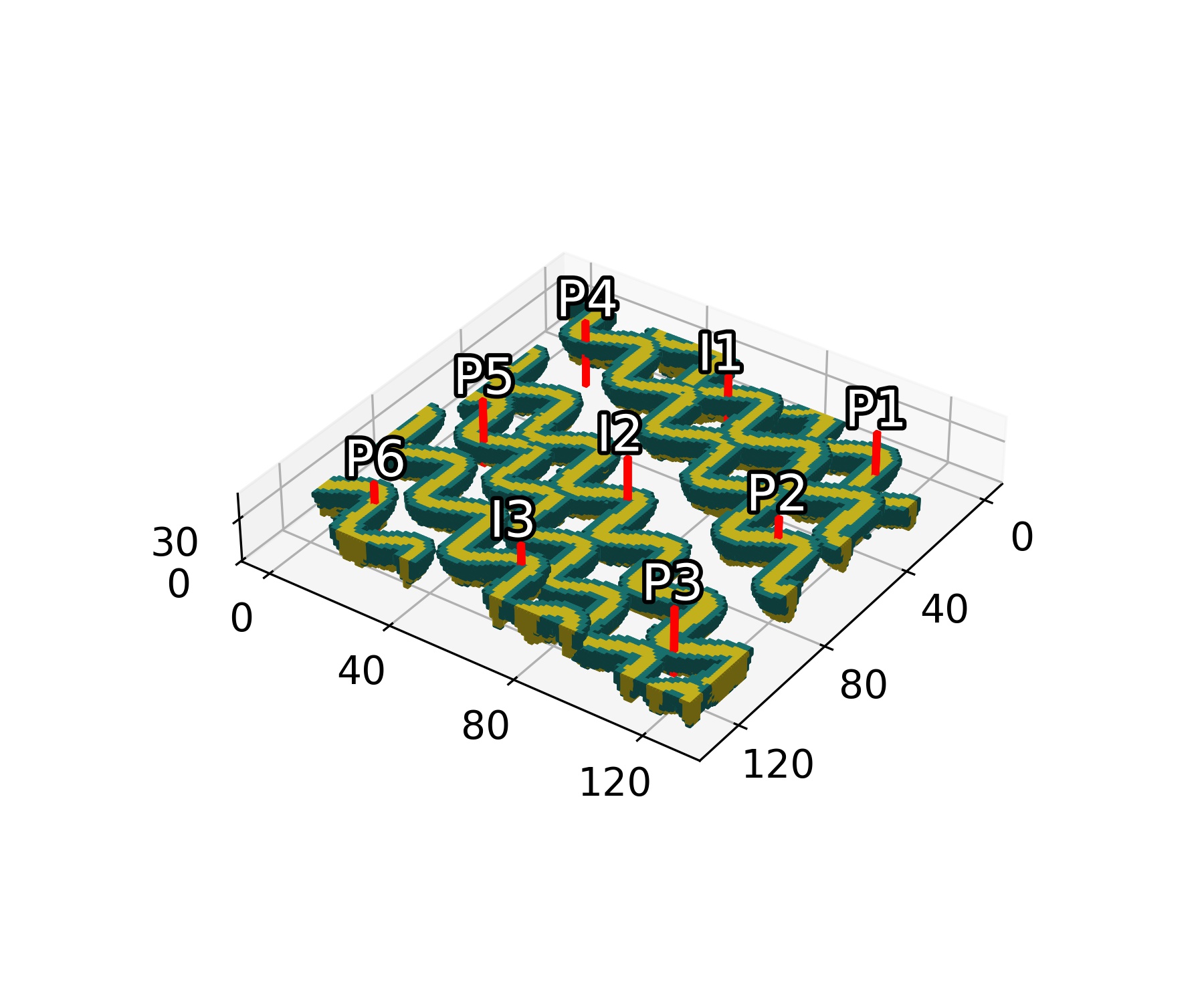}
        \caption{$f_m{=}0.80$, $\theta_{\rm ch}{=}33^\circ$, $w_{\rm ch}{=}5.8$}
    \end{subfigure}
    \hfill
    \begin{subfigure}[b]{0.32\textwidth}
        \centering
        \includegraphics[width=\textwidth, trim=50 50 50 50, clip]{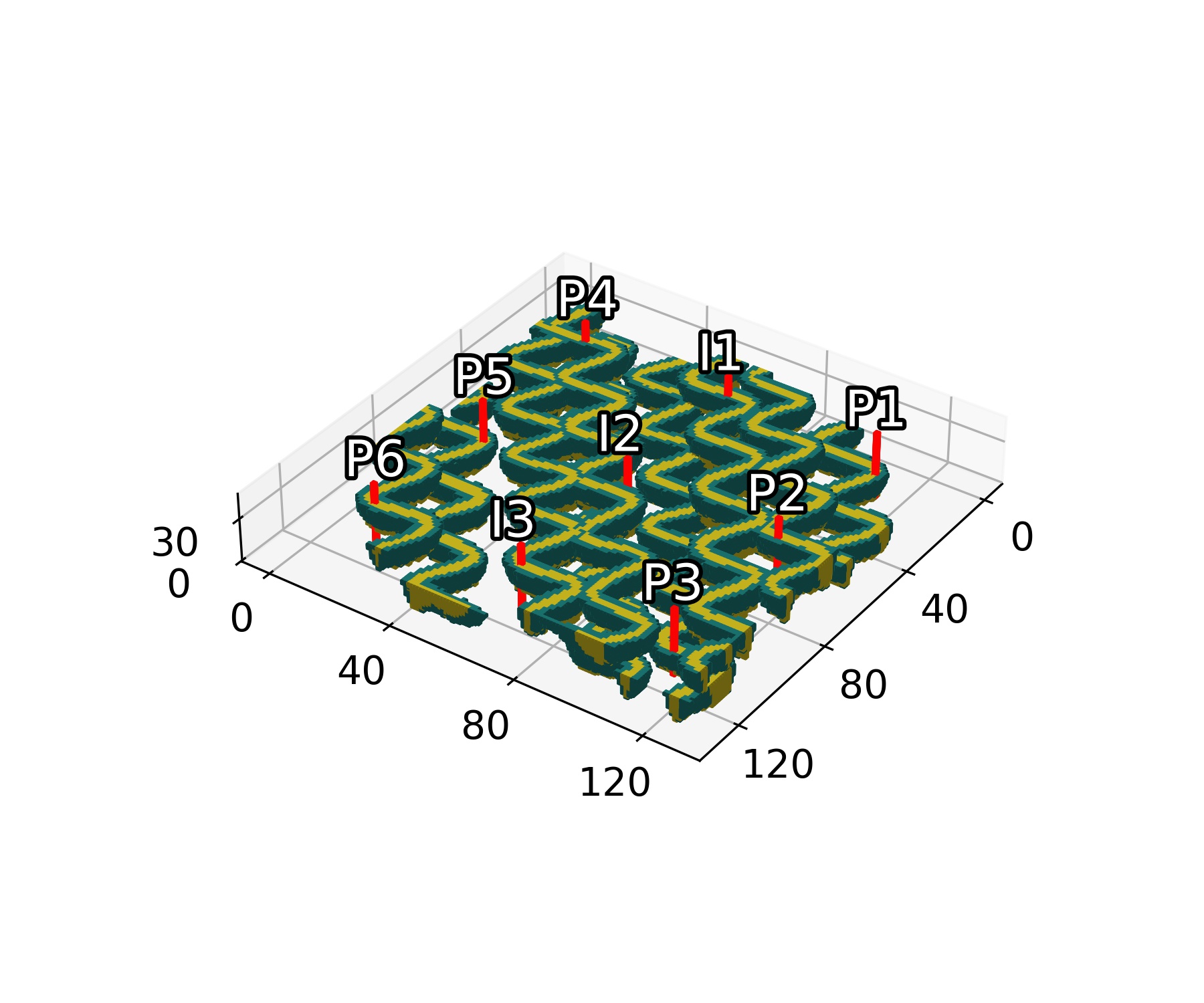}
        \caption{$f_m{=}0.83$, $\theta_{\rm ch}{=}60^\circ$, $w_{\rm ch}{=}4.3$}
    \end{subfigure}
    \caption{Example 3D-LDM generated realizations and associated scenario parameters.} 
    \label{fig:examples_ldm}
\end{figure}

In~\citet{DiFederico2025}, we observed high degrees of consistency between ensembles of reference Petrel geomodels and 3D-LDM-generated models. These assessments were both qualitative (e.g., channel shape, continuity, facies ordering) and quantitative (e.g., geostatistical metrics and flow simulation results). Examples of random 3D-LDM realizations are presented in Figure~\ref{fig:examples_ldm}, along with the corresponding scenario parameters. The mud fraction can be directly computed as the fraction of 0-facies grid blocks. The average channel orientation and width are determined from the 3D-LDM geomodel through application of a lightweight CNN, trained on the Petrel dataset and input OBM labels, as described in~\citet{DiFederico2025}. We observed about 98\% accuracy in hard data conditioning in the generated realizations. 

In total, our previous results clearly demonstrate that the 3D-LDM parameterization can accurately sample from the (prior) distribution represented by the Petrel dataset, indicating it can be effectively used in our DA framework. 
We reiterate that scenario parameters are encoded jointly with the geobody distribution in our 3D-LDM. This approach differs from some of the methods described in Section~\ref{sec:intro}, which utilize additional encoded inputs to condition or guide geomodel generation. Although the geology in this study is hierarchical (and hierarchically uncertain), with our approach we only need a single level of parameterization. This represents a key strength of the 3D-LDM formulation.

\textcolor{black}{It is also important to note that the cases considered in this study share the same general geological features. In other words, the geological ``style'' and the scenario parameter ranges are assumed to be known. The well configuration, which provides conditioning data, is also fixed. If the true geology lies outside the 3D-LDM training set distribution, it cannot be represented by the existing parameterization. In that case, a wider training set and retraining (or the incorporation of model-error treatments) would be required. If new wells are drilled and additional conditioning data become available, the 3D-LDM would require retraining with (new) conditioned OBM geomodels. The computational effort for this retraining step is relatively small compared to the overall computational requirements for the DA workflow.}

\subsection{Data assimilation setup}

We now describe the DA problem considered in this study. In practical settings, observed data derive from field measurements. Here, these data, denoted $\mathbf{d}_{\text{true}}\in \mathbb{R}^{N_d}$, where $N_d$ is the number of observations, are obtained by running the flow simulator on a synthetic true model. The true models, which are part of the testing dataset (not seen during training), are shown in Figure~\ref{fig:examples_petrel}d,e,f. To approximate measurement error, we perturb $\mathbf{d}_{\text{true}}$ with Gaussian noise of zero-mean and standard deviation of 5\% of the corresponding simulated value. \textcolor{black}{No temporal or spatial correlation is introduced, although this effect could be included directly in $C_d$, without any other changes to the workflow, if a means for estimating these correlations is available}. It follows that $\mathbf{d}^{\ast}_{\text{obs}} = \mathbf{d}_{\text{true}} + \mathcal{N}(\mathbf{0}, C_d)$, where $C_d \in \mathbb{R}^{N_d \times N_d}$ is a diagonal covariance matrix. We do not include model error, meaning we assume the simulated forward model is correct and the priors defined by $p(\mathbf{m}_0)$ have been validated (though this could also be readily incorporated in the workflow). 

Observations consist of water injection rates at the $N_{\text{inj}}=3$ injection wells and oil and water production rates at the $N_{\text{prod}}=6$ production wells. These observations provide 15 data streams, or quantities of interest (i.e., $N_{\text{QoI}} = N_{\text{inj}} + 2N_{\text{prod}}=15$). Data are recorded every 50~days during the historical period of 250~days, corresponding to a total of $N_d = 5 \times N_{\text{QoI}} = 75$ observations. The remaining 1250~days are treated as the forecast period. 

In this study we perform DA in both model space and latent space. In model space, permeability ($k_x=k_y$, $k_z=0.2k_x$) is updated in every grid block. We denote the full set of $\log(k_x)$ for a particular model as $\mathbf{K} \in \mathbb{R}^{N_c}$, with $N_c = 524,288$. Given the updated $\log(k_x)_i$, $i=1,\dots, N_c$, we update $\log(k_y)$ and $\log(k_z)$ via $(k_y)_i=(k_x)_i$, $(k_z)_i=0.2(k_x)_i$, $i=1,\dots, N_c$.
Porosity is treated through the porosity-permeability relationship. Specifically, we fit a second-degree polynomial for $\phi_i$ using the facies values for ($\phi$, $\log(k_x)$) given in Section~\ref{sec:geomodels_setup}. This expression can be written as $\phi_i=a_0+a_1 \log(k_x)_i+a_2 [\log(k_x)_i]^2$, for $i=1,\dots, N_c$.
We apply a cutoff (truncation) to assure $(k_x)_i$ (and thus $\phi_i$) is within the prior range, i.e.,  $25~{\rm mD} \leq (k_x)_i \leq 1000~{\rm mD}$. The inverse problem in this case reduces to sampling from the unknown distribution $p(\mathbf{m}_0 \mid \mathbf{d}^{\ast}_{\text{obs}})$, where $\mathbf{m}_0 = \mathbf{K}$.

When the 3D-LDM parameterization is used, DA is performed in the latent space using $\boldsymbol{\xi}_T$ as the model parameter variable. As such, the number of (approximately multi-Gaussian) parameters to calibrate is now $n_c$ (recall $n_c  \ll N_c$). At every DA step, 3D-LDM denoises the updated latent variable $\boldsymbol{\xi}_T$ through the U-net to obtain $\boldsymbol{\xi}_0$ (Eq.~\ref{eqn:denoising}), which is then decoded into an updated, full facies model, $\mathbf{m}_0^{\text{LDM}} = \mathcal{D}(\boldsymbol{\xi}_0)$. Grid block porosity and permeability values are assigned based on facies, as in Section~\ref{sec:geomodels_setup}. The inverse problem in this case reduces to sampling from the unknown distribution $p(\boldsymbol{\xi}_T \mid \mathbf{d}^{\ast}_{\text{obs}})$. 

For both model and latent-space DA, the input to the flow simulator is the full geomodel $\mathbf{m}_0$. We represent the numerical flow simulator as a nonlinear operator $F(\mathbf{m}_0): \mathbb{R}^{N_c} \rightarrow \mathbb{R}^{N_t \times N_{\text{QoI}}}$ that maps $\mathbf{m}_0$ to a set of time-varying injection rates for each injector and a set of time-varying oil and water production rates for each producer. These quantities are denoted, respectively, as $\mathbf{q}_{\text{inj}}^{i} \in \mathbb{R}^{N_t}$ ($i=1,\dots,N_{\text{inj}}$), $\mathbf{q}_{\text{oil}}^{j}\in \mathbb{R}^{N_t}$, and $\mathbf{q}_{\text{wat}}^{j} \in \mathbb{R}^{N_t}$ ($j=1,\dots,N_{\text{prod}}$). 
The full set of injection and production rate data is thus
\begin{equation}
{F}(\mathbf{m}_0) = [\mathbf{q}_{\text{inj}}^{1}, \dots, \mathbf{q}_{\text{inj}}^{N_\text{inj}}, \, 
\mathbf{q}_{\text{oil}}^{1}, \dots, \mathbf{q}_{\text{oil}}^{N_\text{prod}}, \, 
\mathbf{q}_{\text{wat}}^{1}, \dots, \mathbf{q}_{\text{wat}}^{N_\text{prod}}].
\label{eq:sim_output}
\end{equation}
The goal of DA is to tune model parameters to minimize the mismatch between $\mathbf{d}^{\ast}_{\text{obs}}$ and the corresponding simulated data $\mathbf{d} \in \mathbb{R}^{N_d}$ -- that is, the first five time steps of the well rates in $F(\mathbf{m}_{0})$. We express this sub-sampling operation as $\mathbf{d} = F_{\text{obs}}(\mathbf{m}_0)$.

\section{ESMDA in model space and latent space}
\label{sec:esmda_results}

In this section, we apply the popular ESMDA algorithm~\citep{EMERICK20133} both in model space and in latent space (i.e., parameterized with 3D-LDM) for the setup described in Section~\ref{sec:3d_ldm_esmda}. We first provide a brief overview of ESMDA and the localizations applied in this work.

\subsection{ESMDA algorithm with localization}

ESMDA iteratively updates an ensemble of $N_e$ uncertain model parameter fields $\{\mathbf{p}_{n}\}_{n=1}^{N_e} \in \mathbb{R}^{N_p}$, where $N_p$ denotes the number of model parameters,
using ensemble-estimated cross-covariances between parameters and observations. At each iteration $k$ of the $N_a$ assimilation steps, the update equation involves the Kalman gain matrix $K^{(k)} \in \mathbb{R}^{N_p \times N_d}$ and the mismatch between observed and simulated data, 
\begin{equation}
\begin{aligned}
\mathbf{p}^{(k+1)}_n
&= \mathbf{p}^{(k)}_n + K^{(k)} (\mathbf{d}^{\ast}_{\text{obs}} - \mathbf{d}_{n}^{(k)} ) \\
&= \mathbf{p}^{(k)}_n + C_{pd}^{(k)} \left(C_{dd}^{(k)} + \alpha^{(k)} C_d\right)^{-1}
(\mathbf{d}^{\ast}_{\text{obs}} - \mathbf{d}_{n}^{(k)} ), \quad n = 1, \dots, N_e,
\end{aligned}
\label{eqn:esmda_update}
\end{equation}
where $\mathbf{p}^{(k+1)}_n \in \mathbb{R}^{N_p}$ is the updated parameter vector of the ensemble member $n$ and $\mathbf{d}_n^{(k)} \in \mathbb{R}^{N_d}$ is the simulated data vector. The matrix $C_{pd}^{(k)} \in \mathbb{R}^{N_p \times N_d}$ is the cross-covariance between parameters and simulated data and $C_{dd}^{(k)} \in \mathbb{R}^{N_d \times N_d}$ is the autocovariance of the simulated data, both estimated from the ensemble. The scalar $\alpha^{(k)}$ is an inflation coefficient, such that $\mathbf{d}^{\ast}_{\text{obs}} = \mathbf{d}_{\text{true}} + \mathcal{N}(\mathbf{0}, \alpha^{(k)} C_d)$. The set of these coefficients satisfies $\sum_{k=1}^{N_a} \frac{1}{\alpha^{(k)}} = 1$. This distributes the full Bayesian update over multiple linear steps~\citep{EMERICK20133}. 

The updated parameters (and the number of parameters) differ between model space and latent space. Specifically, for model-space ESMDA, $\mathbf{p}^{(k)} = \mathbf{m}_0^{(k)} \in \mathbb{R}^{N_c}$ and $N_p = N_c$. For latent-space ESMDA, $\mathbf{p}^{(k)} = \boldsymbol{\xi}_T^{(k)}$ and $N_p = n_c$.

Due to the finite size of the ensemble, ESMDA, like other algorithms based on Kalman-type updates (e.g., ensemble Kalman filter, EnKF~\citep{enkf_evensen}), can be susceptible to sampling error in $C_{pd}$. The resulting spurious correlations may result in ensemble collapse (underestimated uncertainty) or other problematic effects. To mitigate these issues, localization~\citep{enkf_localiz, Chen2010, EMERICK2016219} is often applied. This entails defining the elements of a correction matrix $L^{(k)} \in \mathbb{R}^{N_p \times N_d}$, such that $K_{\text{loc}}^{{(k)}} = K^{(k)} \odot L^{(k)}$ is used in Eq.~\ref{eqn:esmda_update} instead of $K^{(k)}$. The most common form of localization is distance-based~\citep{gc_localiz}, where correlations are weighted based on physical distance from the data source. This approach, however, is based on the assumption that correlation decreases with distance, which may not be strictly applicable in our setup.

This is because, in model-space DA for channelized systems, parameter-data correlations are governed by geological connectivity rather than proximity~\citep{seabra_score_based_diff}. In addition, in 3D-LDM space, latent variables have no direct spatial interpretation. For these reasons, here we apply a non-distance-based localization method. After experimenting with correlation-based~\citep{luo_cbl_a} and bootstrap methods~\citep{Zhang2010_bootstrap_localiz}, we opted for pseudo-optimal localization (POL)~\citep{FURRER2007227} due to its lack of tuning parameters and robustness, as found in~\citet{LACERDA2019690}. With this treatment, a generic element $l_{ij}$ of $L^{(k)}$ is defined as
\begin{equation}
    l_{ij} = \frac{c_{ij}^2}{c_{ij}^2 + (c_{ij}^2 + c_{ii}c_{jj})/{N_e}}.
    \label{eqn:localiz}
\end{equation}
Here $c_{ij}$ is the correlation between model parameter $i$ and simulated observation $j$. In this way, ensemble-estimated covariances are used to reduce the weight of (potentially) spurious correlations, while preserving those that are statistically significant. The form of Eq.~\ref{eqn:localiz} was derived by minimizing the expected Frobenius-norm error between true and localized covariances~\citep{FURRER2007227}. 

To quantify spurious updates and assess the effectiveness of this localization, flow-independent dummy parameters are updated together with model parameters. We use $3^3=27$ dummy variables (which are readily concatenated to the 3D geomodel or latent variable by increasing all dimensions by 3). Prior dummy variables have a normal distribution with zero mean. In latent space, the standard deviation is set to 1 (as for latent variables). In model space, it is set to 1.26, which represents the standard deviation of all model parameters (in $\log k_x$) across all realizations. Since they have no impact on flow and, therefore, on observations, their posterior distribution is expected to vary very little from the prior. Conversely, significant updates to dummy variables are a sign of spurious correlations. 

Algorithmic details for model and latent-space ESMDA are reported in Algorithm~\ref{alg:esmda}. A comprehensive discussion of best practices for ESMDA can be found in~\citet{emerick_review} and papers referenced therein.

\begin{minipage}[htbp]{0.85\textwidth}
\begin{algorithm}[H]
\caption{ESMDA-based data assimilation with localization}\label{alg:esmda}
\begin{algorithmic}[1]
\State Initialization: set $k=0$
\State Sample $\{\mathbf{p}_n^{(k)}\}_{n=1}^{N_e}$
\State Set inflation coefficients $\{\alpha^{(k)}\}_{k=1}^{N_a}$ such that $\sum_{k=1}^{N_a} \frac{1}{\alpha^{(k)}} = 1$

\For{$k = 0,1,\dots,N_a-1$}
    \For{$n = 1,\dots,N_e$}
        \If{$\mathbf{p} = \boldsymbol{\xi}_T$}
            \State Generate $\mathbf{m}^{\text{LDM}}_0(\mathbf{p}_{n}^{(k)})$
            \State Simulate $\mathbf{d}_n^{(k)} = F_{\text{obs}}\big(\mathbf{m}^{\text{LDM}}_0(\mathbf{p}_{n}^{(k)})\big)$
        \Else
            \State Simulate $\mathbf{d}_n^{(k)} = F_{\text{obs}}\big(\mathbf{p}_n^{(k)}\big)$
        \EndIf
    \EndFor
    \State Apply localization to $K^{(k)}$ using Eq.~\ref{eqn:localiz}

    \For{$n = 1,\dots,N_e$}
        \State Update parameters $\mathbf{p}_n^{(k+1)}$ using Eq.~\ref{eqn:esmda_update}
    \EndFor
\EndFor

\end{algorithmic}
\end{algorithm}
\end{minipage}

\subsection{ESMDA results for model space and latent space}
We now present results for model and latent space ESMDA for Case~1, which is the synthetic true model shown in Figure~\ref{fig:examples_petrel}d. We set $N_e = 1000$, $N_a = 10$, and the same inflation coefficient, $\alpha^{(k)} = 10$, for all iterations. \textcolor{black}{To ensure equal starting points, the prior ensemble is specified to be the same for all cases. This ensemble is generated using 3D-LDM, which is appropriate since 3D-LDM has been shown to accurately sample from the prior distribution (see Section~\ref{sec:3d_ldm_esmda} and~\citet{DiFederico2025}).} We also use POL, as described above. Since ensemble members are independent, flow simulations can be performed in parallel. Running the ensemble in batches of 20 requires approximately 8~hours per ESMDA iteration, for a total of about 80~hours for a full DA run. Model-space and latent-space updates through Eq.~\ref{eqn:esmda_update} (including the 3D-LDM mapping from latent to model space) are negligible compared to the forward simulation time.

\begin{figure}[h!]
    \centering
    \begin{subfigure}[b]{0.46\textwidth}
        \raggedright
        \includegraphics[width=\textwidth, trim=200 0 200 0, clip]{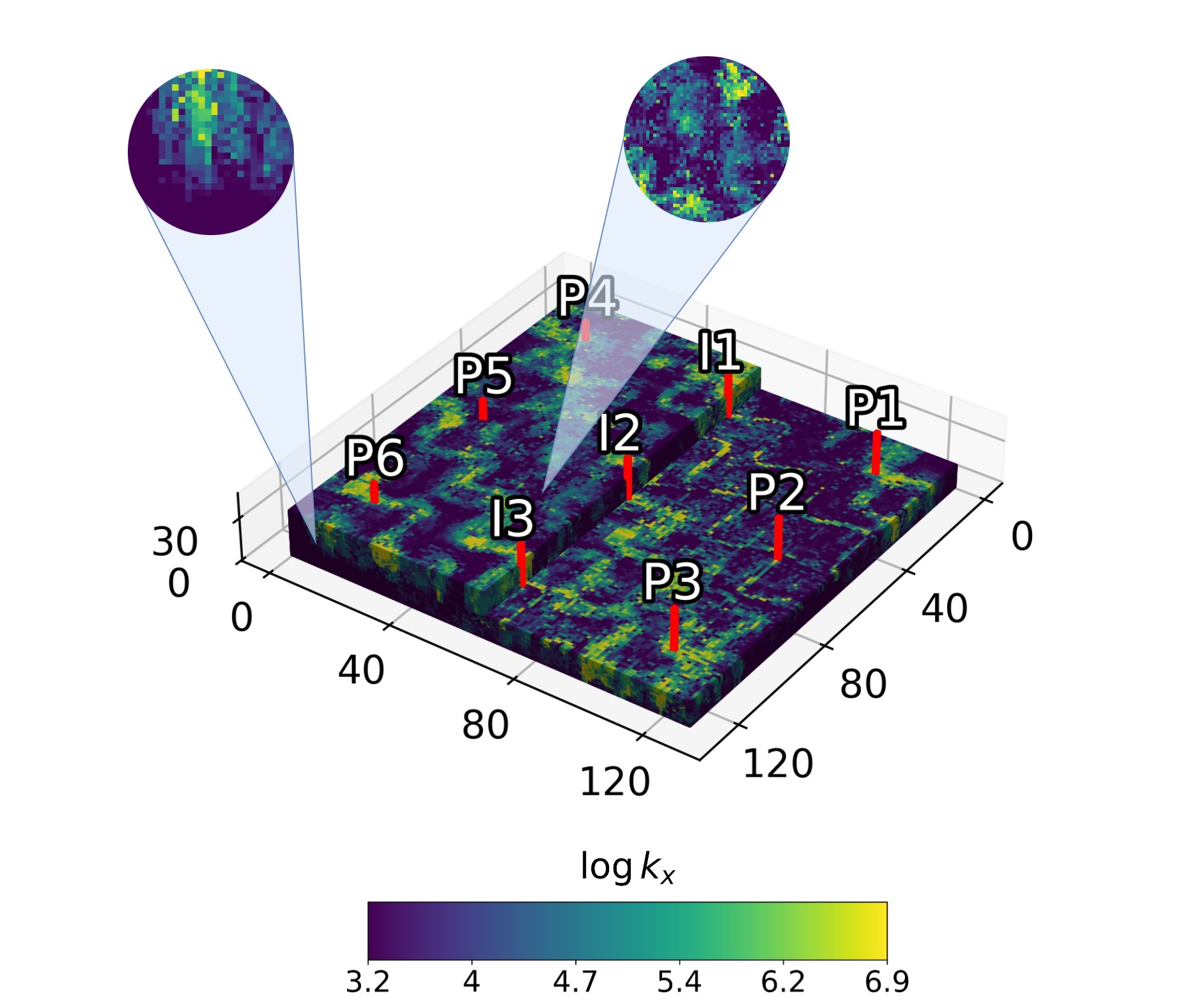}
        \caption{Posterior~1 (model)}
    \end{subfigure}
    \hfill
    \begin{subfigure}[b]{0.46\textwidth}
        \raggedright
        \includegraphics[width=\textwidth, trim=200 0 200 0, clip]{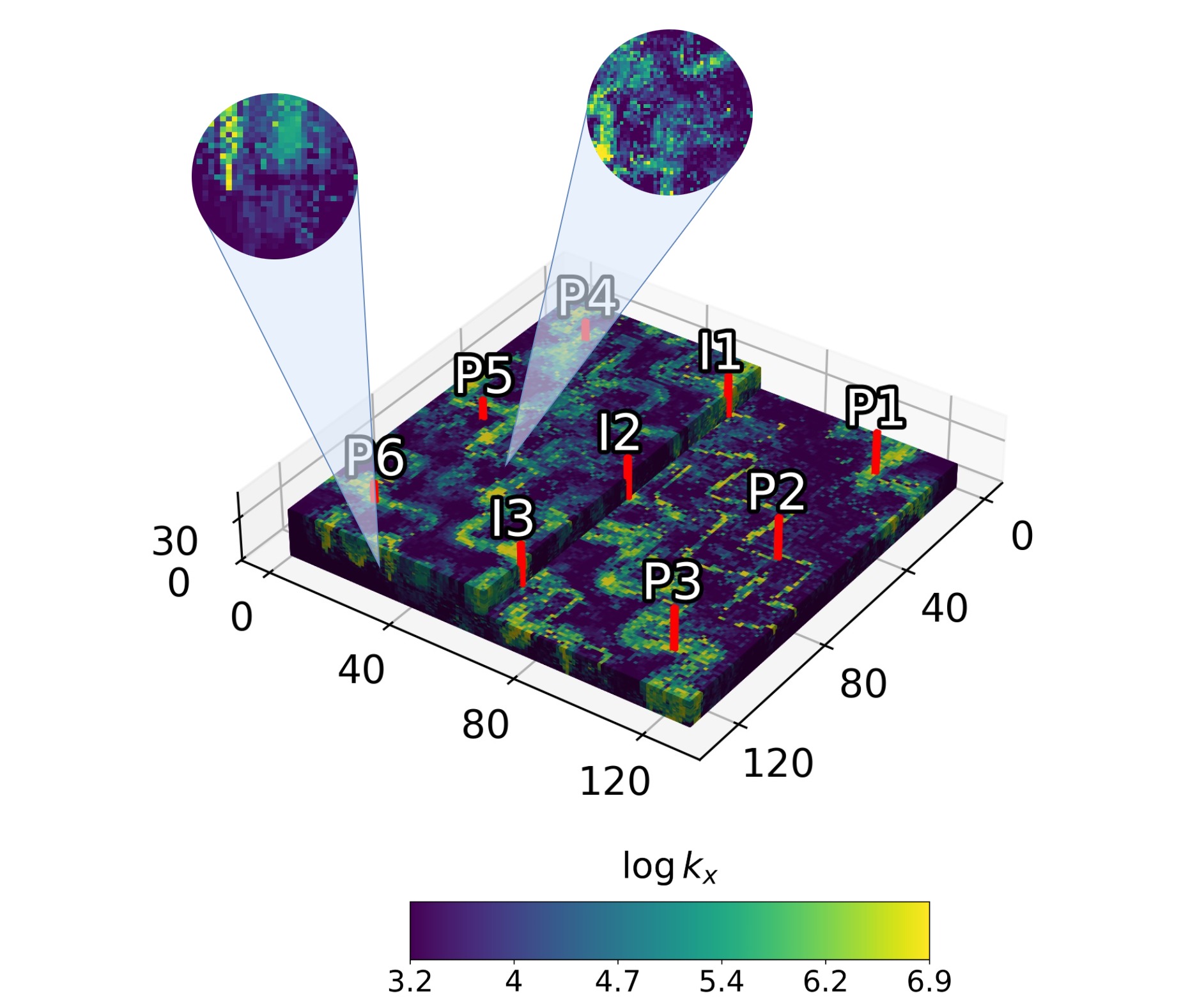}
        \caption{Posterior~2 (model)}
    \end{subfigure}

    \vspace{1em}

    \begin{subfigure}[b]{0.48\textwidth}
        \raggedright
        \includegraphics[width=\textwidth, trim=200 0 200 0, clip]{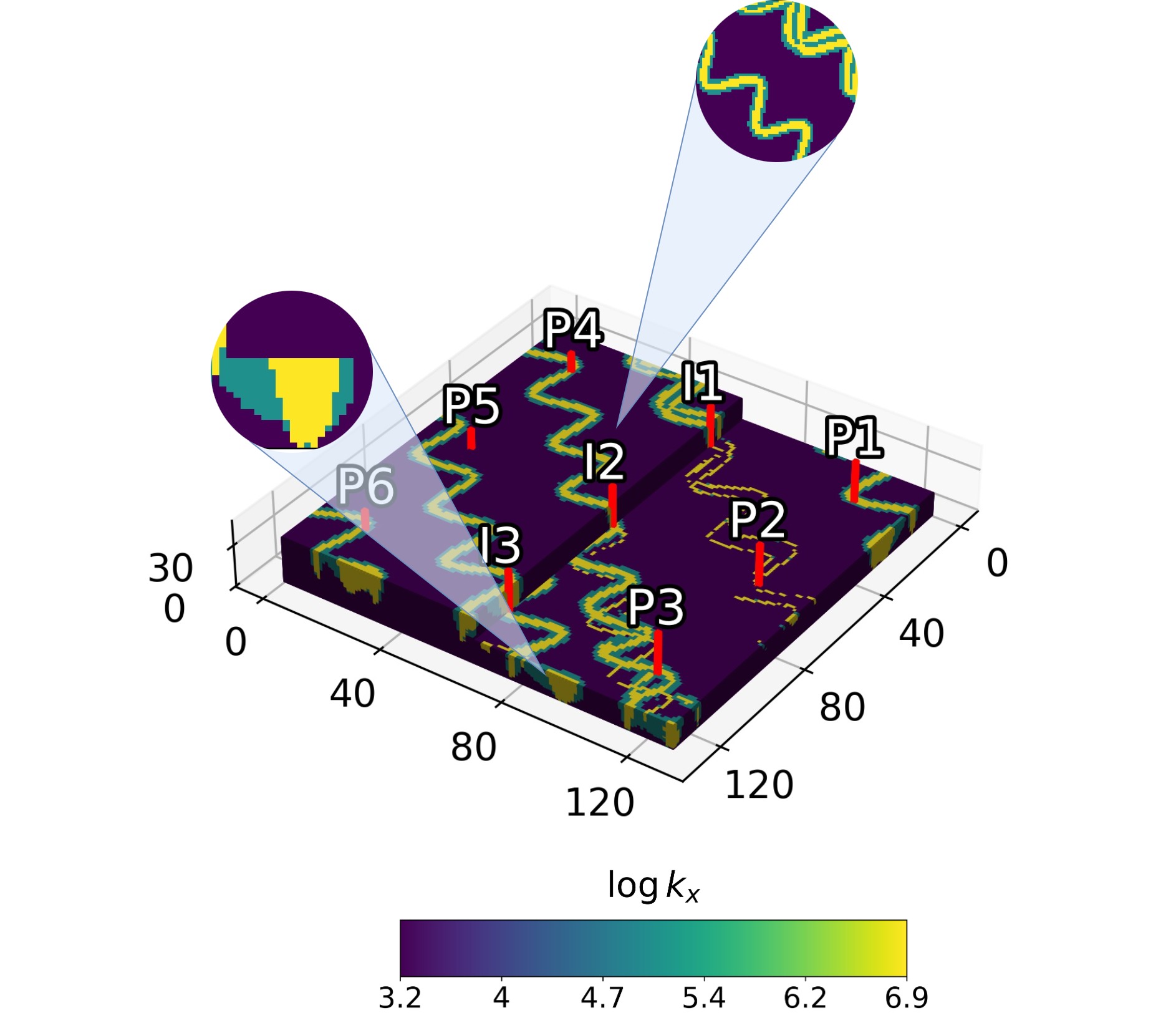}
        \caption{Posterior~1 (latent)}
    \end{subfigure}
    \hfill
    \begin{subfigure}[b]{0.48\textwidth}
        \raggedright
        \includegraphics[width=\textwidth, trim=200 0 200 0, clip]{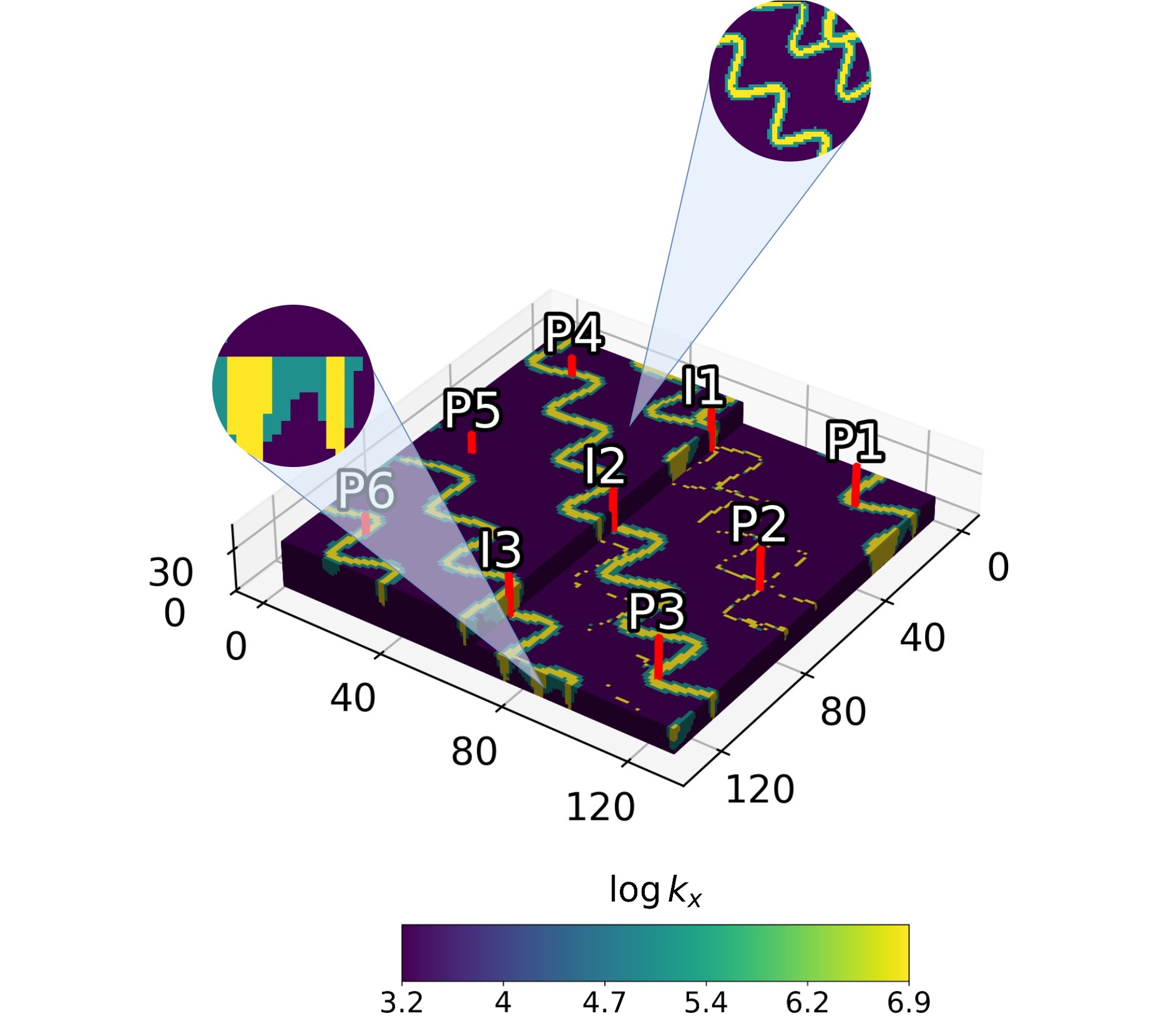}
        \caption{Posterior~2 (latent)}
    \end{subfigure}

    \caption{Case~1: posterior geomodels using model-space ESMDA (top row) and latent-space ESMDA (bottom row). The right-most portion of each model is extracted to enable visualization of the lower half. Circles show magnified views to highlight the details of the geological features.}
    \label{fig:esmda_comparison_models}
\end{figure}

Figure~\ref{fig:esmda_comparison_models} shows posterior (after DA) realizations for model-space (top row) and latent-space (bottom row) ESMDA.
The upper part of the right-most portion of each geomodel has been removed to enable visualization of the lower half of the model. It is evident that model-space updates do not maintain full geological realism in the posterior geomodels. This is highlighted in the circular magnification regions, which show details of the geomodels. Specifically, in Figure~\ref{fig:esmda_comparison_models}a,b, we observe ``smeared'' channels, broken channel bodies, and incorrect facies ordering in many portions of the models.

The 3D-LDM parameterization, by contrast, automatically maintains geological realism and discrete facies for any model parameter $\boldsymbol{\xi}_T$. Therefore, the posterior realizations in Figure~\ref{fig:esmda_comparison_models}c,d exhibit the correct spatial features, with continuous channel bodies and the levee facies surrounding them within the mud background. Such features are clearly in general accordance with the synthetic true model (Figure~\ref{fig:examples_petrel}d), even though there is still ensemble variability. It should be noted that the ``speckling'' observed in the lower halves of Figure~\ref{fig:esmda_comparison_models}c,d is a consequence of the vertical slicing, which in this view leaves the deepest elements of some channels isolated. The ability to maintain geological realism in posterior geomodels is a key reason for using advanced parameterization techniques (such as 3D-LDM) for DA.

\begin{figure}[htbp]
    \centering

    \begin{subfigure}[b]{0.48\textwidth}
        \centering
        \includegraphics[width=\textwidth]{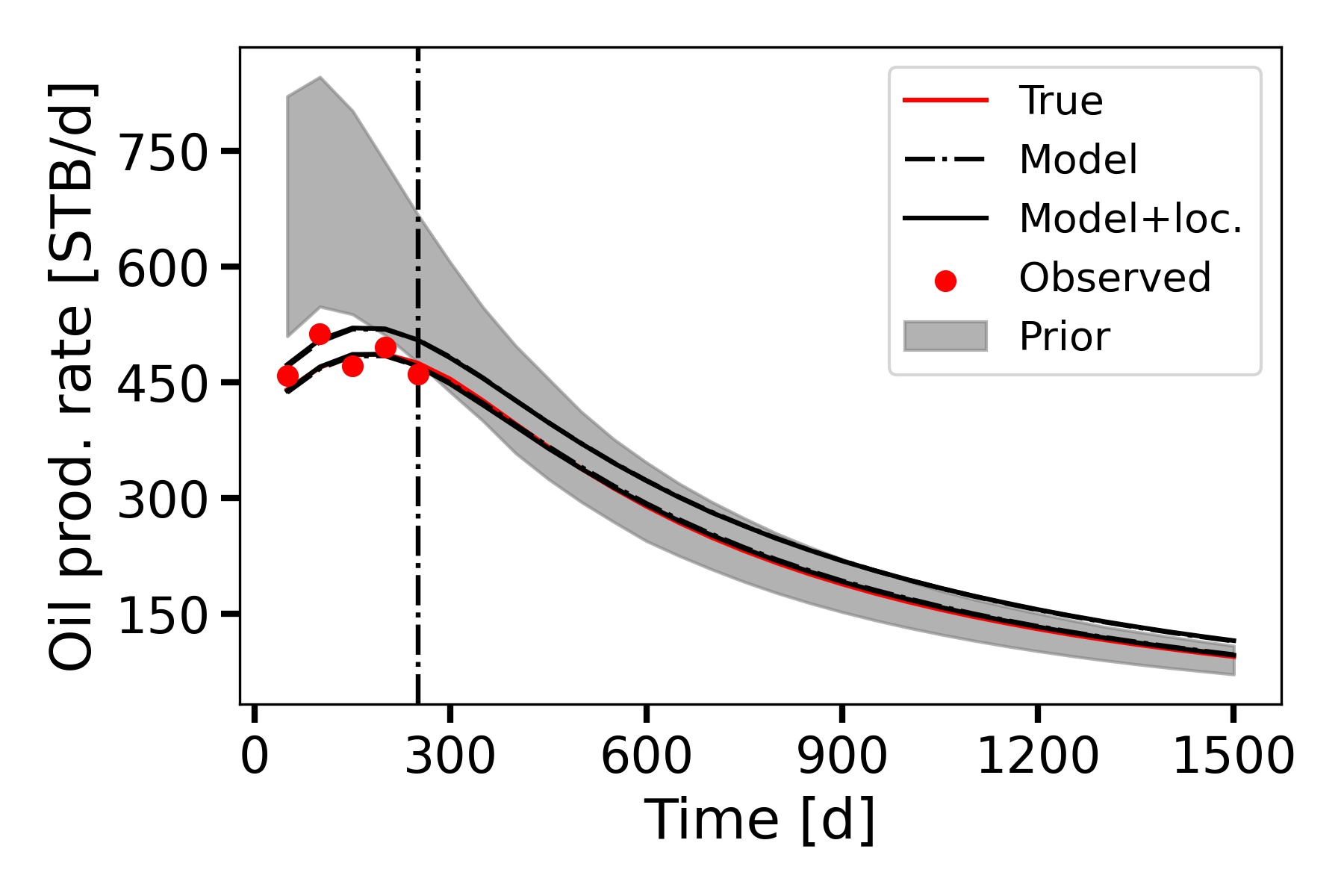}
        \caption{P3 oil production rate}
    \end{subfigure}
    \hfill
    \begin{subfigure}[b]{0.48\textwidth}
        \centering
        \includegraphics[width=\textwidth]{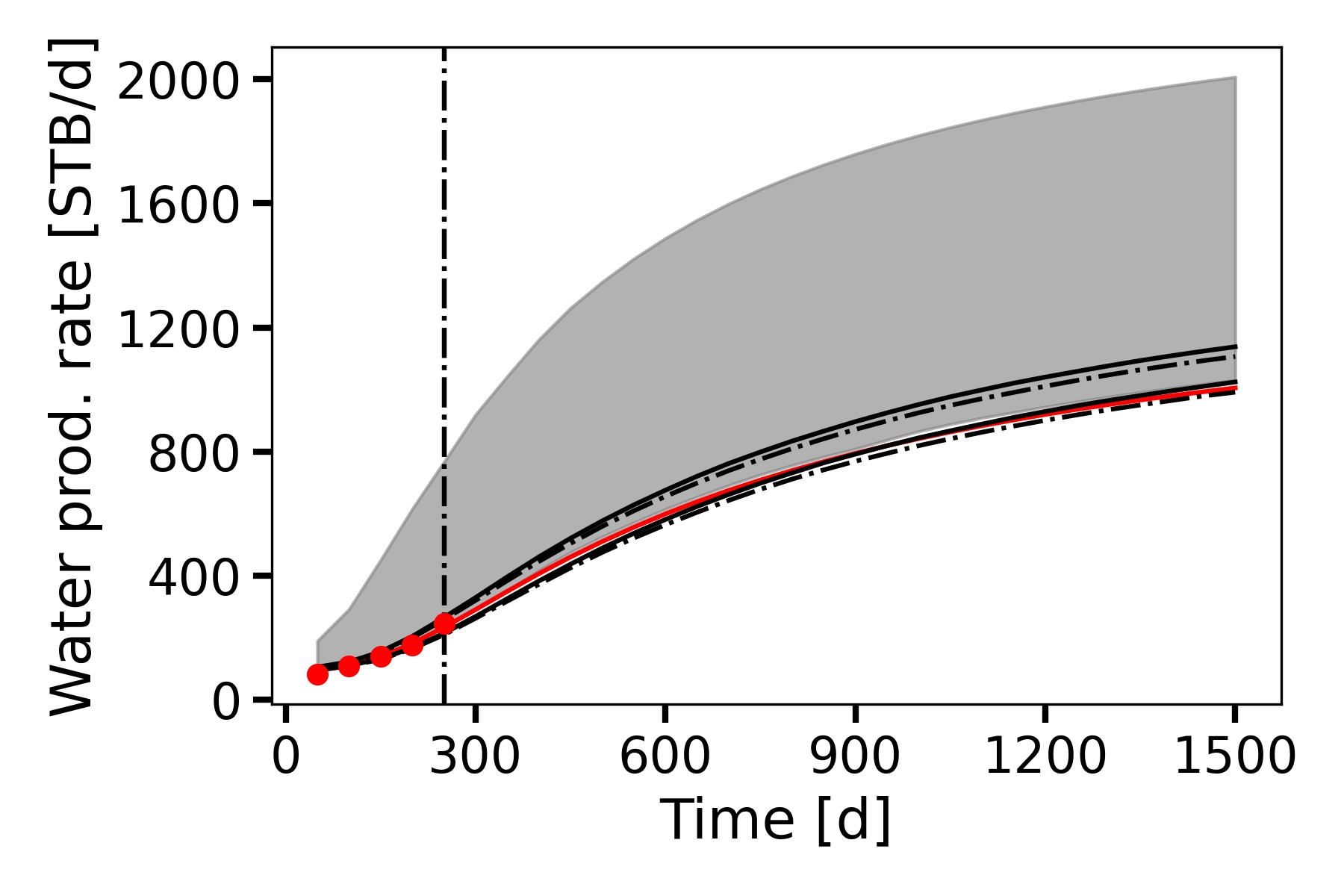}
        \caption{P1 water production rate}
        \label{fig:P1_WAT}
    \end{subfigure}

    \vspace{1em}

    \begin{subfigure}[b]{0.48\textwidth}
        \centering
        \includegraphics[width=\textwidth]{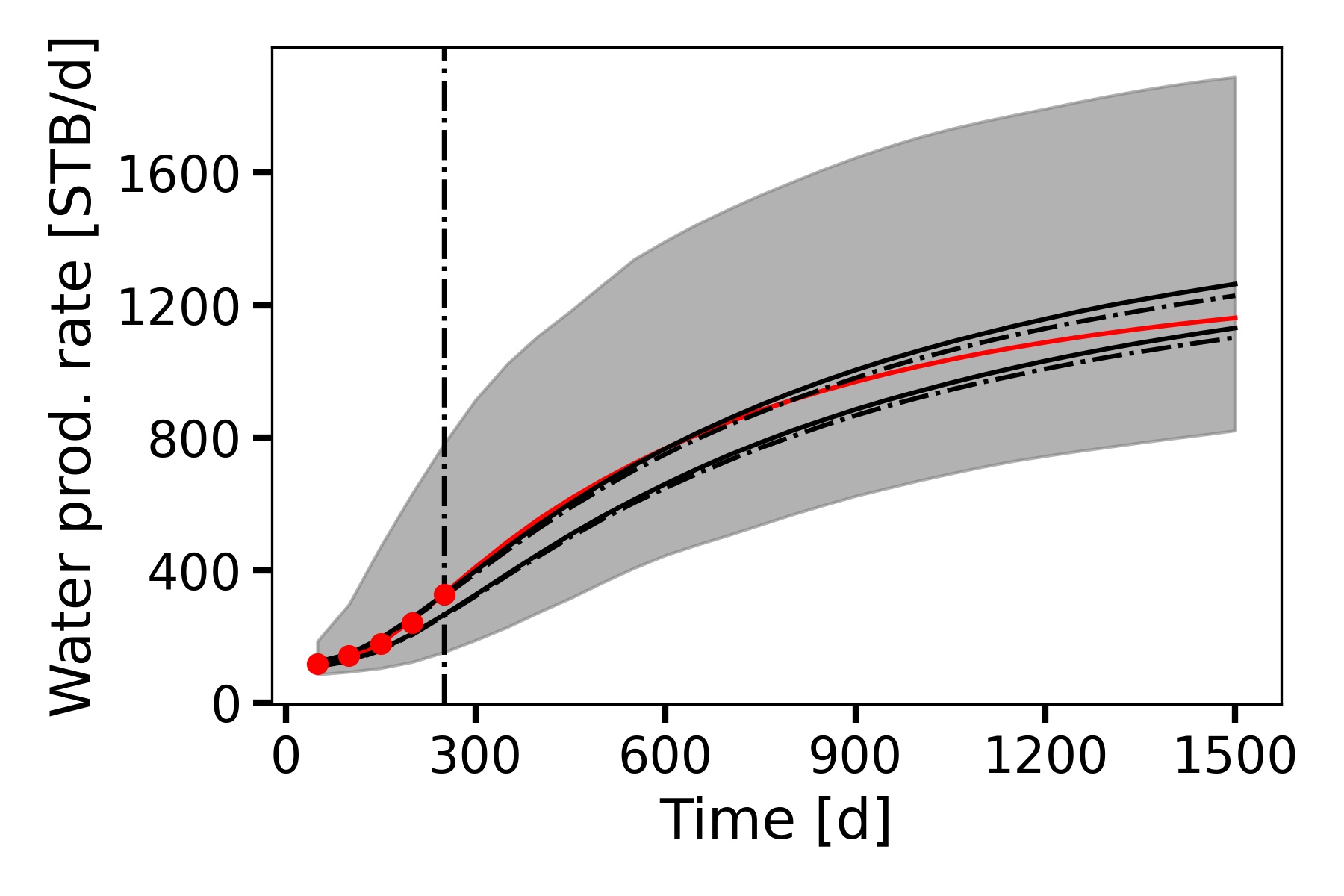}
        \caption{P6 water production rate}
    \end{subfigure}
    \hfill
    \begin{subfigure}[b]{0.48\textwidth}
        \centering
        \includegraphics[width=\textwidth]{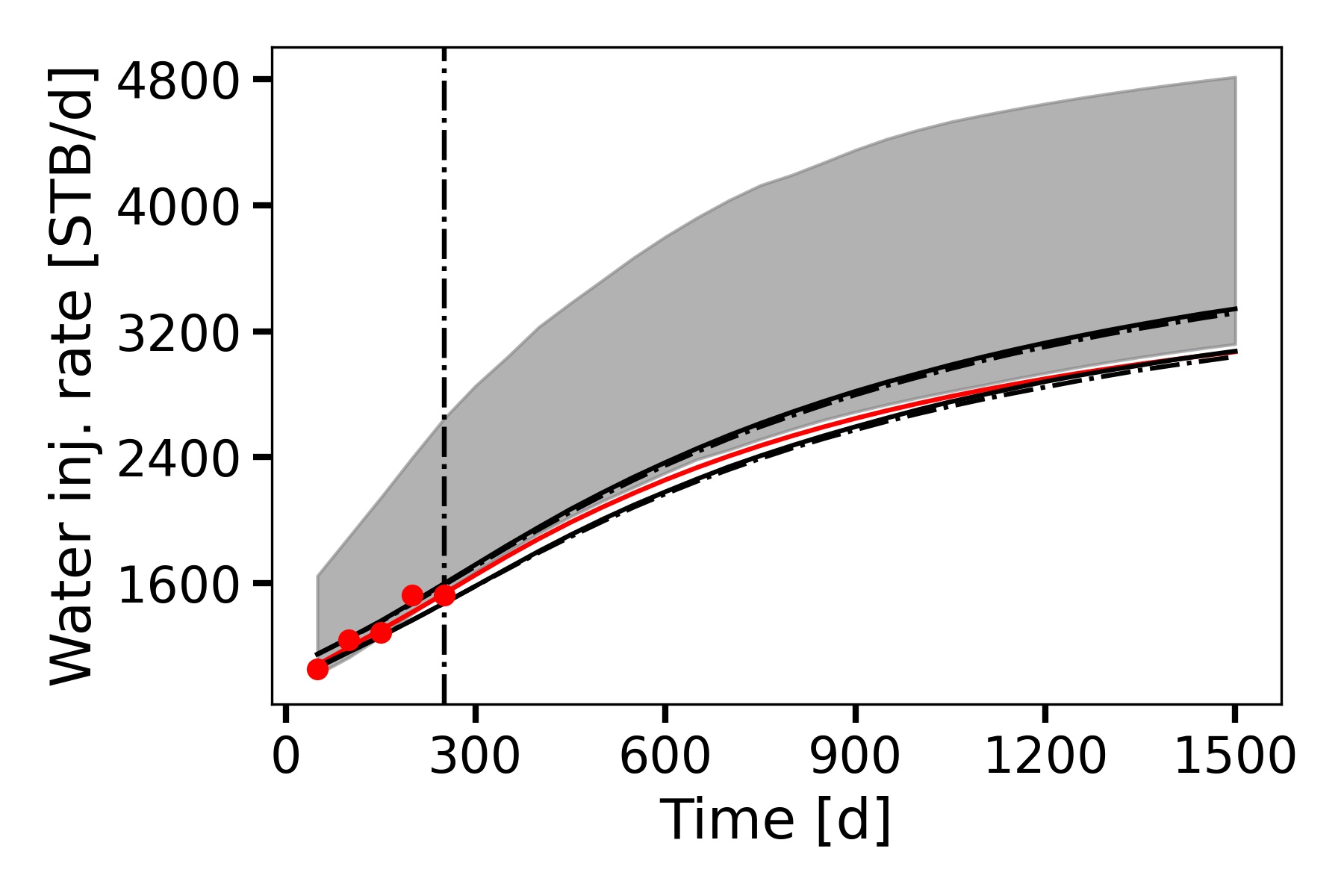}
        \caption{I3 water injection rate}
    \end{subfigure}

    \caption{Case~1: model-space ESMDA results for selected wells, with localization (solid lines) and without localization (dash-dotted lines). Gray regions show the prior P$_{10}$--P$_{90}$ range, black lines denote the posterior P$_{10}$ and P$_{90}$ curves, and red points and red curves represent observed and true data, respectively. The vertical black dot-dash line indicates the end of the history matching period.}
    \label{fig:esmda_comparison_rates_model}
\end{figure}

\begin{figure}[htbp]
    \centering

    \begin{subfigure}[b]{0.48\textwidth}
        \centering
        \includegraphics[width=\textwidth]{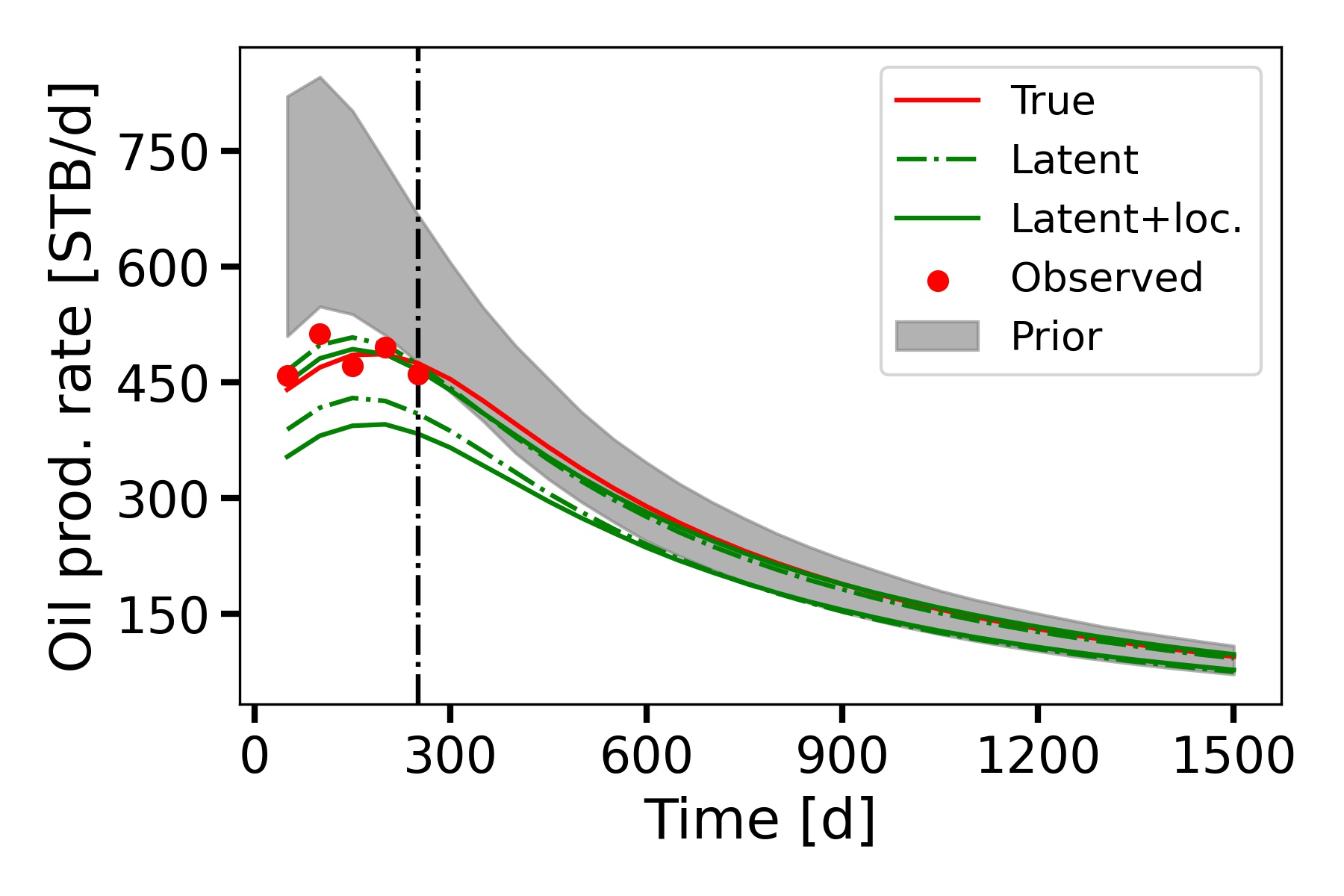}
        \caption{P3 oil production rate}
    \end{subfigure}
    \hfill
    \begin{subfigure}[b]{0.48\textwidth}
        \centering
        \includegraphics[width=\textwidth]{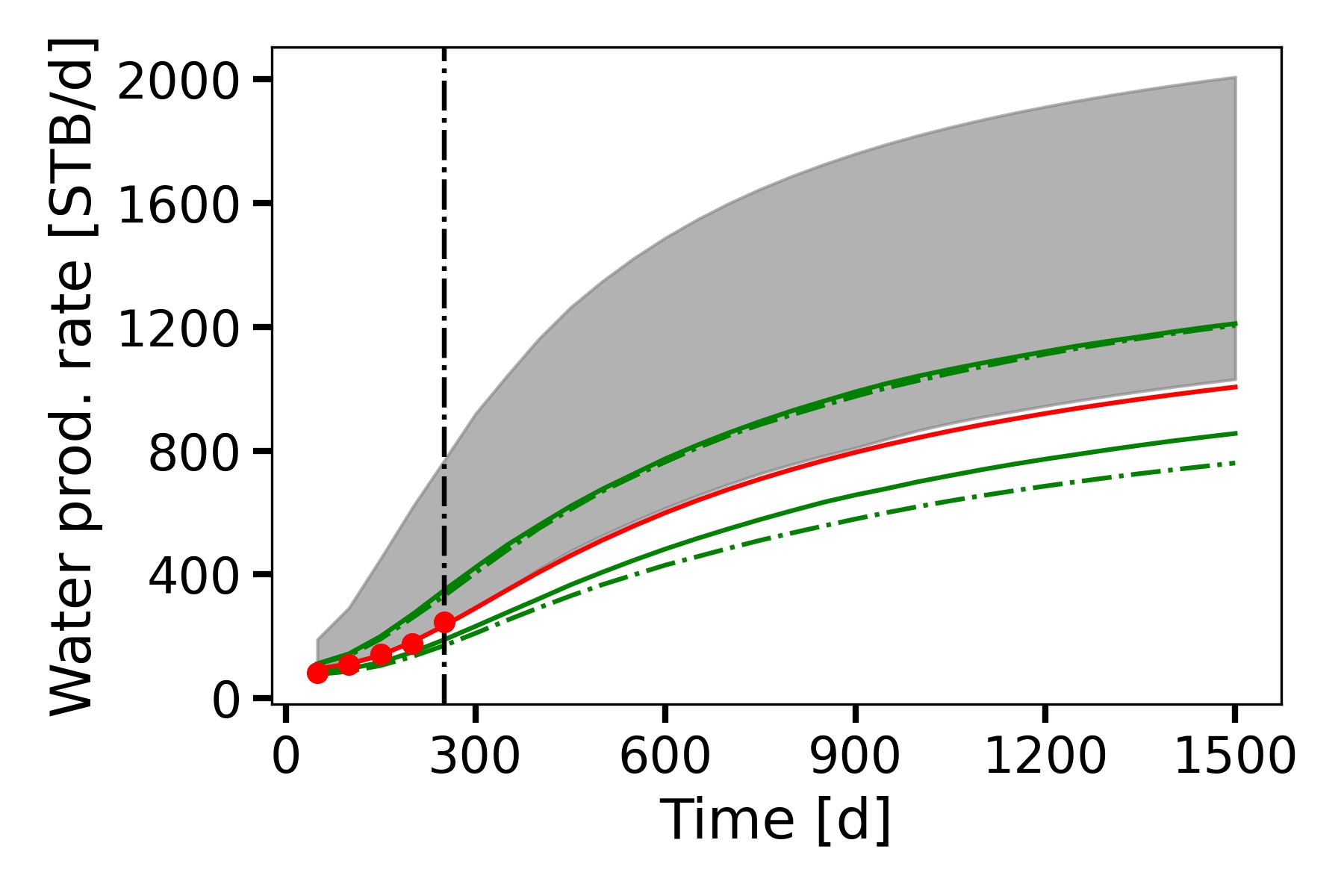}
        \caption{P1 water production rate}
    \end{subfigure}

    \vspace{1em}

    \begin{subfigure}[b]{0.48\textwidth}
        \centering
        \includegraphics[width=\textwidth]{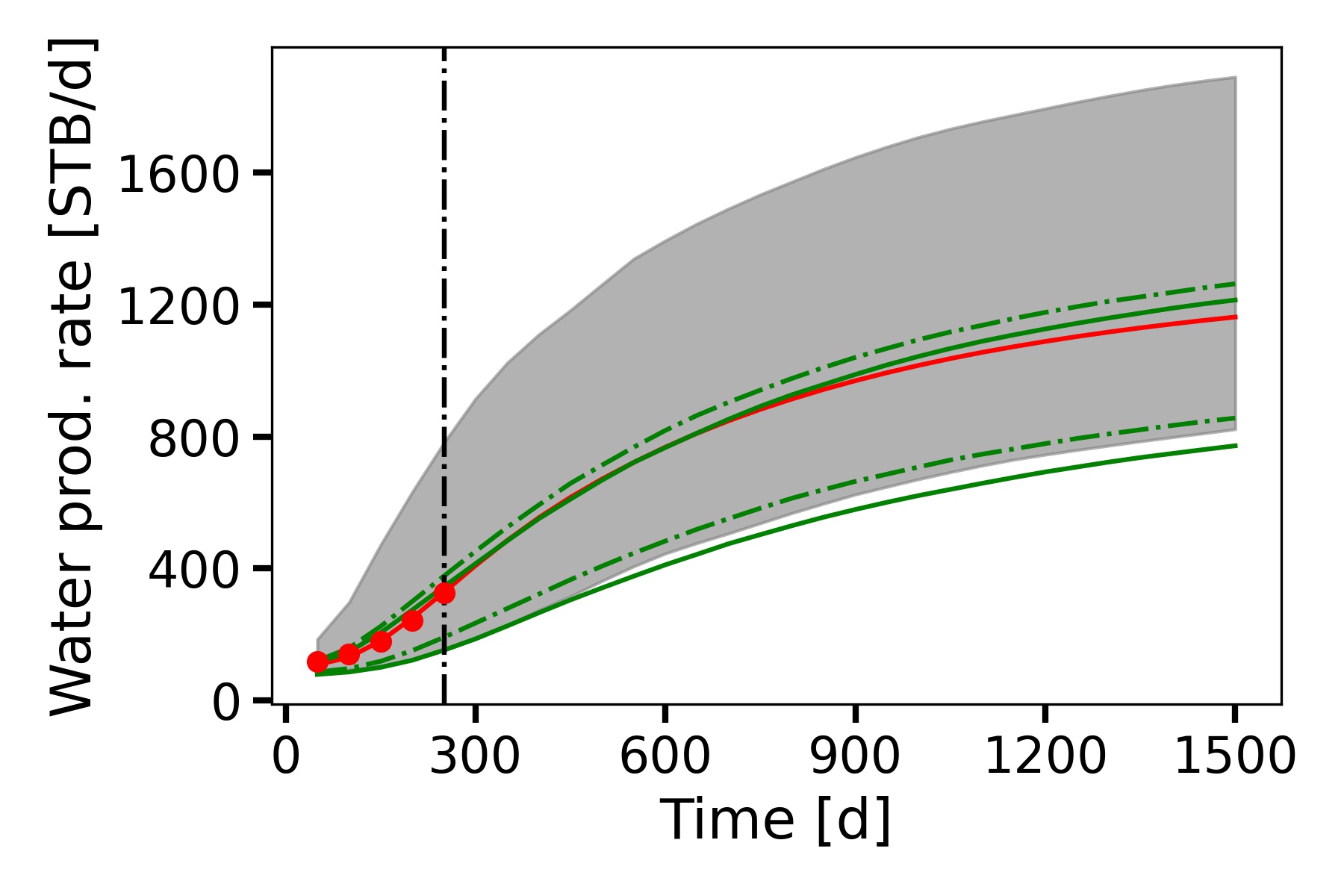}
        \caption{P6 water production rate}
    \end{subfigure}
    \hfill
    \begin{subfigure}[b]{0.48\textwidth}
        \centering
        \includegraphics[width=\textwidth]{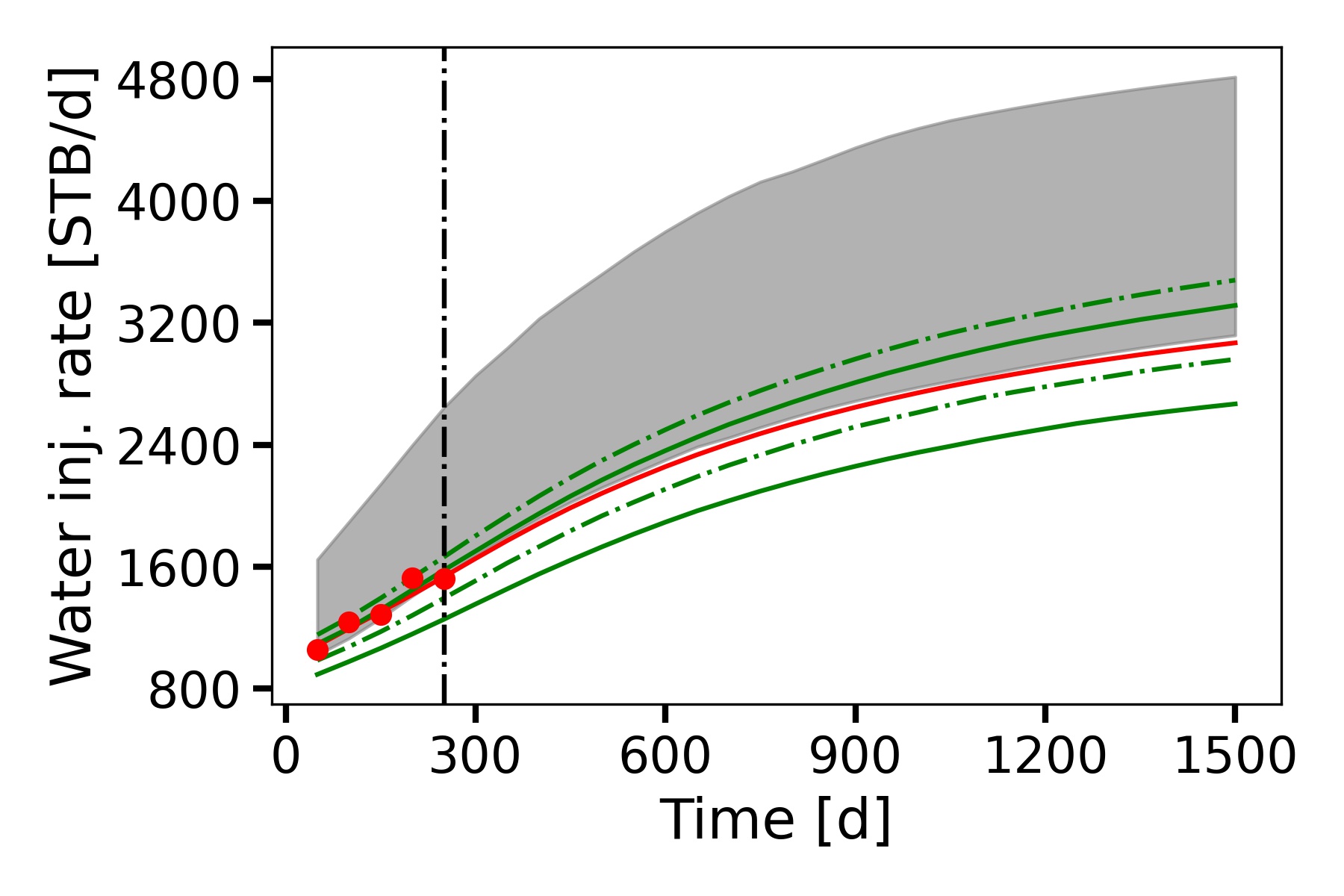}
        \caption{I3 water injection rate}
    \end{subfigure}

    \caption{Case~1: latent-space ESMDA results for selected wells, with localization (solid lines) and without localization (dash-dotted lines). Gray regions show the prior P$_{10}$--P$_{90}$ range, green lines denote the posterior P$_{10}$ and P$_{90}$ curves, and red points and red curves represent observed and true data, respectively. The vertical black dot-dash line indicates the end of the history matching period.}
    \label{fig:esmda_comparison_rates_latent}
\end{figure}

Prior and posterior flow responses are shown in Figure~\ref{fig:esmda_comparison_rates_model} (model-space) and Figure~\ref{fig:esmda_comparison_rates_latent} (latent-space). The prior P$_{10}$--P$_{90}$ (percentile) range appears in gray, while the P$_{10}$ and P$_{90}$ posterior curves across the ensemble are shown in black and green. The vertical black line at 250~days indicates the end of the historical period and the start of the forecast period. The data points used as observations are plotted as red circles, and the true model simulation results are shown as the solid red line. In these figures we also show posterior flow responses for the same case without localization (dash-dotted lines). Given the large ensemble size ($N_e=1000$), the impact of localization is very small in model space (Figure~\ref{fig:esmda_comparison_rates_model}). There are more noticeable differences in latent space (Figure~\ref{fig:esmda_comparison_rates_latent}), though the size of the P$_{10}$--P$_{90}$ ranges are similar with and without localization. 

\begin{figure}[htbp]
    \centering

    \begin{subfigure}[b]{0.48\textwidth}
        \centering
        \includegraphics[width=\textwidth]{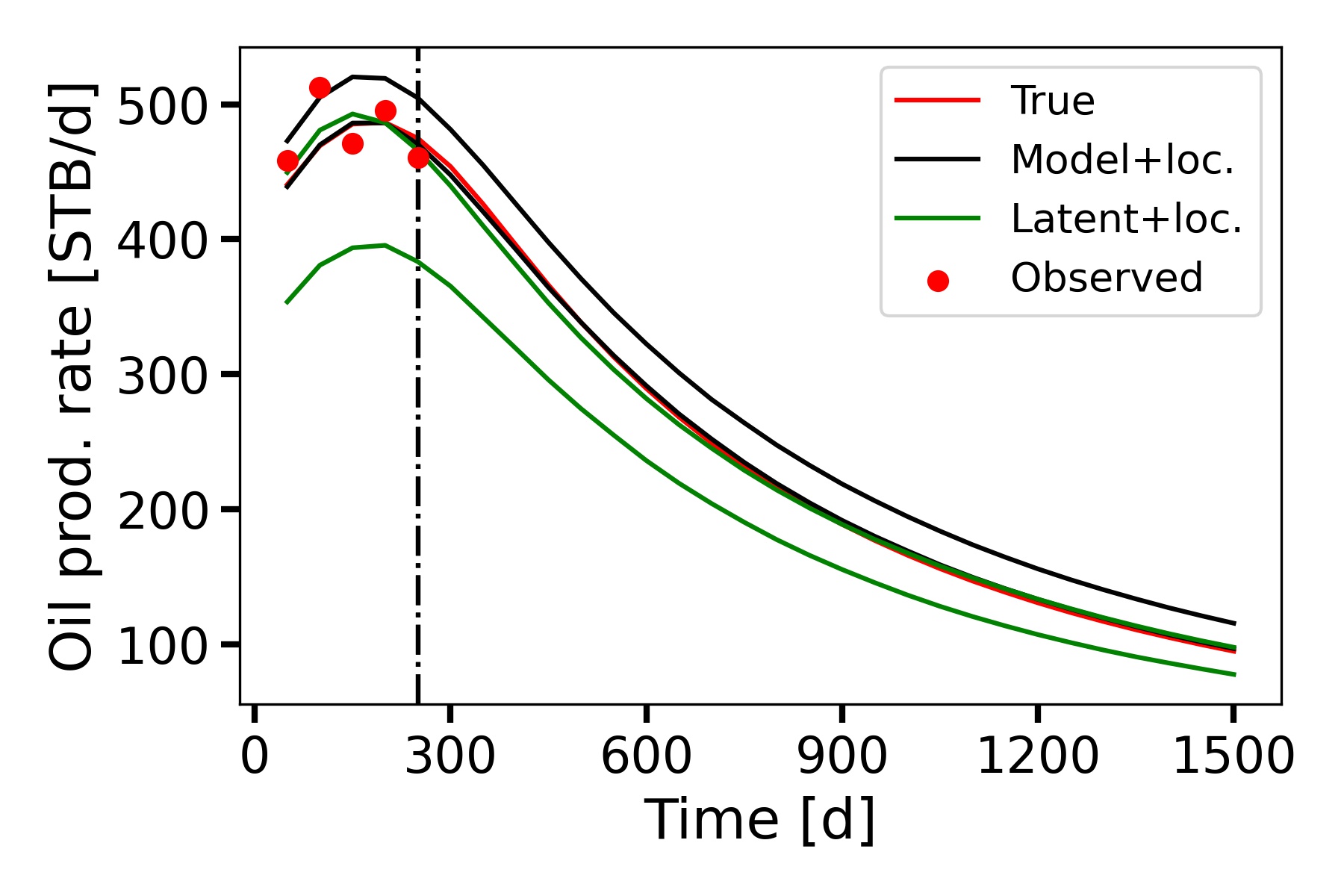}
        \caption{P3 oil production rate}
    \end{subfigure}
    \hfill
    \begin{subfigure}[b]{0.48\textwidth}
        \centering
        \includegraphics[width=\textwidth]{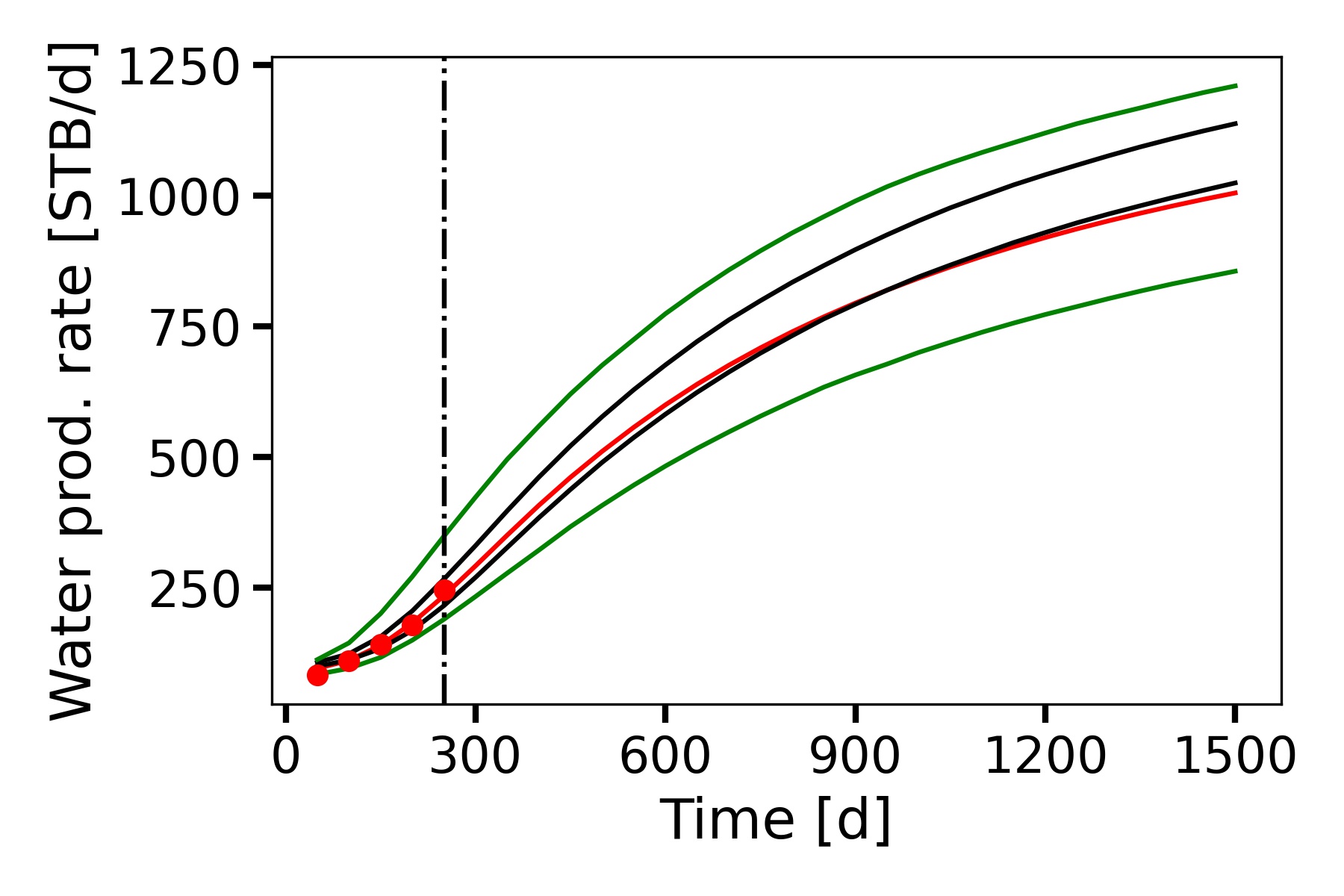}
        \caption{P1 water production rate}
    \end{subfigure}

    \vspace{1em}

    \begin{subfigure}[b]{0.48\textwidth}
        \centering
        \includegraphics[width=\textwidth]{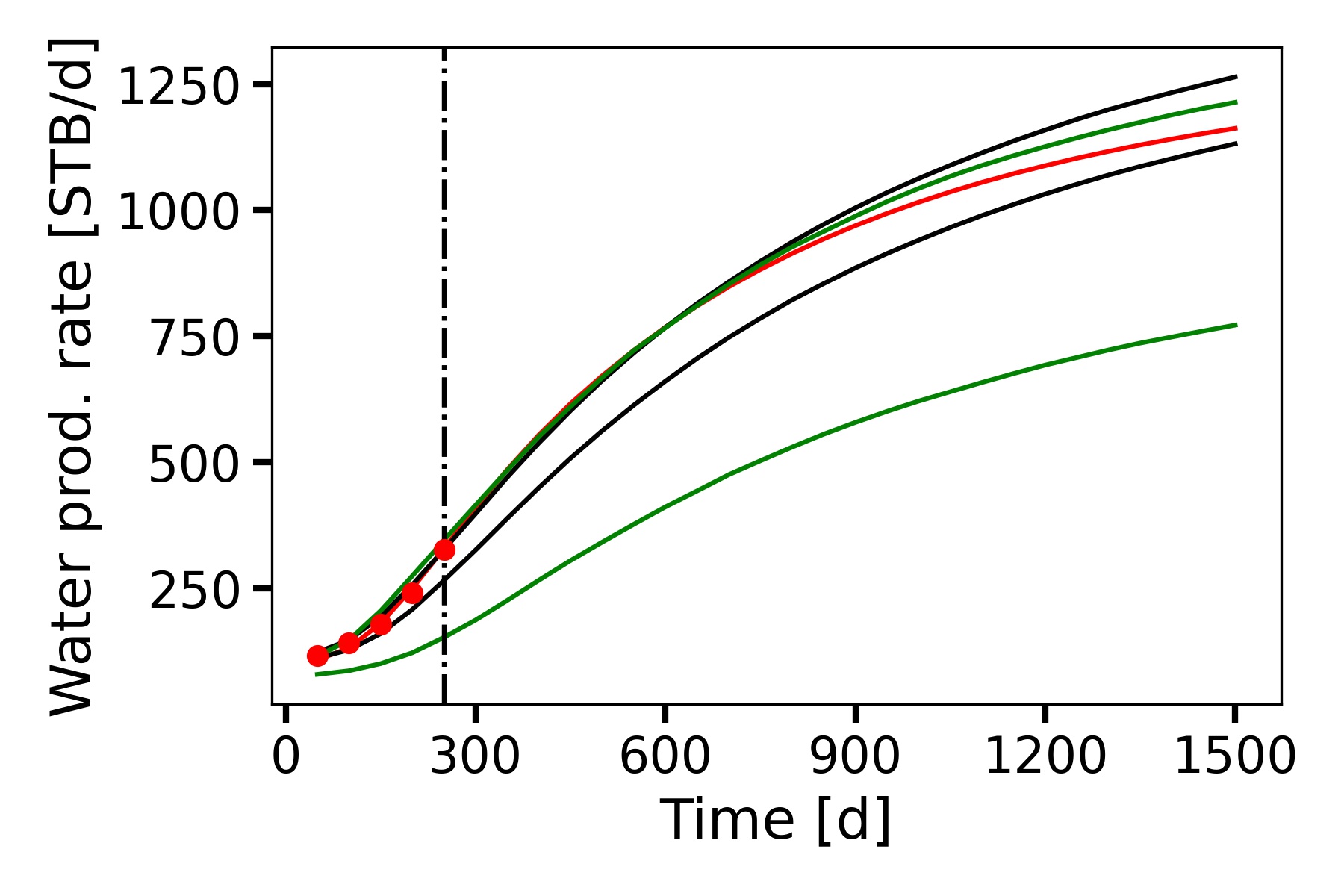}
        \caption{P6 water production rate}
        \label{fig:P4_OIL}
    \end{subfigure}
    \hfill
    \begin{subfigure}[b]{0.48\textwidth}
        \centering
        \includegraphics[width=\textwidth]{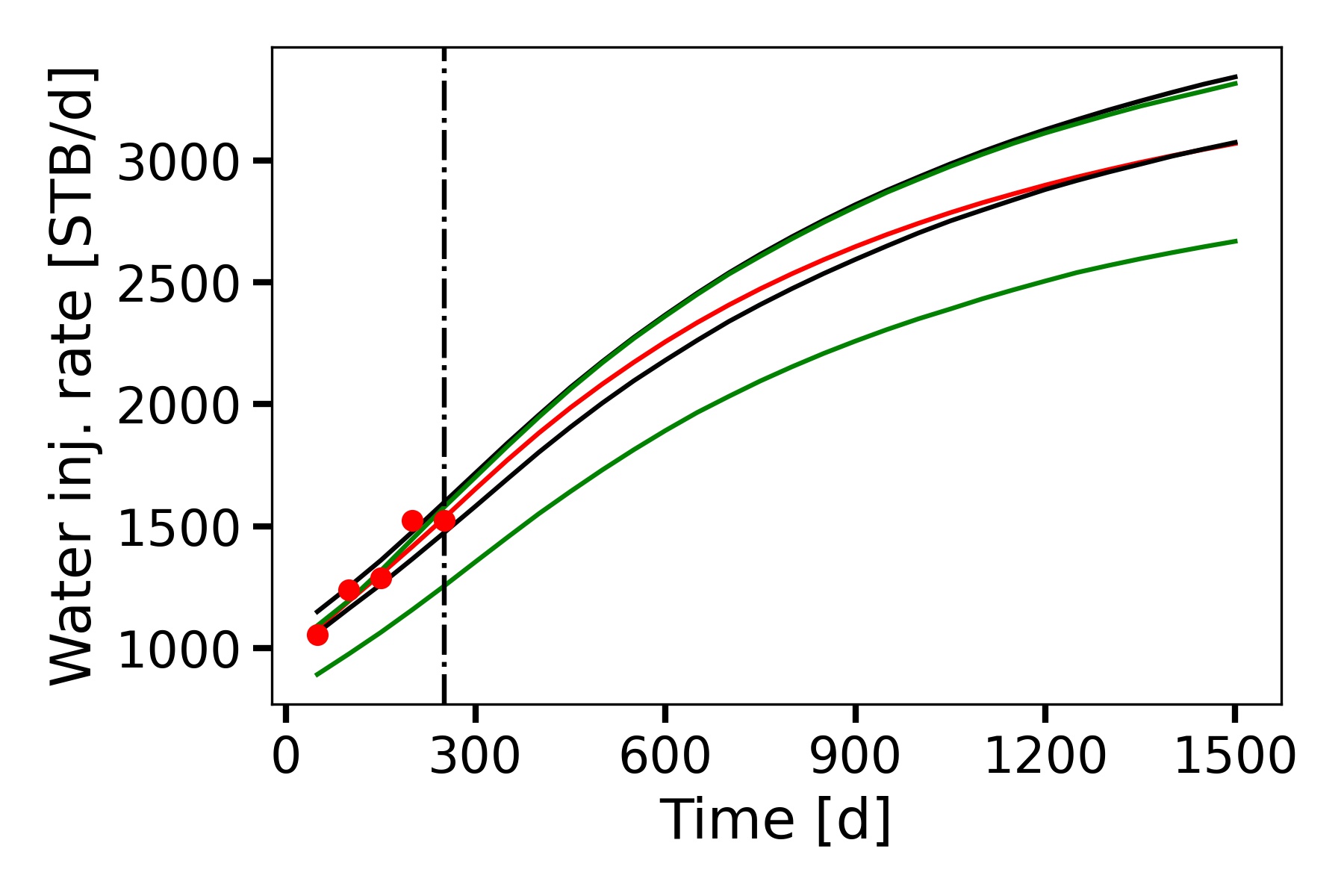}
        \caption{I3 water injection rate}
    \end{subfigure}

    \caption{Case~1: model-space (black) and latent-space (green) ESMDA results, with localization, for selected wells. Black and green lines denote the posterior P$_{10}$ and P$_{90}$ curves, and red points and red curves represent observed and true data, respectively. The vertical black dot-dash line indicates the end of the history matching period.}
    \label{fig:esmda_comparison_rates_both}
\end{figure}

In all cases, we observe narrower spreads in the posterior predictions than in the prior results, indicating uncertainty reduction after DA. There is, however, substantially more uncertainty reduction in the model space than in latent space. To directly compare model-space and latent-space predictions (both with localization), we present results for the same wells in Figure~\ref{fig:esmda_comparison_rates_both}. Here the prior results have been omitted for better visualization. These plots clearly highlight large differences in the amount of uncertainty reduction achieved and in the degree of correspondence of the P$_{10}$--P$_{90}$ ranges with the observed data. 
Data matching accuracy is consistently higher when performing DA in model space.  

To quantify the accuracy of data matching, we use the root mean squared error (RMSE) between observed and simulated posterior data
\begin{equation}
\mathrm{RMSE}
=
\frac{1}{N_e}
\sum_{n=1}^{N_e}
\bigg\{
\frac{1}{N_d}
\sum_{i=1}^{N_d}
\left(
\mathbf{d}^{(N_a)}_{n}(i)
- 
\mathbf{d}_{\mathrm{true}}(i)
\right)^2
\bigg\}^{\frac {1} {2}}.
\label{eqn:rmse}
\end{equation}
We quantify the degree of uncertainty reduction on model parameters using the normalized variance (NV), as in~\citet{Lacerda2021}
\begin{equation}
\text{NV} = \frac{\operatorname{tr}({C}_{pp}^{(N_a)})}
{\operatorname{tr}({C}_{pp}^{(0)})}.
\label{eqn:NV}
\end{equation}
NV essentially measures the relative change from prior (superscript 0) to posterior (superscript $N_a$) variance across model parameters. It is computed in all cases in the geological space. An NV of 1 indicates virtually no variance loss from the prior to posterior ensemble, while an NV of 0 indicates ensemble collapse. NV is also computed separately for the dummy variables (this quantity is denoted $\text{NV}_{\text{dum}}$).

The uncertainty reduction in well rates is quantified as the ratio of posterior and prior P$_{90}$--P$_{10}$ ranges, i.e., 
\begin{equation}
\Delta \text{P}_{90-10}
=
\frac{1}{N_d}
\sum_{i=1}^{N_d}
\frac{
\mathrm{P}_{90}\!\left(\mathbf{d}^{(N_a)}_{n}(i)\right)
-
\mathrm{P}_{10}\!\left(\mathbf{d}^{(N_a)}_{n}(i)\right)
}{
\mathrm{P}_{90}\!\left(\mathbf{d}^{(0)}_{n}(i)\right)
-
\mathrm{P}_{10}\!\left(\mathbf{d}^{(0)}_{n}(i)\right)
}.
\label{eq:p9010_ratio}
\end{equation}
This quantity corresponds to the uncertainty reduction between prior and posterior results shown in the well rate plots (Figures~\ref{fig:esmda_comparison_rates_model} and~\ref{fig:esmda_comparison_rates_latent}). Like NV, it spans from 0 (ensemble collapse), to 1 (no uncertainty reduction).

Values for these metrics, for both methods with (``+ loc.'') and without localization, are reported in Table~\ref{table:esmda_metrics_da}.  Model-space ESMDA achieves better data fitting than latent-space ESMDA, with RMSE values of $\sim$33~STB/d and $\sim$90~STB/d, respectively, for the cases with localization. The $\text{NV}_{\text{dum}}$ values are all close to 1, confirming that spurious correlations have limited impact on the ESMDA updates with these 1000-member ensembles. Slight improvements in the $\text{NV}_{\text{dum}}$ values are observed for the cases with localization, as would be expected. NV values for the model parameters are consistently lower in model space, indicating more uncertainty reduction in model space than in latent space.
Uncertainty reduction in the flow responses reflects the same trend. For the cases with localization, the $\Delta \text{P}_{90-10}\ \text{ratio}$ is $2.6\times$ higher (0.31 versus 0.12) in latent space than in model space.

\begin{table}[htbp]
\centering
\caption{Performance metrics for ESMDA in model and latent space for Case~1, with and without localization.}
\label{table:esmda_metrics_da}
\begin{tabular}{@{}lcccc@{}}
\toprule
\textbf{Algorithm} 
& \textbf{RMSE} [STB/d]
& \textbf{NV} 
& \textbf{NV$_{\text{dum}}$} 
& $\mathbf{\Delta \text{P}_{90-10}}$ \\
\midrule
ESMDA (model)         & 32.97 & 0.43 & 0.93 & 0.11 \\
ESMDA (model + loc.)  & 33.37 & 0.43 & 0.97 & 0.12 \\
ESMDA (latent)        & 68.45 & 0.51 & 0.90 & 0.28 \\
ESMDA (latent + loc.) & 90.42 & 0.56 & 0.93 & 0.31 \\
\bottomrule
\end{tabular}
\end{table}

The results presented in this section highlight the differences between model-space and latent-space ESMDA. On the one hand, more realistic uncertainty reduction and better data matching are achieved in model space (with or without localization) than in latent space. On the other hand, latent-space parameterization ensures plausible, geologically realistic posterior models, in contrast to the channel smear and loss of continuity observed in model space. The higher level of uncertainty reduction in flow results in model space may be due in part to the larger number of parameters available to achieve the data match, as well as to the lack of geological plausibility constraints. However, latent-space ESMDA seems to exhibit an inability to achieve more significant uncertainty reduction. Importantly, this behavior is opposite to the classical problem of ensemble collapse with ensemble-Kalman methods. 

\textcolor{black}{The model-space ESMDA results presented here can be viewed as ``baseline'' results that are indicative of the basic performance of ensemble-Kalman methods applied to channelized systems. The ESMDA option applied for this assessment is a standard-type procedure, and it is possible that other treatments, such as those involving thresholding or transforms, would perform differently (potentially better in some metrics). We reiterate that the reason for presenting model-space ESMDA results here is to highlight key performance differences with latent-space ESMDA. The overestimation of posterior uncertainty with latent-space ESMDA will be confirmed in the detailed results presented in Section~\ref{sec:hm_results}.}

In order to test the robustness of this finding, we performed additional tests with higher numbers of ESMDA iterations (up to 40) and we evaluated other ensemble Kalman filter methods (EnKF) in place of ESMDA. The same general trends as in the results presented above were consistently observed. This suggests that Kalman-gain-based algorithms (Eq.~\ref{eqn:esmda_update}) may not provide accurate posterior estimates when combined with highly nonlinear parameterizations, such as 3D-LDM. \textcolor{black}{This issue has been observed, to a limited extent, in previous studies on subsurface flow~\citep{meng_esmda_conference, DIFEDERICO2025105755, fan2026deep_fracture}. To our knowledge, however, this behavior has not been systematically assessed through detailed comparisons to Monte Carlo sampling methods.} Motivated by these findings, our goal is now to achieve both appropriate uncertainty reduction and the geological realism captured by the 3D-LDM parameterization. Along these lines, in the next section we describe a latent-space DA workflow that utilizes rigorous Monte Carlo-based sampling algorithms. Since these methods typically require ${\mathcal O}(10^5-10^6)$ function evaluations, we also develop a fast and accurate surrogate for the flow simulator. 

\section{Flow surrogate and MC-based data assimilation methods}
\label{sec:surr_and_mc}

In this section, we present the methods required for rigorous DA performed in the 3D-LDM latent space. First, we describe our flow surrogate model, which enables the evaluation of large numbers of forward runs at greatly reduced computational cost. Then, we present the Markov chain Monte Carlo (MCMC) and Sequential Monte Carlo (SMC) posterior sampling methods, applied on the latent variable $\boldsymbol{\xi}_T$.

\subsection{Flow surrogate model}

We develop a neural network-based surrogate, denoted $\hat{F}: \mathbb{R}^{N_c} \rightarrow \mathbb{R}^{N_t \times N_{\text{QoI}}}$ ($\hat{F} \approx F$), to learn the mapping in Eq.~\ref{eq:sim_output} from $\mathbf{m}_0$ to the well flow rates, i.e., 
\begin{equation} 
\hat{F}(\mathbf{m}_0, \theta) = [\hat{\mathbf{q}}_{\text{inj}}^{1}, \dots, \hat{\mathbf{q}}_{\text{inj}}^{N_\text{inj}}, \, \hat{\mathbf{q}}_{\text{oil}}^{1}, \dots, \hat{\mathbf{q}}_{\text{oil}}^{N_\text{prod}}, \, \hat{\mathbf{q}}_{\text{wat}}^{1}, \dots, \hat{\mathbf{q}}_{\text{wat}}^{N_\text{prod}}]. \label{eq:surr_output} \end{equation}
Here $\theta$ represents the trainable parameters of the surrogate and the ``hat'' superscript indicates a surrogate-predicted output. Unlike many surrogate models, our model does not predict reservoir state variables (e.g., pressure and saturation) and then compute well responses. Instead, we adopt a fit-for-purpose strategy that directly predicts the full well-rate time series. The surrogate is therefore not intended to replicate the full output of the numerical simulator, but rather to provide fast and accurate estimates of the quantities required for DA. This greatly simplifies the surrogate architecture and training requirements. 

\begin{figure}[htbp]
    \centering
    \includegraphics[width=\textwidth, trim=7cm 0cm 15cm 8cm, clip]{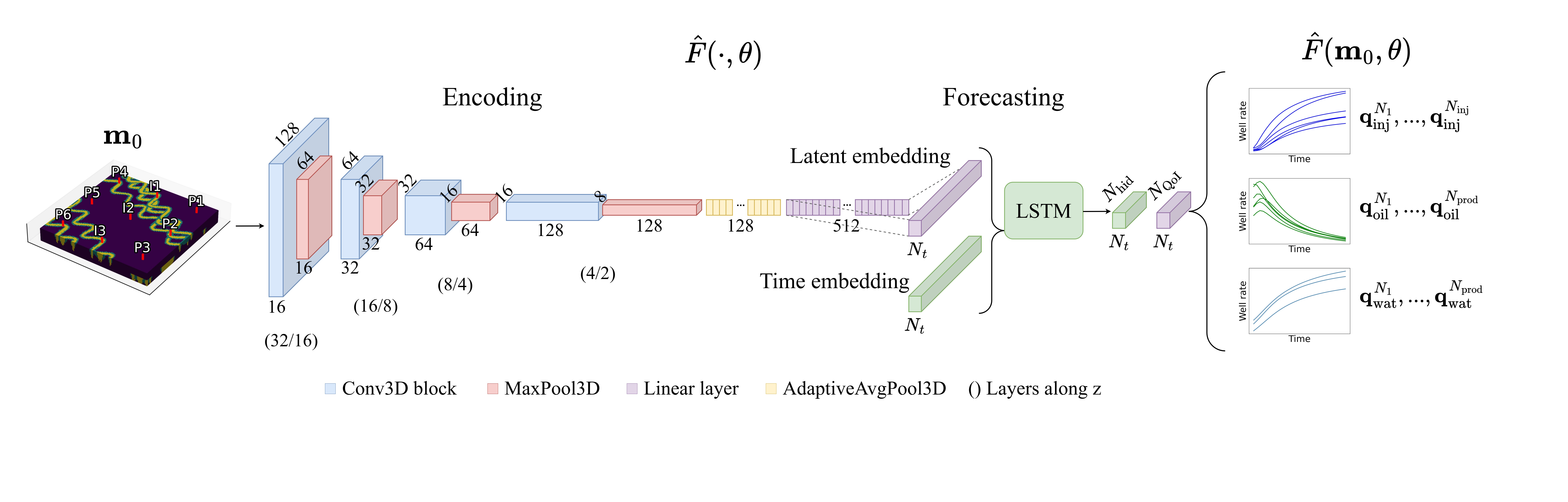}
    \caption{Schematic of the flow surrogate model. Each block represents the output of the corresponding layer(s),
    with values indicating the number of channels and dimension. The $z$-axis is hidden (and reported in brackets) to simplify the stacked 3D feature maps.}
    \label{fig:flow_surr_scheme}
\end{figure}

A schematic representation of the surrogate model is shown in Figure~\ref{fig:flow_surr_scheme}. The architecture combines 3D convolutional layers to learn geological spatial features, and recurrent layers (long short-term memory, LSTM~\citep{hochreiter1997long}), to learn the temporal evolution of well rates. The geomodel $\mathbf{m}_0$ is first encoded by the convolutional layers into a lower-dimensional representation, which is then decoded by the LSTM unit to predict the time-varying well rates. The detailed architecture of the surrogate model is provided in Table~\ref{tab:geotorates_architecture}. Each Conv3D Block in the encoder consists of Conv3D $\rightarrow$ BatchNorm3D $\rightarrow$ ReLU. The latent representation is passed through an LSTM with time embedding, and the output layer provides $N_{\text{QoI}}$ time series.

\begin{table}[htbp]
\centering
\caption{
Architecture of the flow surrogate model, with input shape $(1,N_x,N_y,N_z)$.}

\footnotesize
\renewcommand{\arraystretch}{1.5}\begin{tabular}{lllc}
\hline
\textbf{Stage} & \textbf{Operation} & \textbf{Parameters} & \textbf{Output} \\
\hline
Input & Input & - & $(1,N_x,N_y,N_z)$ \\
\hline
\multirow{8}{*}{Encoder}
& Conv3D Block 1 & 16 filters, $3\times3\times3$, s=1, p=1 & $(16,N_x,N_y,N_z)$ \\
& MaxPool3D & $2\times2\times2$, s=2 & $(16,N_x/2,N_y/2,N_z/2)$ \\
& Conv3D Block 2 & 32 filters, $3\times3\times3$, s=1, p=1 & $(32,N_x/2,N_y/2,N_z/2)$ \\
& MaxPool3D & $2\times2\times2$, s=2 & $(32,N_x/4,N_y/4,N_z/4)$ \\
& Conv3D Block 3 & 64 filters, $3\times3\times3$, s=1, p=1 & $(64,N_x/4,N_y/4,N_z/4)$ \\
& MaxPool3D & $2\times2\times2$, s=2 & $(64,N_x/8,N_y/8,N_z/8)$ \\
& Conv3D Block 4 & 128 filters, $3\times3\times3$, s=1, p=1 & $(128,N_x/8,N_y/8,N_z/8)$ \\
& MaxPool3D & $2\times2\times2$, s=2 & $(128,N_x/16,N_y/16,N_z/16)$ \\
\hline
Spatial Pooling & AdaptiveAvgPool3d & - & $(128,1,1,1)$ \\
Flatten & Flatten & - & $(128)$ \\
\hline
Latent Vector & Linear + ReLU 
&  - & $(512)$ \\
\hline
Time Embedding & Embedding & $30 \times 512$ & $(N_t, 512)$ \\
Concatenation & Cat([latent, time\_emb]) & - & $(N_t, 1024)$ \\
\hline
Decoder & LSTM (1 layer) & $N_{\text{hid}}=64$ & $(N_t, 64)$ \\
\hline
Output & Linear & - & $(N_t, N_{\text{QoI}})$ \\
\hline
\end{tabular}
\label{tab:geotorates_architecture}
\end{table}


\textcolor{black}{The flow surrogate is trained on numerical simulation results for the same set of 3000 geomodels that were used to construct the 3D-LDM (described in Section~\ref{sec:geomodels_setup}). The same 80-10-10\% split for training, validation, and testing is applied.} The simulation setup is the same as in Section~\ref{sec:geomodels_setup}. Training simulations are performed 20~runs at a time, resulting in a total elapsed runtime of approximately 25~hours. For each sample, the training loss is evaluated on a per-well and per-time-step basis by computing mean-squared error (MSE) between simulator and surrogate well-rate quantities, i.e.,
\begin{equation}
    L_{\text{surr}} = \text{MSE}({\hat{F}}(\mathbf{m}_0, \theta), {F}(\mathbf{m}_0)).
    \label{eq:train_loss_surr}
\end{equation}
The loss is computed and averaged over batches of size \(N_b\). 

The model is trained using the Adam optimizer~\citep{adam} with a batch size of 16 and a learning rate of $5 \times 10^{-3}$. The total training time is approximately 35~minutes on a single NVIDIA A100 GPU. Once trained, a single forward evaluation requires approximately 0.001~s on the same GPU, or about 0.2~s on a CPU. This represents significant speedup relative to the numerical simulator (which requires about 10~minutes per run on a CPU).

\subsection{Surrogate model performance}

We now evaluate the accuracy of well rate predictions for the trained flow surrogate model. All metrics refer to the testing set of $N_{\text{test}} =300$ samples. We define the following relative errors for water injection, oil production, and water production
\begin{equation}
    \delta_{\text{inj}} = \frac{1}{N_t N_{\text{inj}}} \sum_{i=1}^{N_{\text{inj}}} \sum_{t=1}^{N_t}
\frac{\left| \hat{\mathbf{q}}^{i}_{\text{inj}}(t) - \mathbf{q}^{i}_{\text{inj}}(t) \right|}
{\mathbf{q}^{i}_{\text{inj}}(t) + \epsilon_d},
\label{eq:test_loss_inj}
\end{equation}
\begin{equation}
    \delta_{\text{oil}} = \frac{1}{N_t N_{\text{prod}}} \sum_{j=1}^{N_{\text{prod}}} \sum_{t=1}^{N_t}
\frac{\left| \hat{\mathbf{q}}^{j}_{\text{oil}}(t) - \mathbf{q}^{j}_{\text{oil}}(t) \right|}
{\mathbf{q}^{j}_{\text{oil}}(t) + \epsilon_d},
\label{eq:test_loss_oil}
\end{equation}
\begin{equation}
    \delta_{\text{wat}} = \frac{1}{N_t N_{\text{prod}}} \sum_{j=1}^{N_{\text{prod}}} \sum_{t=1}^{N_t}
\frac{\left| \hat{\mathbf{q}}^{j}_{\text{wat}}(t) - \mathbf{q}^{j}_{\text{wat}}(t) \right|}
{\mathbf{q}^{j}_{\text{wat}}(t) + \epsilon_d},
\label{eq:test_loss_wat}
\end{equation}
where $\epsilon_d=0.1$ is a constant to prevent division by very small values and $t$ is the physical simulation time (not to be confused with the diffusion step index introduced in Section~\ref{sec:3d_ldm_esmda}). We considered a range of values for $\epsilon_d$ and found the error results to be quite insensitive to this specification.

\begin{figure}[htbp]
    \centering
    \includegraphics[width=0.65\textwidth, trim=0cm 0cm 0 0, clip]{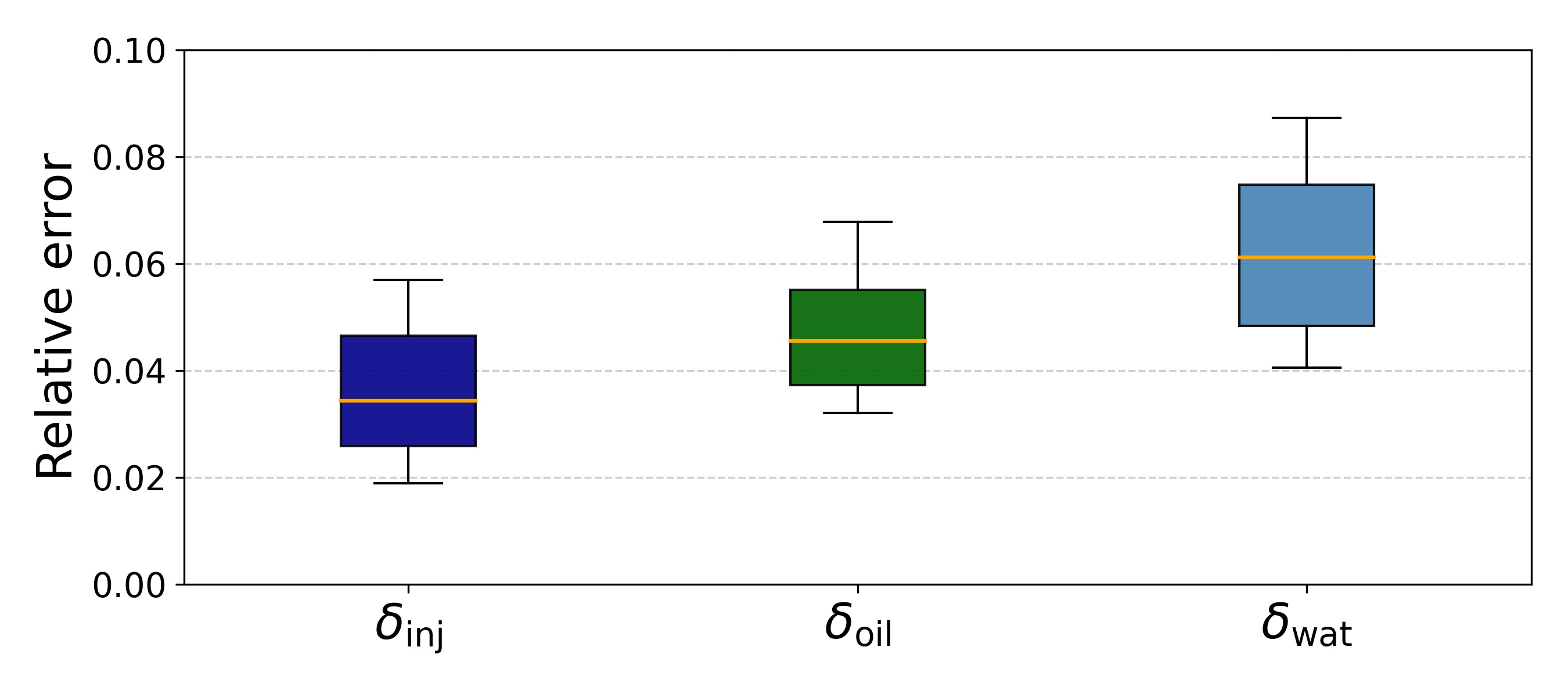}
    \caption{Relative errors for water injection, oil production, and water production rates for the surrogate model relative to the simulator for the testing set.}
    \label{fig:box_errors_surr}
\end{figure}

Box plots for $\delta_{\text{inj}}$, $\delta_{\text{oil}}$, and $\delta_{\text{wat}}$ are presented in Figure~\ref{fig:box_errors_surr}. The lines extending beyond the boxes show the P$_{10}$ and P$_{90}$ percentile errors, and the box edges indicate the P$_{25}$ and P$_{75}$ percentile errors. The orange lines inside the boxes denote the P$_{50}$ errors. The mean values for $\delta_{\text{inj}}$,  $\delta_{\text{oil}}$, and $\delta_{\text{wat}}$ are 3.7\%, 4.7\%, and 6.3\%, respectively. These errors are comparable to those reported in~\citet{TANG2022103692} and~\citet{han_surrogate_mcmc}, who also used RNN-based surrogate models for DA in subsurface flow problems.

\begin{figure}[htbp]
    \centering
    \begin{subfigure}[b]{0.32\textwidth}
        \centering
        \includegraphics[width=\textwidth, trim=0 0 0 0, clip]{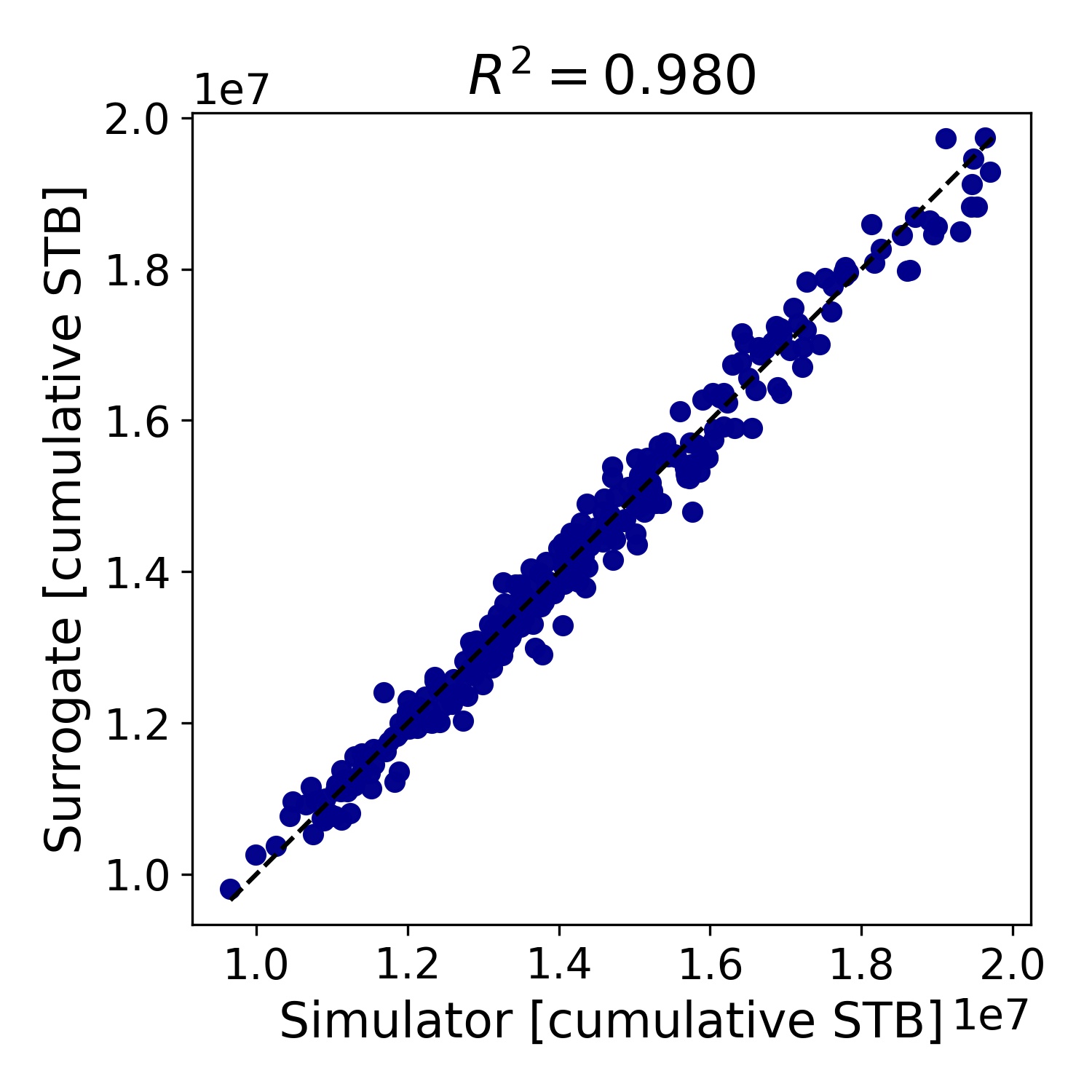}
        \caption{Field water inj. rate}
        \label{fig:cumul_err_inj}
    \end{subfigure}
    \hfill
    \begin{subfigure}[b]{0.32\textwidth}
        \centering
        \includegraphics[width=\textwidth, trim=0 0 0 0, clip]{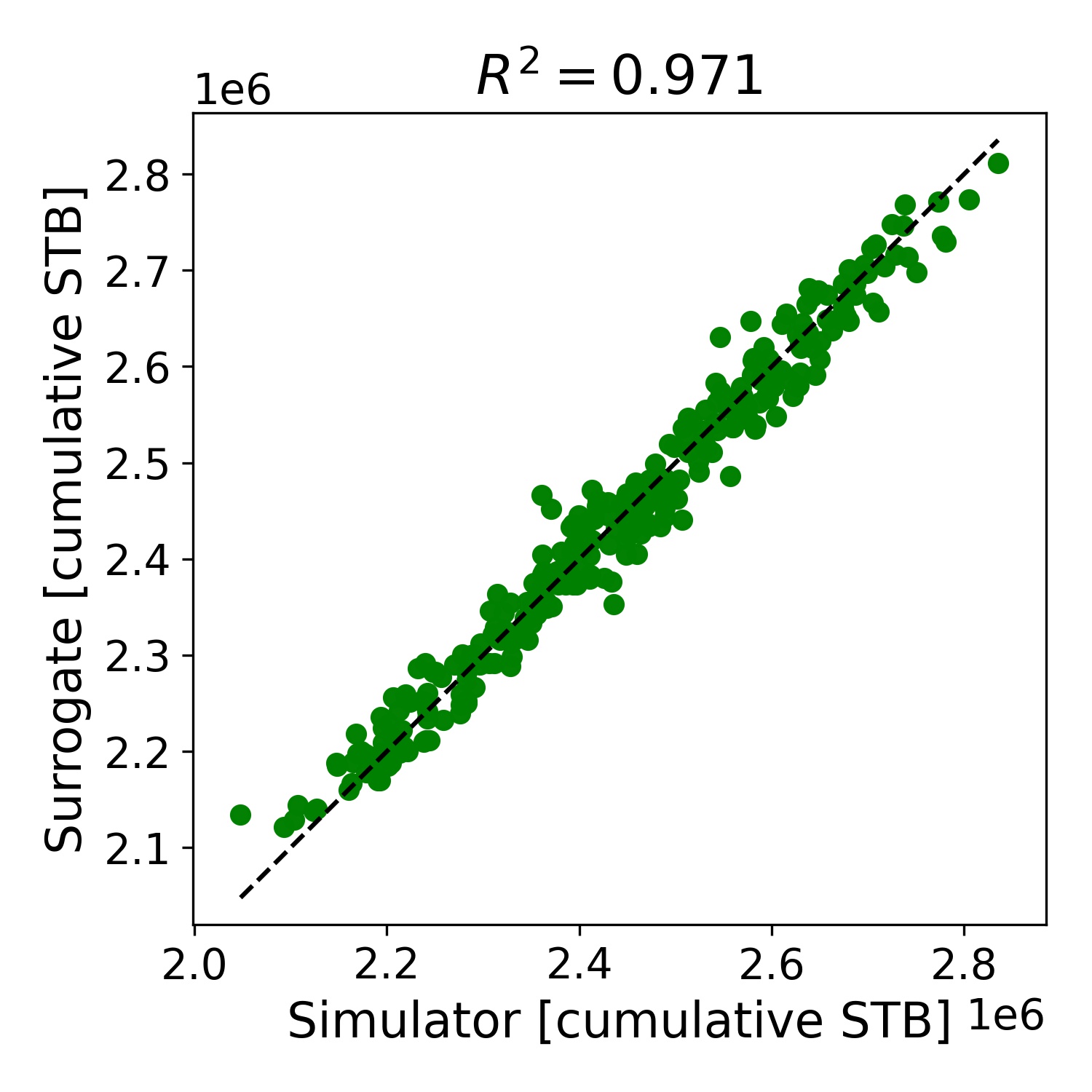}
        \caption{Field oil prod. rate}
        \label{fig:cumul_err_oil}
    \end{subfigure}
    \hfill
    \begin{subfigure}[b]{0.32\textwidth}
        \centering
        \includegraphics[width=\textwidth, trim=0 0 0 0, clip]{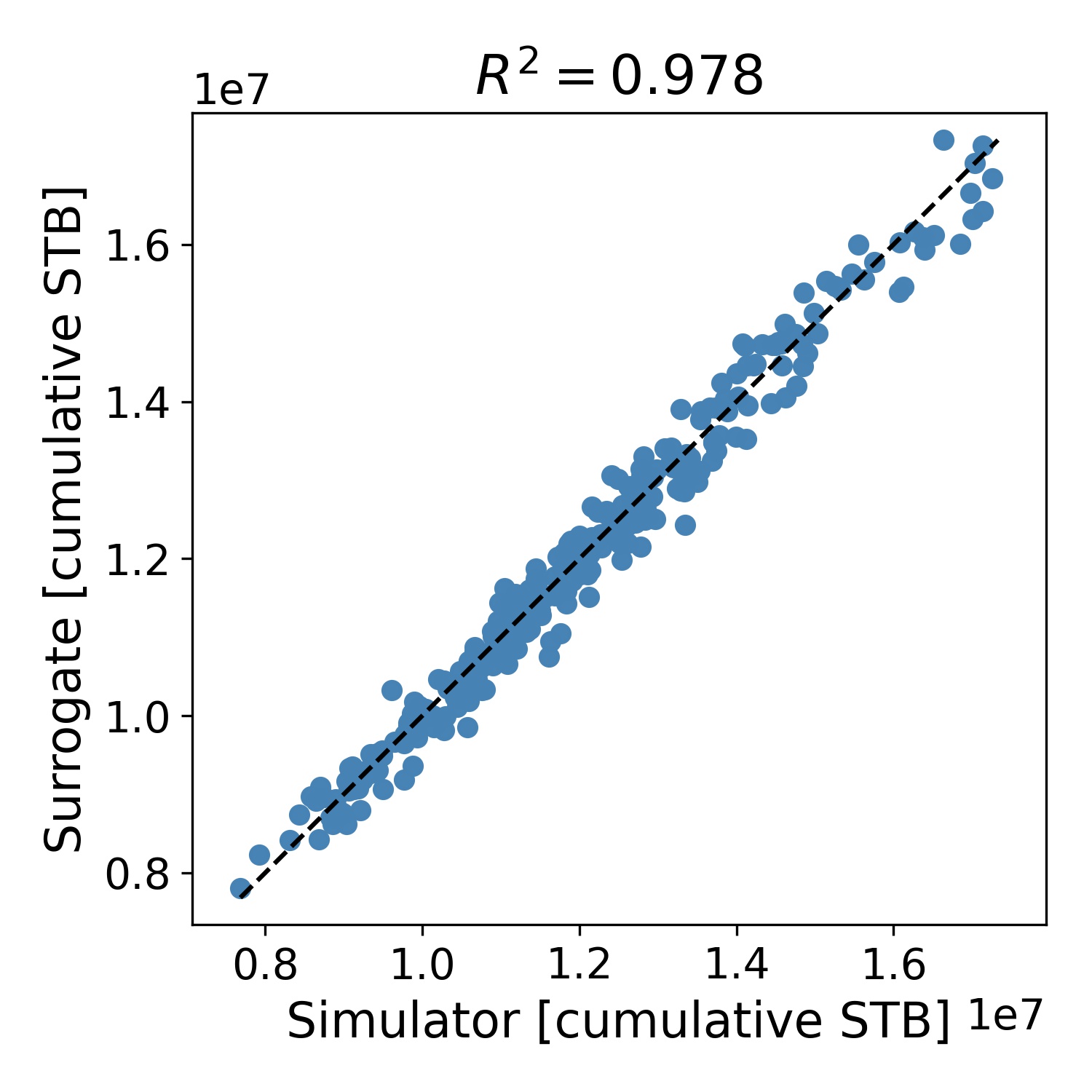}
        \caption{Field water prod. rate}
        \label{fig:cumul_err_water}
    \end{subfigure}
    \caption{Scatter plots comparing surrogate model and simulator results for cumulative water injection, oil production, and water production over the full simulation time frame. The $R^2$ coefficient is reported for each plot.}
    \label{fig:cumul_errs}
\end{figure}

\begin{figure}[htbp]
    \centering
    \begin{subfigure}[b]{0.32\textwidth}
        \centering
        \includegraphics[width=\textwidth, trim=15 0 0 0, clip]{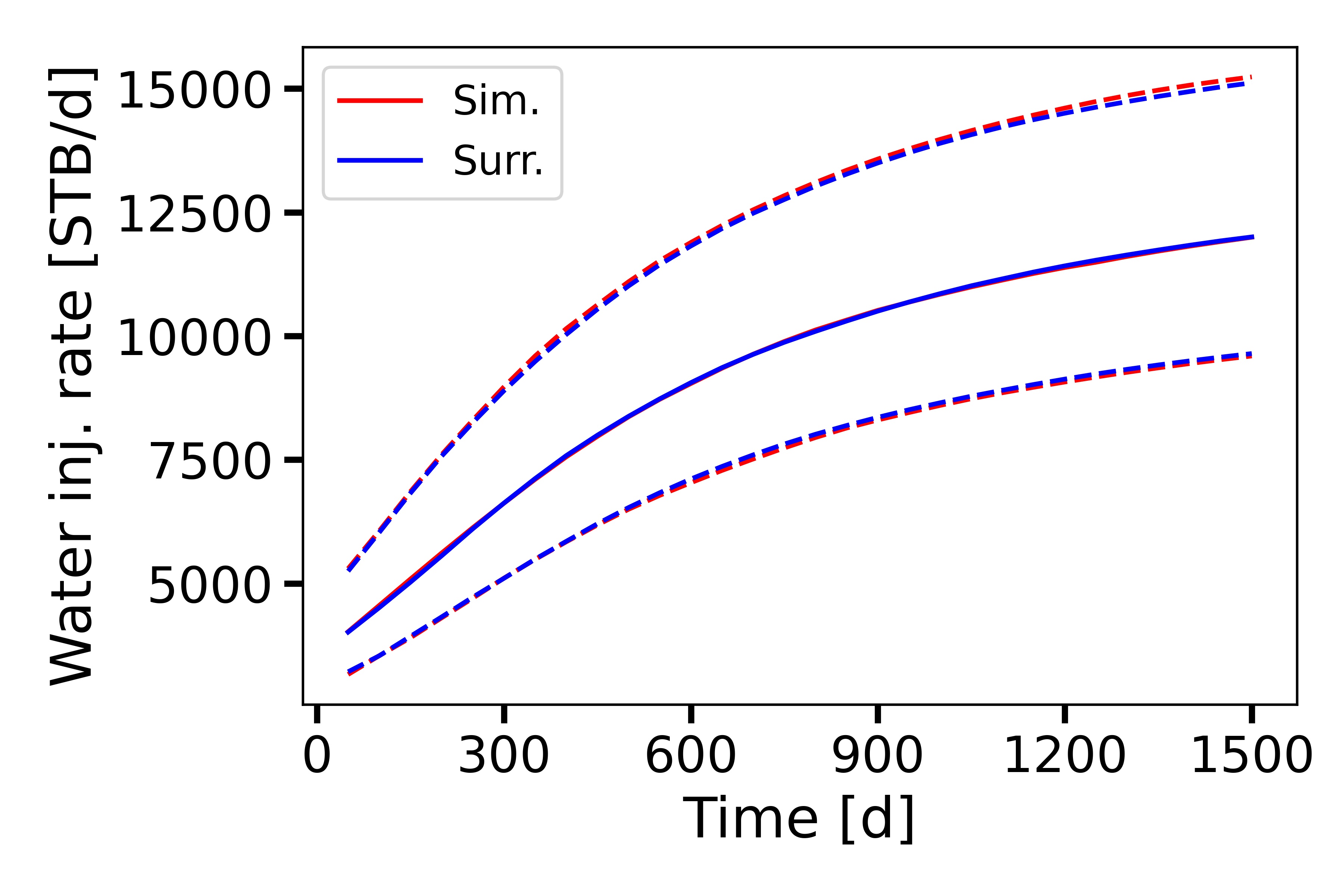}
        \caption{Field water inj. rate}
        \label{fig:field_inj}
    \end{subfigure}
    \hfill
    \begin{subfigure}[b]{0.32\textwidth}
        \centering
        \includegraphics[width=\textwidth, trim=15 0 0 0, clip]{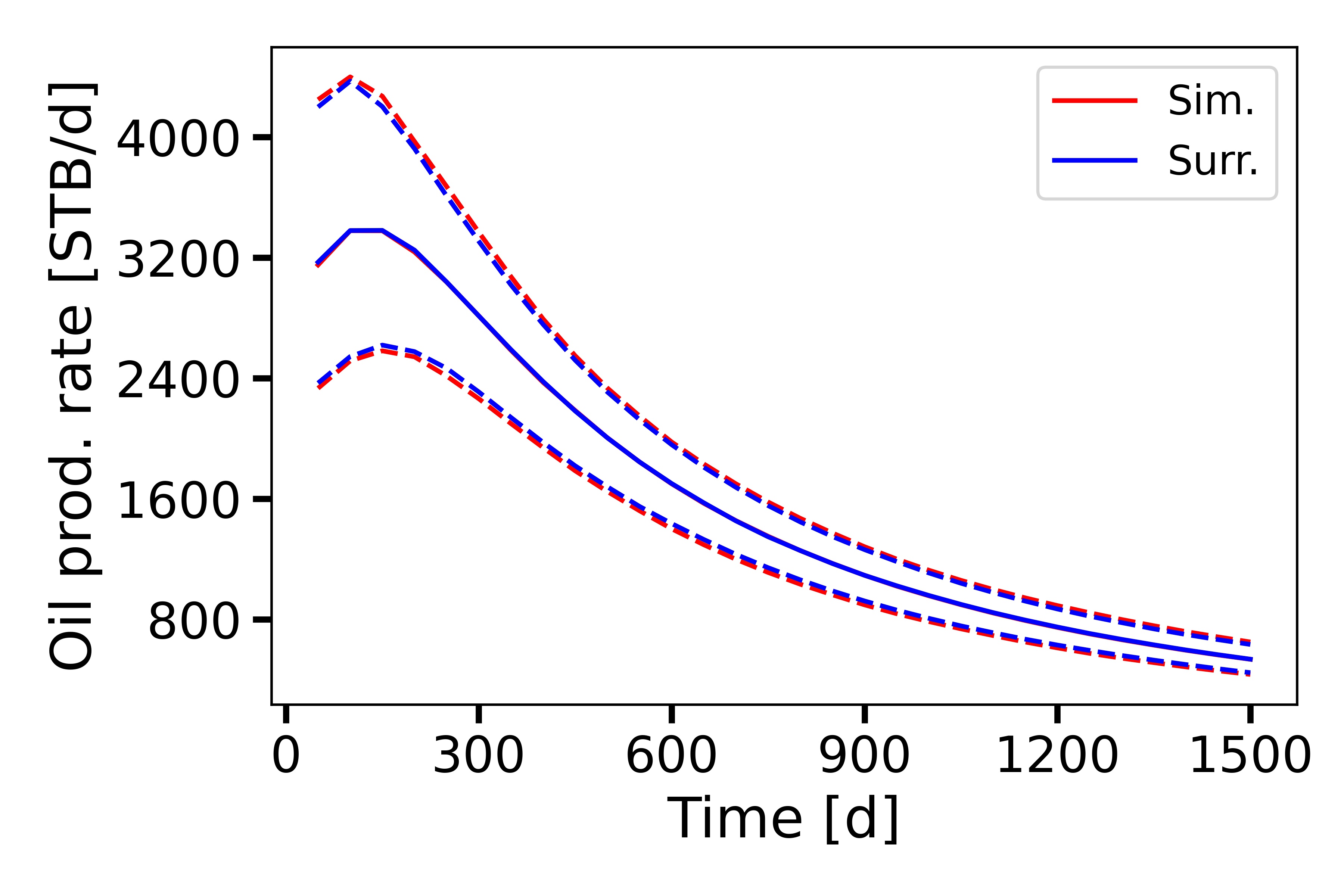}
        \caption{Field oil prod. rate}
        \label{fig:field_oil}
    \end{subfigure}
    \hfill
    \begin{subfigure}[b]{0.32\textwidth}
        \centering
        \includegraphics[width=\textwidth, trim=15 0 0 0, clip]{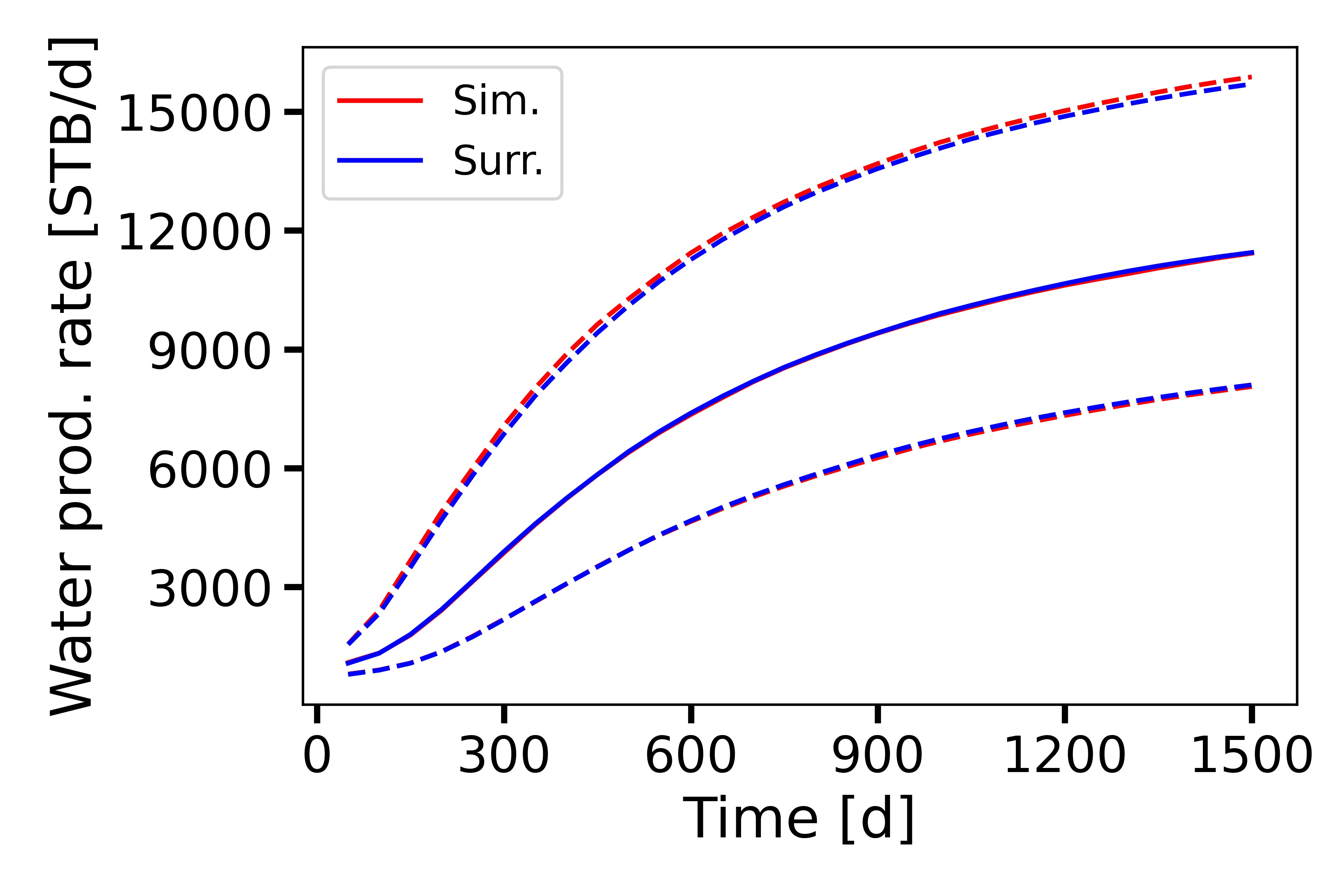}
        \caption{Field water prod. rate}
        \label{fig:field_wat}
    \end{subfigure}
    \caption{Field-level flow statistics from the simulator (red curves) and surrogate model (blue curves) over the testing set of 300 geomodels. The upper, middle (solid) and lower curves correspond to the P$_{90}$, P$_{50}$ and P$_{10}$ percentile results.}
    \label{fig:field_rates}
\end{figure}

Scatter plots for cumulative quantities at the end of the simulation time frame (1500~days) are shown in Figure~\ref{fig:cumul_errs}. We observe a high degree of surrogate model accuracy for all three quantities, with the points closely aligned with the 45-degree line (i.e., $R^2$ near 1).  As a final assessment, in Figure~\ref{fig:field_rates} we present flow statistics for field-level flow rates across the testing set. The dashed lines indicate the P$_{10}$ and P$_{90}$ responses, and solid lines show the P$_{50}$ results. Very close visual agreement between the two sets of curves is clearly achieved. Although not reported for purposes of brevity, similar surrogate model performance is observed at the well level.

Although the results presented above demonstrate that the surrogate model is quite accurate, it is (technically) an imperfect model relative to the reference numerical simulator. Thus, it is important to account for this error within the DA workflow. To accomplish this we follow the approach used in~\citet{han_surrogate_mcmc}, where the surrogate model error is quantified for each of the $N_{\text{test}}$ samples from the testing set. The error for each sample $n$, denoted $\boldsymbol{\epsilon}_{\text{surr},n}  \in \mathbb{R}^{n_{d}}$, is given by
\begin{equation}
\boldsymbol{\epsilon}_{\text{surr},n} = \mathbf{d}_{n} - \hat{\mathbf{d}}_{n}, 
\label{eq:eps_surr}
\end{equation}
where $\mathbf{d}_{n}$ and $\hat{\mathbf{d}}_{n}$ represent the simulator and surrogate model observations (that is, $F_{\text{obs}}(\mathbf{m}_{0,n})$ and $\hat{F}_{\text{obs}}(\mathbf{m}_{0,n}, \theta)$).
The full centered matrix $E_{\text{surr}} \in \mathbb{R}^{n_{d} \times N_{\text{test}}}$ can be constructed as follows
\begin{equation}
E_{\text{surr}} = \frac{1}{\sqrt{N_{\text{test}} - 1}} \, 
\left[
\boldsymbol{\epsilon}_{\text{surr},1} - \bar{\boldsymbol{\epsilon}}_{\text{surr}}, \;
\boldsymbol{\epsilon}_{\text{surr},2} - \bar{\boldsymbol{\epsilon}}_{\text{surr}}, \;
\cdots, \;
\boldsymbol{\epsilon}_{\text{surr},N_{\text{test}}} - \bar{\boldsymbol{\epsilon}}_{\text{surr}}
\right],
\label{eq:E_surr}
\end{equation}
where $\bar{\boldsymbol{\epsilon}}_{\text{surr}} \in \mathbb{R}^{n_{d}}$ is the mean of the surrogate model error vector over the $N_{\text{test}}$ samples. 

The mean and standard deviation of the rate errors across all observation points are 4.3~STB/d and 62.4~STB/d, respectively. 
We also compute the bias for each observation normalized by the corresponding standard deviation of the same quantity across the testing set. The average value obtained is 0.007, meaning the bias is on average less than 1\% of one standard deviation of the observations. It is therefore appropriate to treat the trained flow surrogate model as unbiased, in which case the model error can be represented in terms of a covariance.

Following~\citet{Oliver_Reynolds_Liu_2008}, this model error covariance, denoted $C_{\text{surr}} \in \mathbb{R}^{n_{d} \times n_{d}}$, is given by
\begin{equation}
C_{\text{surr}} = E_{\text{surr}} \, E_{\text{surr}}^\top .
\label{eq:C_surr}
\end{equation}
This matrix is, in general, nondiagonal and can thus account for both spatial and temporal correlations. We combine $C_{\text{surr}}$ with measurement error to give the total covariance matrix $C_{\text{tot}}$, 
\begin{equation}
    C_{\text{tot}} = C_d + C_{\text{surr}}. 
    \label{eq:total_model_err_cov}
\end{equation}
The low (and essentially unbiased) surrogate model error, combined with our accounting for imperfect-model effects through $C_{\text{surr}}$, enable the use of the surrogate for DA. This is an essential capability given the very high computational demands of the Monte Carlo-based methods we now describe.

\textcolor{black}{In summary, the surrogate constructed in this work is designed to provide well rates for wells at fixed locations operating under specified BHPs. Retraining would be required if the well locations or controls are changed. If necessary, the surrogate could be generalized to treat time-varying BHPs by adding well controls as surrogate inputs and modifying the architecture accordingly, along the lines described in~\citep{yongdo_spej}.}

\subsection{Markov chain Monte Carlo algorithm}
\label{sec:mcmc}
The Markov chain Monte Carlo (MCMC) algorithm entails simulating a Markov chain, or a few chains in parallel, to generate a sequence of samples. With a sufficient number of iterations, MCMC is considered to approximate the posterior distribution. In this work we use the preconditioned Crank-Nicolson (pCN) MCMC algorithm~\citep{cotter_pcn, chen2019dimensionrobustmcmcbayesianinverse}, which is appropriate for Gaussian variables. At every new iteration $k+1$, the current latent sample $\boldsymbol{\xi}_T^{(k)}$ is perturbed to generate a new candidate sample, denoted $\boldsymbol{\xi}_T'$, through application of
\begin{equation}
\boldsymbol{\xi}^{\prime}_T = \{1 - \beta^2_{\text{MCMC}}\}^{\frac {1}{2}} \, \boldsymbol{\xi}^{(k)}_T + \beta_{\text{MCMC}} \, \boldsymbol{\epsilon}, \quad \text{with }  \boldsymbol{\epsilon} \sim \mathcal{N}(\mathbf{0}, \mathbf{I}_{n_c}).
\label{eq:mcmc_update}
\end{equation}
Here $\beta_{\text{MCMC}}$ is a scalar coefficient that controls the update step size. This parameter is typically adjusted to achieve an acceptance rate in the range of 10-40\%.
The pCN proposal preserves the Gaussian prior for $\boldsymbol{\xi}_T^{\prime}$ and is dimension-robust, enabling stable performance even for a relatively high-dimensional latent space.

The likelihood is taken to be of the usual form, i.e., 
\begin{equation}
p(\mathbf{d}^{\ast}_{\text{obs}} \mid \boldsymbol{\xi}_T  ) \propto \exp \left(-\frac{1}{2} \, (\mathbf{d} - \mathbf{d}^{\ast}_{\text{obs}})^\top 
\, {C}_{\rm tot}^{-1} \, (\mathbf{d} - \mathbf{d}^{\ast}_{\text{obs}}) \right).
\label{eq:likelihood}
\end{equation}
Here $\mathbf{d}$ denotes the simulated data, obtained by first generating the corresponding geomodel $\mathbf{m}_0^{\text{LDM}}(\boldsymbol{\xi}_T)$ through 3D-LDM, and then evaluating the flow response using the surrogate, $\mathbf{d} = {\hat F_{\text{obs}}}\big(\mathbf{m}^{\text{LDM}}_0(\boldsymbol{\xi}_T)\big)$.
We apply the Metropolis-Hastings acceptance criterion~\citep{Hastings197097}, where the probability of accepting the proposed latent variable $\boldsymbol{\xi}_T^{\prime}$ is given by
\begin{equation}
\alpha = \min\left(1, 
\frac{p(\mathbf{d}_{\text{obs}}^{\ast} \mid \boldsymbol{\xi}_T^{\prime})}{p(\mathbf{d}_{\text{obs}}^{\ast} \mid \boldsymbol{\xi}_T^{(k)})}
\right).
\label{eq:alpha_min_mcmc}
\end{equation}
The maximum number of iterations, $N_{\text{max}}$, is set to 500,000. The first 10,000 iterations, which constitute the burn-in period, are discarded to mitigate initialization bias. Algorithmic details are provided in Algorithm~\ref{alg:latent_mcmc}.

\begin{minipage}[t]{0.85\textwidth}
\begin{algorithm}[H]
\caption{MCMC-based data assimilation (single chain)}
\label{alg:latent_mcmc}
\begin{algorithmic}[1]
\State Initialization: $k = 0$
\State Sample $\boldsymbol{\xi}_T^{(k)} \sim \mathcal{N}(\mathbf{0}, \mathbf{I}_{n_c})$
\State Generate $\mathbf{m}^{\text{LDM}}_0(\boldsymbol{\xi}_T^{(k)})$
\State Simulate $\mathbf{d}^{(k)} = {\hat F}_{\text{obs}}\big(\mathbf{m}^{\text{LDM}}_0(\boldsymbol{\xi}_T^{(k)})\big)$
\State Compute $p(\mathbf{d}^{\ast}_{\text{obs}} \mid \boldsymbol{\xi}_T^{(k)})$ using Eq.~\ref{eq:likelihood}
\While{not converged \textbf{and} $k < N_{\text{max}}$}  
\State Propose $\boldsymbol{\xi}_T'$ using Eq.~\ref{eq:mcmc_update}
    \State Repeat steps~3--5 using $\boldsymbol{\xi}_T'$ to obtain $\mathbf{d}'$ and $p(\mathbf{d}_{\text{obs}}^{\ast} \mid \boldsymbol{\xi}_T')$
    \State Compute acceptance probability $\alpha$ using Eq.~\ref{eq:alpha_min_mcmc}
    \State Draw $u \sim \mathcal{U}(0,1)$
    
    \If{$u < \alpha$}
        \State $\boldsymbol{\xi}_T^{(k+1)} = \boldsymbol{\xi}_T'$
    \Else
        \State $\boldsymbol{\xi}_T^{(k+1)} = \boldsymbol{\xi}_T^{(k)}$
    \EndIf
    \State $k = k + 1$
\EndWhile
\end{algorithmic}
\end{algorithm}
\end{minipage}

\bigskip

\textcolor{black}{In our implementation, we run five chains in parallel. We use the $\hat{R}$ diagnostic~\citep{Vehtari_2021} to assess the overall convergence of the DA procedure. This metric quantifies the within-chain and between-chain variance for latent model parameters. Convergence is taken to occur when all parameters have reached $\hat{R} < 1.01$, as recommended by~\citet{Vehtari_2021}, or when the practical limit $N_{\text{max}}=$~500,000 is reached. Since the $\hat{R} < 1.01$ condition can be difficult to achieve in high-dimensional problems, a threshold of 1.05 is often used in practice. We will show in Section~\ref{sec:impact} that the use of $\hat{R} < 1.05$ appears to suffice for our problem (though we reiterate that $\hat{R} < 1.01$ is used except where otherwise specified).}

\subsection{Tempered Sequential Monte Carlo algorithm (SMC)}

The tempered SMC algorithm~\citep{del_moral, Dai03072022} combines the computational efficiency of ensemble-based methods with rigorous Monte Carlo posterior sampling. For this method, at each iteration $k$, the target density is defined as
\begin{equation}
p_k(\boldsymbol{\xi}_T \mid \mathbf{d}_{\mathrm{obs}}^{\ast}) \propto\ p(\boldsymbol{\xi}_T)\,
p\!\left(\mathbf{d}_{\mathrm{obs}}^{\ast} \mid \boldsymbol{\xi}_T \right)^{\beta_{k}},
\label{eq:smc_likelihood}
\end{equation}
where $\beta_k \in [0,1]$ denotes the tempering parameter. An ensemble of $N_e$ samples, denoted $\{\boldsymbol{\xi}_{T,n}^{(k)}\}_{n=1}^{N_e}$, is propagated from $p_{k}(\boldsymbol{\xi}_T \mid \mathbf{d}_{\mathrm{obs}}^{\ast})$ to $p_{k+1}(\boldsymbol{\xi}_T \mid \mathbf{d}_{\mathrm{obs}}^{\ast})$ through three steps -- importance weighting, resampling, and short-chain MCMC mutation. When $\beta_k=0$, the target is the prior distribution $p(\boldsymbol{\xi}_T)$, and when $\beta_k=1$, the target is the posterior distribution $p(\boldsymbol{\xi}_T \mid \mathbf{d}_{\mathrm{obs}}^{\ast})$. The goal of tempering is to gradually introduce the posterior distribution via the adaptive $\beta_k$ schedule. Likelihoods are computed through Eq.~\ref{eq:likelihood}, as in the MCMC method. 

Our tempered SMC implementation proceeds as follows. First, we apply 3D-LDM to provide the geomodel corresponding to the latent variable, $\mathbf{m}_0^{\text{LDM}}(\boldsymbol{\xi}_T)$. The surrogate flow model then predicts the flow response, $\mathbf{d} = {\hat F}_{\text{obs}}\big(\mathbf{m}^{\text{LDM}}_0(\boldsymbol{\xi}_T)\big)$.
The next tempering parameter $\beta_{k+1}$ is selected adaptively to satisfy a prescribed effective sample size (ESS) target. This is achieved by determining $\Delta \beta_k  \in [0, 1 - \beta_k]$ (and thus $\beta_{k+1} = \beta_{k} + \Delta \beta_k$) through application of
\begin{equation}
\mathrm{ESS}(\Delta\beta) =
\frac{\left(\sum_{n=1}^{N_e} p(\mathbf{d}_{\mathrm{obs}}^{\ast} \mid \boldsymbol{\xi}_{T,n}^{(k)})^{\Delta\beta}\right)^2}
{\sum_{n=1}^{N_e} p(\mathbf{d}_{\mathrm{obs}}^{\ast} \mid \boldsymbol{\xi}_{T,n}^{(k)})^{2\Delta\beta}}.
\label{eq:update_beta_smc}
\end{equation}
This follows the standard ESS definition from importance weighting. It measures how evenly the likelihood is spread across the ensemble as a function of $\Delta\beta_k$. Eq.~\ref{eq:update_beta_smc} is solved, using a numerical root-finding method, to find the largest $\Delta \beta_{k}$ that keeps the ESS above a specified threshold ($0.5 N_{e}$ in this work). An excessively large $\Delta \beta_{k}$ would cause the likelihood to change abruptly, concentrating weight on a few members, while the rest collapse to zero. This would result in the loss of diversity within the ensemble.

Once the next $\beta_{k+1}$ is computed, weights are assigned as
\begin{equation}
\omega^{(k+1)}_{n}
=
\omega^{(k)}_{n}\,
p\!\left(
\mathbf{d}_{\mathrm{obs}}^{\ast}
\mid
\boldsymbol{\xi}_{T,n}^{(k)}
\right)^{\Delta \beta_k},
\quad \text{with} \quad
\Delta \beta_k = \beta_{k+1} - \beta_{k},
\label{eq:smc_weights}
\end{equation}
which leads to higher-likelihood samples having higher weights. The initial weights are set to $1/N_e$. After normalization, the normalized weights $\{\tilde{\omega}_n^{(k+1)}\}_{n=1}^{N_e}$ define a discrete probability distribution over the current ensemble. Systematic resampling is then used to draw a new ensemble of $N_e$ equally weighted samples, which acts to reduce or eliminate low-probability samples. The resampled ensemble is then mutated with MCMC steps using the pCN proposal (and constant step size $\beta_{\text{mut}}$), as described in Section~\ref{sec:mcmc}. In contrast to the standalone MCMC procedure, here these are very short chains involving only $N_{\text{mut}}=35$~iterations. The acceptance probability in Eq.~\ref{eq:alpha_min_mcmc} is modified to
\begin{equation}
\alpha
=
\min\!\left(
1,\;
\frac{
p\!\left(\mathbf{d}_{\mathrm{obs}}^{\ast} \mid \boldsymbol{\xi}_{T,n}^{\prime}\right)^{\beta_{k+1}}
}{
p\!\left(\mathbf{d}_{\mathrm{obs}}^{\ast} \mid \boldsymbol{\xi}_{T,n}^{(k+1)}\right)^{\beta_{k+1}}
}
\right).
\label{eq:alpha_min_smc}
\end{equation}
This modification is introduced to ensure sampling is performed from the tempered target distribution. The above steps are repeated until $\beta_{k+1} = 1$. \textcolor{black}{Since the final step is a reweighting, posterior statistics are computed using the final normalized weights $\{\tilde{\omega}_n\}_{n=1}^{N_e}$.}
Algorithmic details of our SMC procedure are provided in Algorithm~\ref{alg:latent_smc}.

\begin{minipage}[t]{0.85\textwidth}
\begin{algorithm}[H]
\caption{Tempered SMC-based data assimilation}
\label{alg:latent_smc}
\begin{algorithmic}[1]
\State Initialization: set $k=0$, $\beta_0 = 0$, $\{\omega_n^{(k+1)}\}_{n=1}^{N_e} =
\{\frac{1}{N_e}\}_{n=1}^{N_e}$
\State Sample $\{\boldsymbol{\xi}_{T,n}^{(k)}\}_{n=1}^{N_e} \sim \mathcal{N}(\mathbf{0},\mathbf{I}_{n_c})$
\For{$n = 1,\dots,N_e$}
\State Generate $\mathbf{m}^{\text{LDM}}_0(\boldsymbol{\xi}_{T,n}^{(k)})$
\State Simulate $\mathbf{d}_n^{(k)} = {\hat F}_{\text{obs}}\big(\mathbf{m}^{\text{LDM}}_0(\boldsymbol{\xi}_{T,n}^{(k)})\big)$
\State Compute $p(\mathbf{d}_{\mathrm{obs}}^{\ast} \mid \boldsymbol{\xi}_{T,n}^{(k)})$
\EndFor
\While{$\beta_k < 1$}
\State Determine $\beta_{k+1}$ using Eq.~\ref{eq:update_beta_smc}
\For{$n = 1,\dots,N_e$}
\State Update weights $\omega_n^{(k+1)}$ using Eq.~\ref{eq:smc_weights}
\EndFor
\State Normalize weights $\{\tilde{\omega}_n^{(k+1)}\}_{n=1}^{N_e}$
\If{$\beta_{k+1} < 1$}
\State Resample $\{\boldsymbol{\xi}_{T,n}^{(k+1)}\}_{n=1}^{N_e}$ according to $\{\tilde{\omega}_n^{(k+1)}\}$
\For{$n = 1,\dots,N_e$}
\For{$j = 1,\dots,N_{\text{mut}}$}
\State Propose $\boldsymbol{\xi}_{T,n}'$ using Eq.~\ref{eq:mcmc_update}
\State Repeat steps 4--6 for $\boldsymbol{\xi}_{T,n}'$
\State Compute $\alpha$ using Eq.~\ref{eq:alpha_min_smc}
\State Draw $u \sim \mathcal{U}(0,1)$
\If{$u < \alpha$}
\State $\boldsymbol{\xi}_{T,n}^{(k+1)} = \boldsymbol{\xi}_{T,n}'$
\Else
\State $\boldsymbol{\xi}_{T,n}^{(k+1)} = \boldsymbol{\xi}_{T,n}^{(k+1)}$
\EndIf
\EndFor
\EndFor
\EndIf
\State $k = k+1$
\EndWhile
\end{algorithmic}
\end{algorithm}
\end{minipage}

\section{Latent-space DA results using ESMDA, MCMC and SMC}
\label{sec:hm_results}

\textcolor{black}{In this section, we compare ESMDA, MCMC, and SMC, all operating in latent space, and all using the flow surrogate described in Section~\ref{sec:surr_and_mc} as the forward model. Function evaluations using the surrogate, as well as the 3D-LDM mappings, are performed on the same NVIDIA A100 GPU as is used for training.} The DA and flow problems are as presented in Section~\ref{sec:3d_ldm_esmda}. For SMC, we specify $N_e = 1000$. To assess the robustness of the workflow, we test three different true models (shown in Figure~\ref{fig:examples_petrel}). Case~1 is the same as that used in Section~\ref{sec:esmda_results}. For ESMDA, we use the same parameters and apply the localization procedure described earlier. 

\textcolor{black}{Estimates for wall-clock time and the number of function evaluations, for a typical run for each of the three DA methods (with all evaluations performed using the surrogate), are provided in Table~\ref{tab:algo_params}. Values for MCMC are given for two different stop criteria. The DA algorithms are parallelized, which means ESMDA ensemble members, the individual candidates in each MCMC chain, and the SMC particles can be independently evaluated. The evaluations required for a single ESMDA or SMC iteration require multiple calls to the GPU (i.e., all 1000 3D-LDM reconstructions required in an iteration cannot be performed simultaneously). The five MCMC chains, by contrast, can all be run in parallel on a single GPU. The computational costs of MCMC and SMC are dominated by the high number of forward evaluations, while most of the wall-clock time for ESMDA is spent on the Kalman update (Eq.~\ref{eqn:esmda_update}). Even with a large ensemble size, ESMDA requires by far the lowest number of function evaluations. Although the five MCMC chains run in parallel, this algorithm has the highest computational cost because each chain operates sequentially. SMC has characteristics of both ESMDA and MCMC, so its wall-clock time and function evaluations rank between the other two algorithms.}

\begin{table}[htbp]
\caption{Comparison of DA algorithms for Cases~1--3 using flow surrogate (DA performed in 3D-LDM latent space).}
\label{tab:algo_params}
\centering
\renewcommand{\arraystretch}{1.4}
\begin{tabular}{@{}lccc@{}}
\toprule
\textbf{Algorithm} & \textbf{Ensemble size} & \textbf{\# of function evals.} & \textbf{Wall-clock time [h]} \\
\midrule
ESMDA & 1000 & $10^4$ & 0.5 \\
MCMC ($\hat{R} < 1.05$) & 5~chains & $\sim 1 \times 10^6$ & 14 \\
MCMC ($\hat{R} < 1.01$) & 5~chains & $\sim 2 \times 10^6$ & 30   \\
SMC   & 1000 & $\sim 2 \times 10^5$ & 3   \\
\bottomrule
\end{tabular}
\end{table}

\subsection{Case~1 results}
\label{sec:case1_results}

For this case, we set $\beta_{\text{mut}}=0.12$ for SMC and $\beta_{\text{MCMC}} = 0.1$. These values are determined based on numerical experiments, involving low numbers of iterations, in which performance (e.g., acceptance rate) was tracked. MCMC requires approximately 346,000~iterations 
of Algorithm~\ref{alg:latent_mcmc} to reach $\hat{R} < 1.01$, and has a final acceptance rate of 37\%. With $N_{\text{chains}}=5$, this equates to a total of $\sim$1,730,000 function evaluations. SMC requires 6~iterations of Algorithm~\ref{alg:latent_smc}, for a total of 210,000 function evaluations, and has a final acceptance rate of 25\%. \textcolor{black}{The sensitivity of the DA results to the MCMC convergence criterion ($\hat{R}$ value) and to the number of SMC mutation steps ($N_{\text{mut}}$) are assessed in Section~\ref{sec:impact}.}

\begin{figure}[htbp]
    \centering
    \begin{subfigure}[b]{0.32\textwidth}
        \centering
        \includegraphics[width=\textwidth, trim=50 50 50 50, clip]{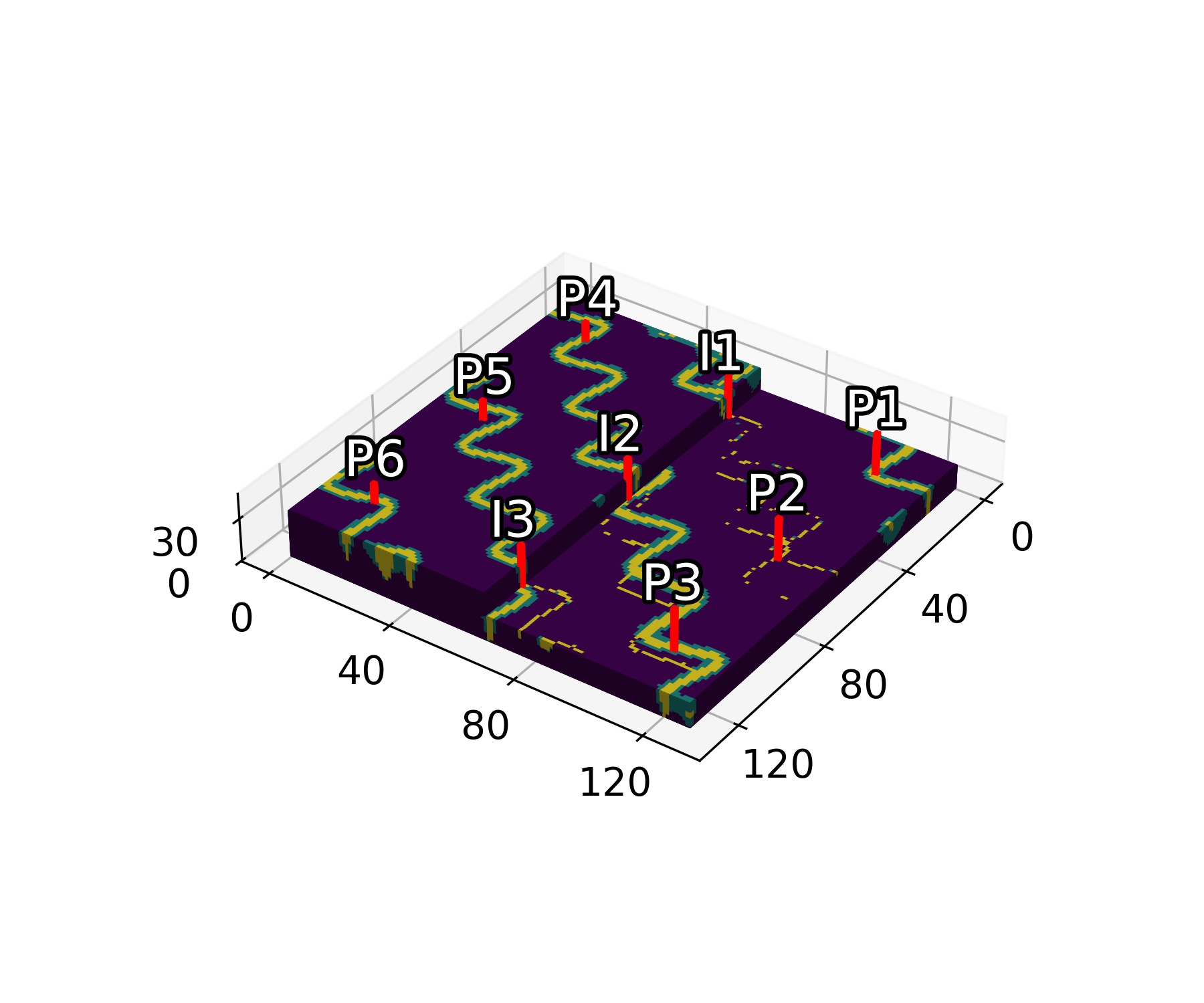}
        \caption{ESMDA}
        \label{fig:esmda_latent_1}
    \end{subfigure}
    \hfill
    \begin{subfigure}[b]{0.32\textwidth}
        \centering
        \includegraphics[width=\textwidth, trim=50 50 50 50, clip]{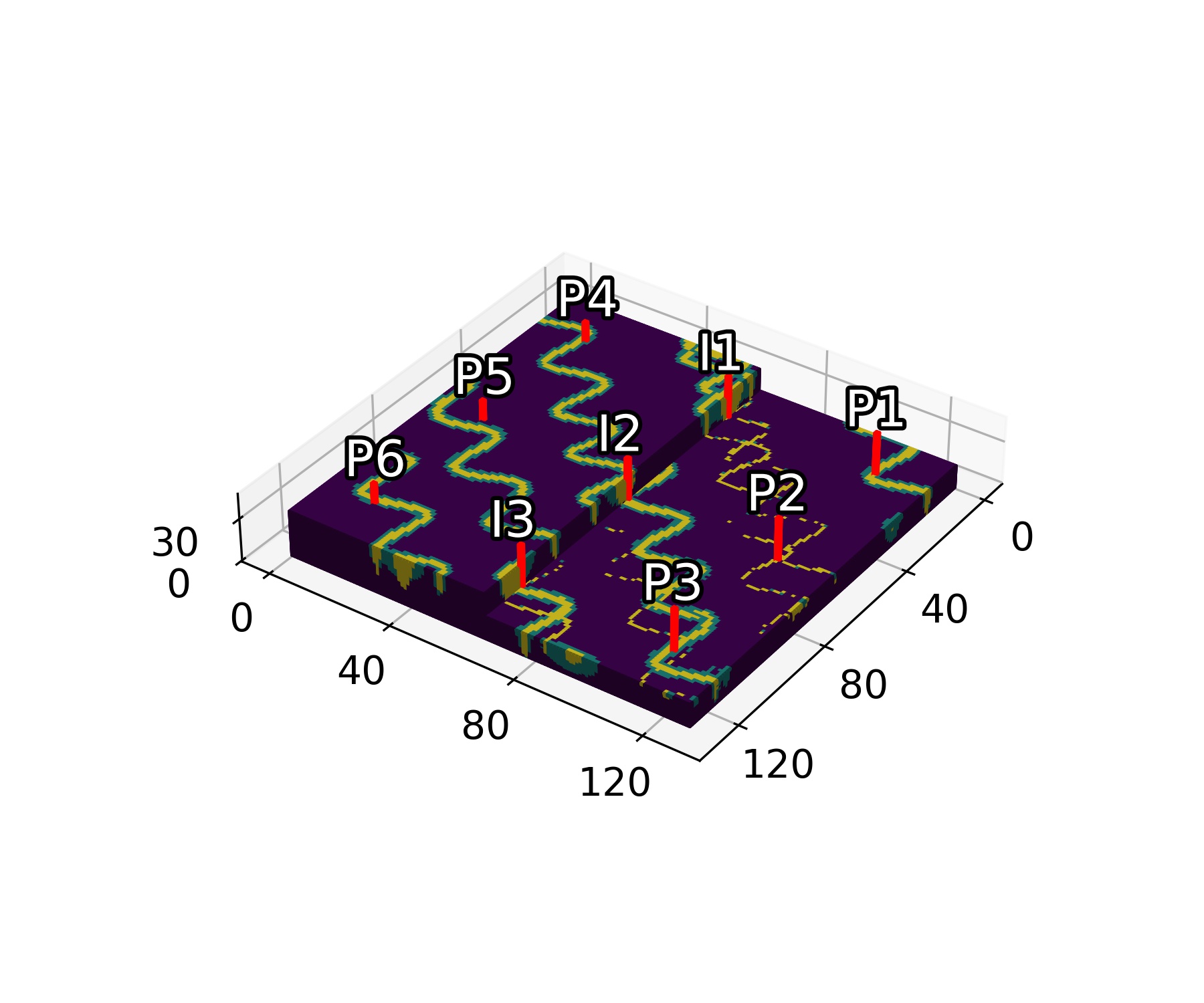}
        \caption{MCMC}
        \label{fig:mcmc_latent_1}
    \end{subfigure}
    \hfill
    \begin{subfigure}[b]{0.32\textwidth}
        \centering
        \includegraphics[width=\textwidth, trim=50 50 50 50, clip]{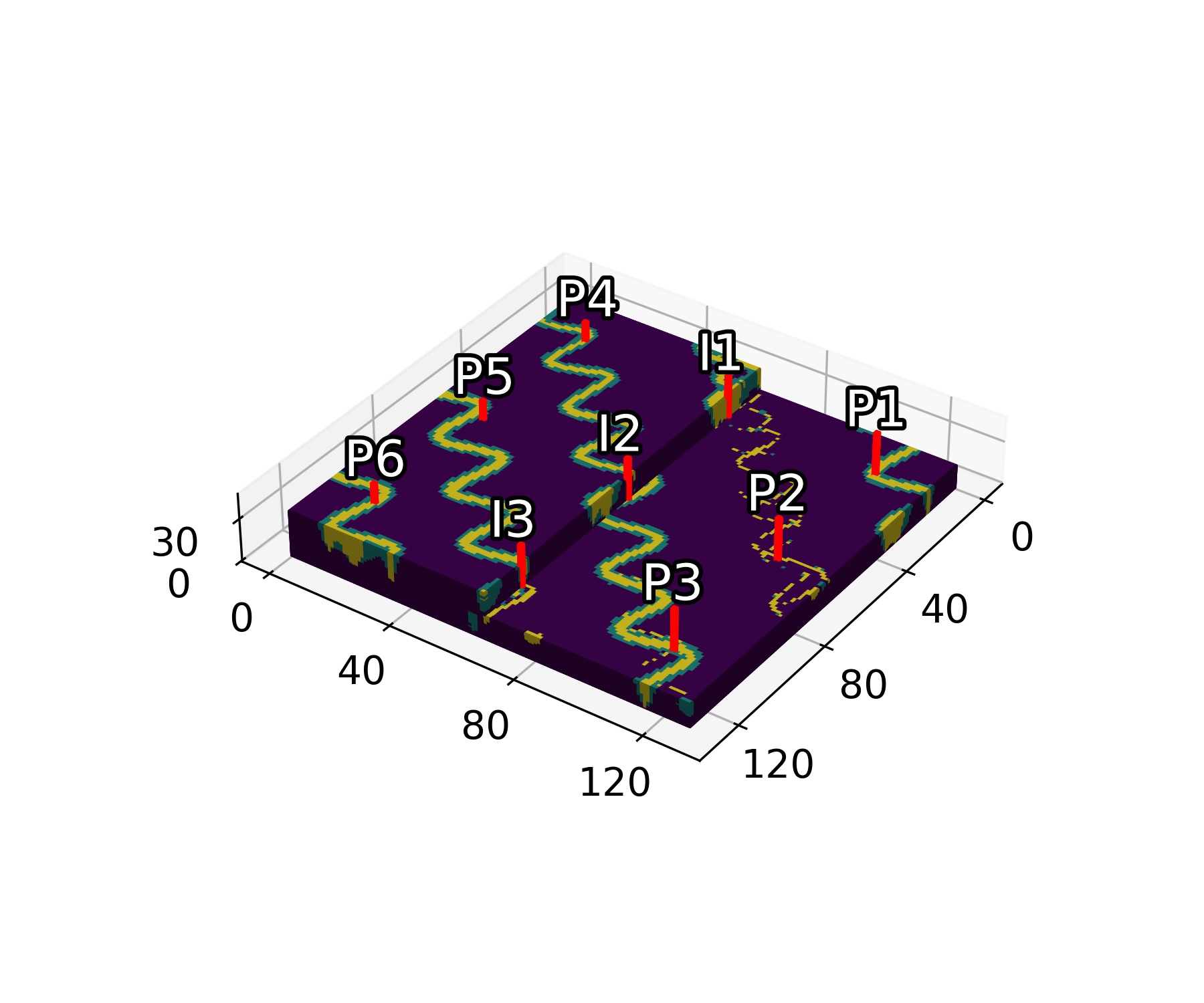}
        \caption{SMC}
        \label{fig:smc_latent_1}
    \end{subfigure}

    \caption{Randomly selected posterior realizations obtained using ESMDA, MCMC, and SMC (Case~1). DA performed in 3D-LDM latent space. All models maintain geological realism and display features consistent with the true model (Figure~\ref{fig:examples_petrel}d).}
    
    \label{fig:case1_post_models}
\end{figure}

In Figure~\ref{fig:case1_post_models}, we show one randomly selected posterior realization for each of the three algorithms. Compared to the high variability of the priors (see Figure~\ref{fig:examples_petrel}), the posterior models show geological features consistent with the synthetic true model (Figure~\ref{fig:examples_petrel}d), e.g., two distinct parallel channels along the main diagonal. Since DA is performed in the 3D-LDM latent space, geological realism is preserved with all algorithms. 

\begin{figure}[htbp]
    \centering
    \begin{subfigure}[b]{0.99\textwidth}
        \centering
        \includegraphics[width=\textwidth]{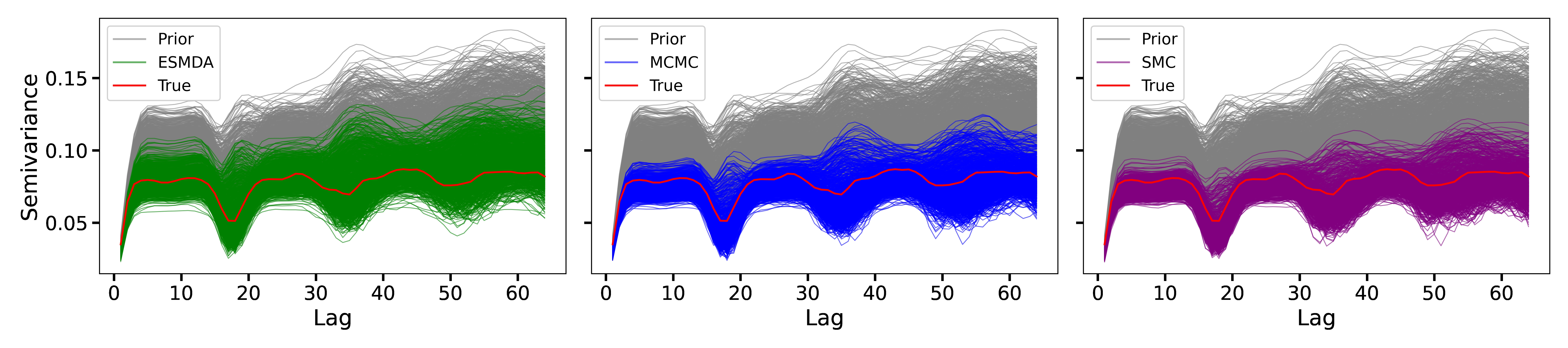}
        \caption{Variograms, along diagonal}
        \label{fig:prop1_dir1}
    \end{subfigure}

    \vspace{1em}

    \begin{subfigure}[b]{0.99\textwidth}
        \centering
        \includegraphics[width=\textwidth]{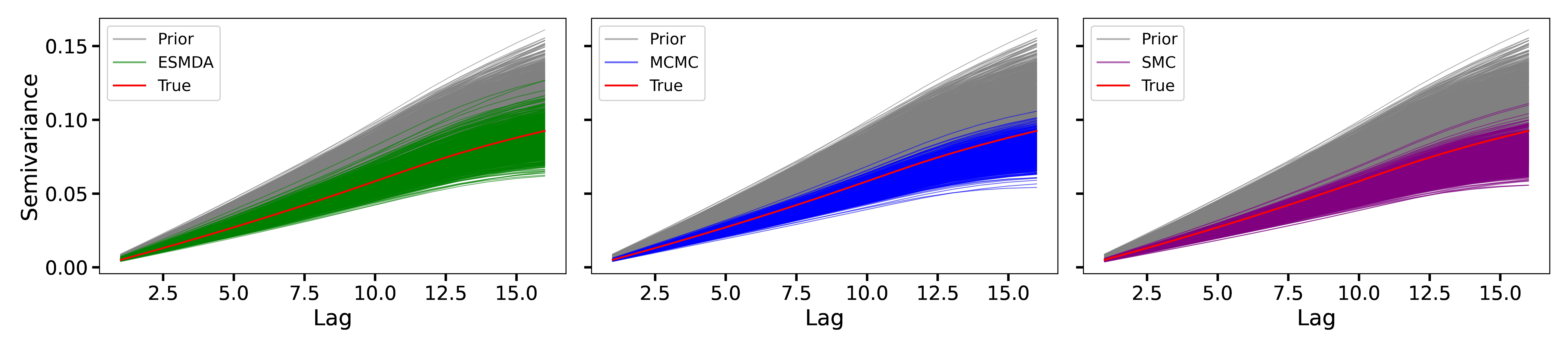}
        \caption{Variograms, along vertical axis}
        \label{fig:props1_dir2}
    \end{subfigure}

    \caption{Variograms for prior and posterior realizations obtained using ESMDA, MCMC, and SMC (Case~1), with lag distance in grid blocks. Gray curves show results for prior realizations. Red solid curves denote results for the true model (Figure~\ref{fig:examples_petrel}d), while green, blue, and purple curves correspond to individual posterior realizations. Posterior curves appear as a subset of the priors (preserved realism) and show similar trends compared to the true model.}
    \label{fig:realism_props_1}
\end{figure}

\begin{figure}[htbp]
    \centering
    \begin{subfigure}[b]{0.99\textwidth}
        \centering
        \includegraphics[width=\textwidth]{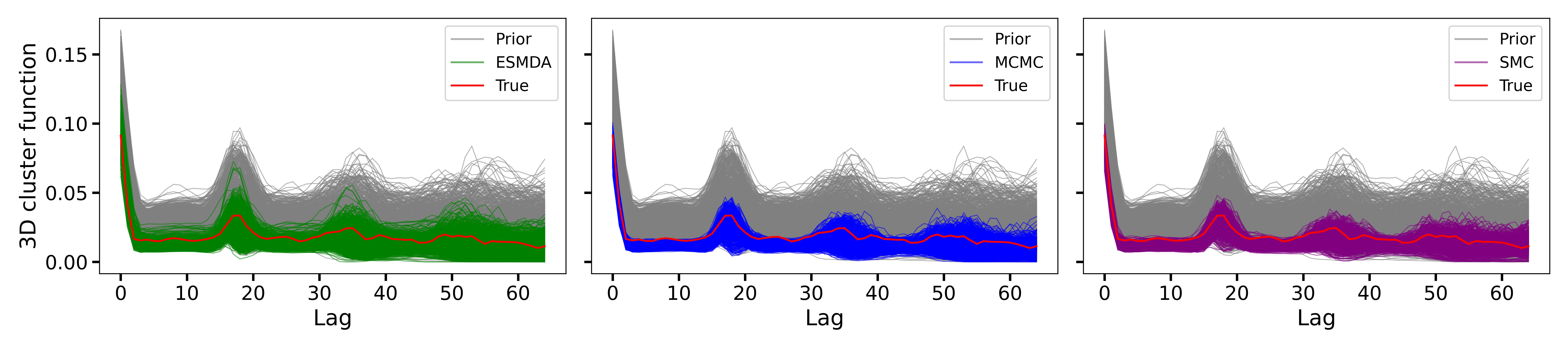}
        \caption{Cluster functions, along diagonal}
        \label{fig:props2_dir1}
    \end{subfigure}

    \vspace{1em}

    \begin{subfigure}[b]{0.99\textwidth}
        \centering
        \includegraphics[width=\textwidth]{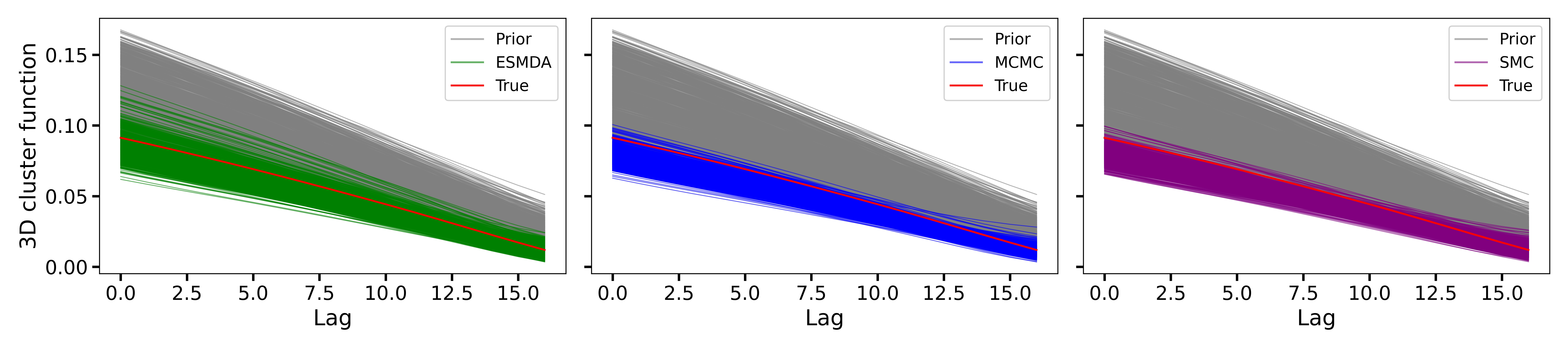}
        \caption{Cluster functions, along vertical axis}
        \label{fig:props2_dir2}
    \end{subfigure}

    \caption{Cluster functions for prior and posterior realizations obtained using ESMDA, MCMC, and SMC (Case~1), with lag distance in grid blocks. Gray curves show results for prior realizations. Red solid curves denote results for the true model (Figure~\ref{fig:examples_petrel}d), while green, blue, and purple curves correspond to individual posterior realizations.} 
    \label{fig:realism_props_2}
\end{figure}

\textcolor{black}{It is also useful to assess the spatial statistics of the posterior geomodels. Many different metrics can be computed, though here we focus on variograms (shown in Figure~\ref{fig:realism_props_1}) and 3D cluster functions (Figure~\ref{fig:realism_props_2}). The 3D variogram curves~\citep{pyrcz2014geostatistical,DiFederico2025} are computed directly on the facies fields. The mud, levee, and channel facies correspond to facies values of 0, 0.5 and 1, respectively. These variograms describe how quickly the facies become dissimilar as the distance (lag) between two locations increases. Variogram results are presented in Figure~\ref{fig:realism_props_1} for both the main diagonal (45$^\circ$, along the P3--P4 line in Figure~\ref{fig:case1_post_models}) and for the vertical (along the $z$-axis) direction, up to a lag of half of the maximum distance in the target direction. The gray curves are samples from the prior distribution, while the green, blue, and purple lines represent 1000 posterior samples from ESMDA, MCMC, and SMC, respectively. The red solid line denotes the true model. For all cases, the posteriors occupy a distinct portion (subset) of the prior range, demonstrating both uncertainty reduction and the ability of 3D-LDM to retain geological realism during DA. Importantly, the true result lies within the posterior ensemble in all cases. We observe a degree of consistency between the three sets of results, though the ESMDA results show more spread than the MCMC and SMC results.}

\textcolor{black}{We next consider 3D cluster functions~\citep{torquato2002}. This function describes how likely it is for two locations at a given lag distance to belong to the same connected channel body.
This requires us to first identify and label the connected channel bodies in each model. Two channel cells belong to the same body (cluster) if there is a continuous path of channel cells, moving only through face neighbors in any ($i,j,k$) direction, between them. The value at lag $h$ is the fraction of all cell pairs separated by $h$ (in the target direction) where both cells belong to the same channel body. Results are shown in Figure~\ref{fig:realism_props_2}, for the same directions as were considered in Figure~\ref{fig:realism_props_1}. Analogous conclusions can be drawn from these results as for those in Figure~\ref{fig:realism_props_1}. Specifically, we observe clear uncertainty reduction, the true result falls within the posterior, and ESMDA provides wider posteriors than MCMC and SMC. We note finally that similar results were observed, for both the variogram and 3D cluster functions, along other directions.} 


Table~\ref{tab:case1_post_params} reports mean and standard deviation values for the geological scenario parameters for the prior, posterior and true models. As noted earlier, the mud fraction $f_m$ is computed as the fraction of mud-facies grid blocks, while the average channel orientation $\theta_{\text{ch}}$ and width $w_{\text{ch}}$ are extracted using a CNN trained on the prior dataset and labeled with the input parameters from the Petrel OBM tool~\citep{DiFederico2025}. We reiterate that the scenario parameters are not separately encoded into our 3D-LDM parameterization or directly updated during DA. 

Across almost all metrics and methods, we observe mean values shifting in the direction of the true scenario parameters  ($f_m = 0.85$, $\theta_{\rm ch} = 45^\circ$, $w_{\rm ch} = 4$). Uncertainty reduction is reflected in the lower standard deviation of the posterior distributions relative to the priors. The performance of MCMC and SMC is very consistent, with relative errors on mean values (compared to true values) within 5\%. ESMDA, by contrast, exhibits slightly lower accuracy and less uncertainty reduction, with posterior standard deviations about 1.5$\times$ larger than those of MCMC and SMC.

\begin{table}[htbp]
\centering
\caption{DA algorithm performance for scenario parameters (Case~1, DA performed in 3D-LDM latent space).}
\label{tab:case1_post_params}
\begin{tabular}{@{}lcccccc@{}}
\toprule
& \multicolumn{2}{c}{$\boldsymbol{f_m}$} 
& \multicolumn{2}{c}{$\boldsymbol{\theta_{\rm ch}}$} 
& \multicolumn{2}{c}{$\boldsymbol{w_{\rm ch}}$} \\
\cmidrule(lr){2-3} \cmidrule(lr){4-5} \cmidrule(lr){6-7}
& Mean & Std 
& Mean & Std 
& Mean & Std \\
\midrule
{Prior}        
& 0.795 & 0.043
& 45.0 & 8.66
& 5.00 & 0.577 \\
\midrule
{ESMDA}        
& 0.841 & 0.014 
& 48.4 & 5.86 
& 4.29 & 0.292 \\
{MCMC}         
& 0.853 & 0.009 
& 44.9 & 4.27 
& 4.15 & 0.187 \\
{SMC}          
& 0.854 & 0.009 
& 45.7 & 4.13 
& 4.20 & 0.204 \\
\midrule
{True}         
& \multicolumn{2}{c}{0.85}
& \multicolumn{2}{c}{45$^\circ$}
& \multicolumn{2}{c}{4} \\
\bottomrule
\end{tabular}
\end{table}

\begin{figure}[htbp]
    \centering

    \begin{subfigure}[b]{0.48\textwidth}
        \centering
        \includegraphics[width=\textwidth]{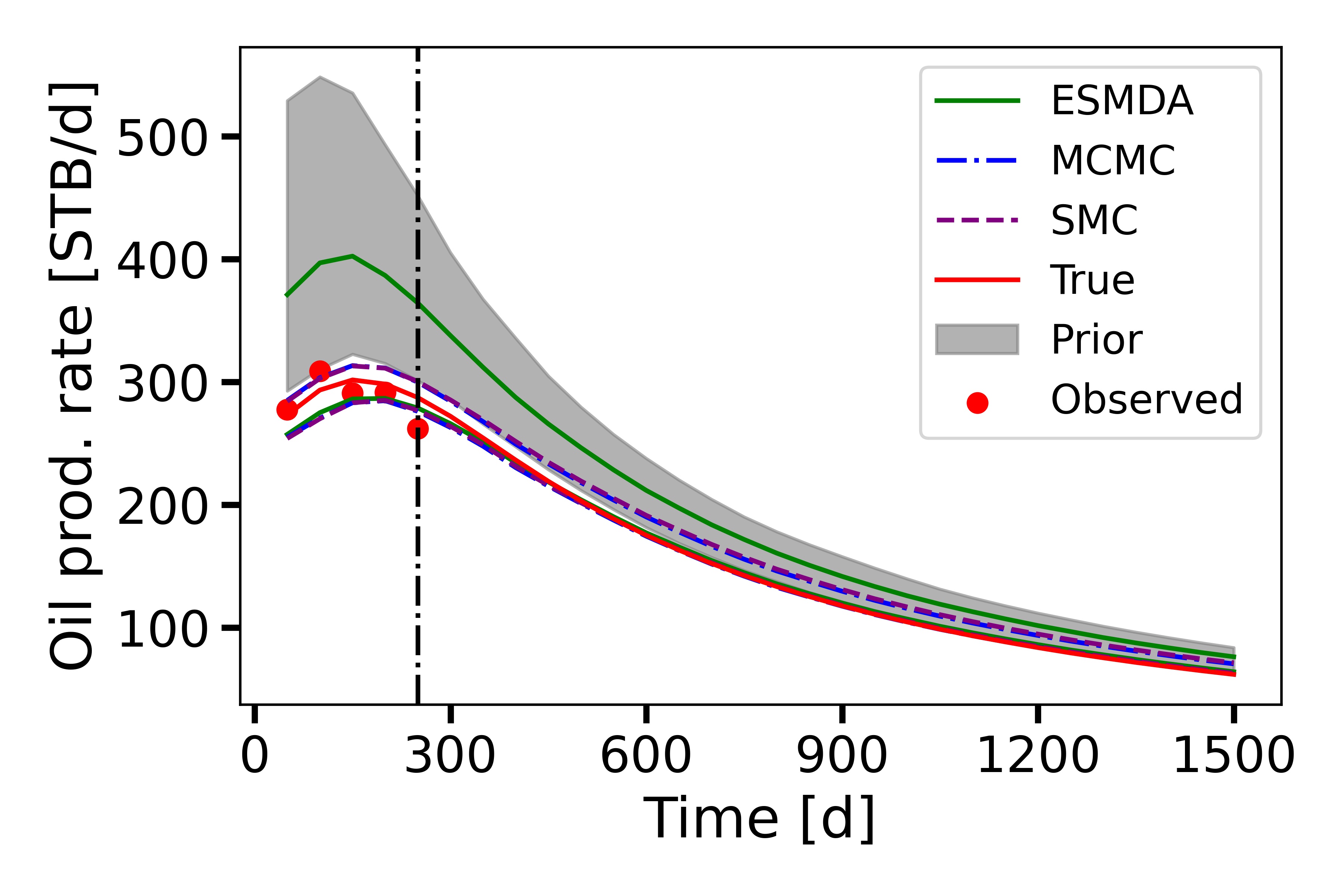}
        \caption{P1 oil production rate}
    \end{subfigure}
    \hfill
    \begin{subfigure}[b]{0.48\textwidth}
        \centering
        \includegraphics[width=\textwidth]{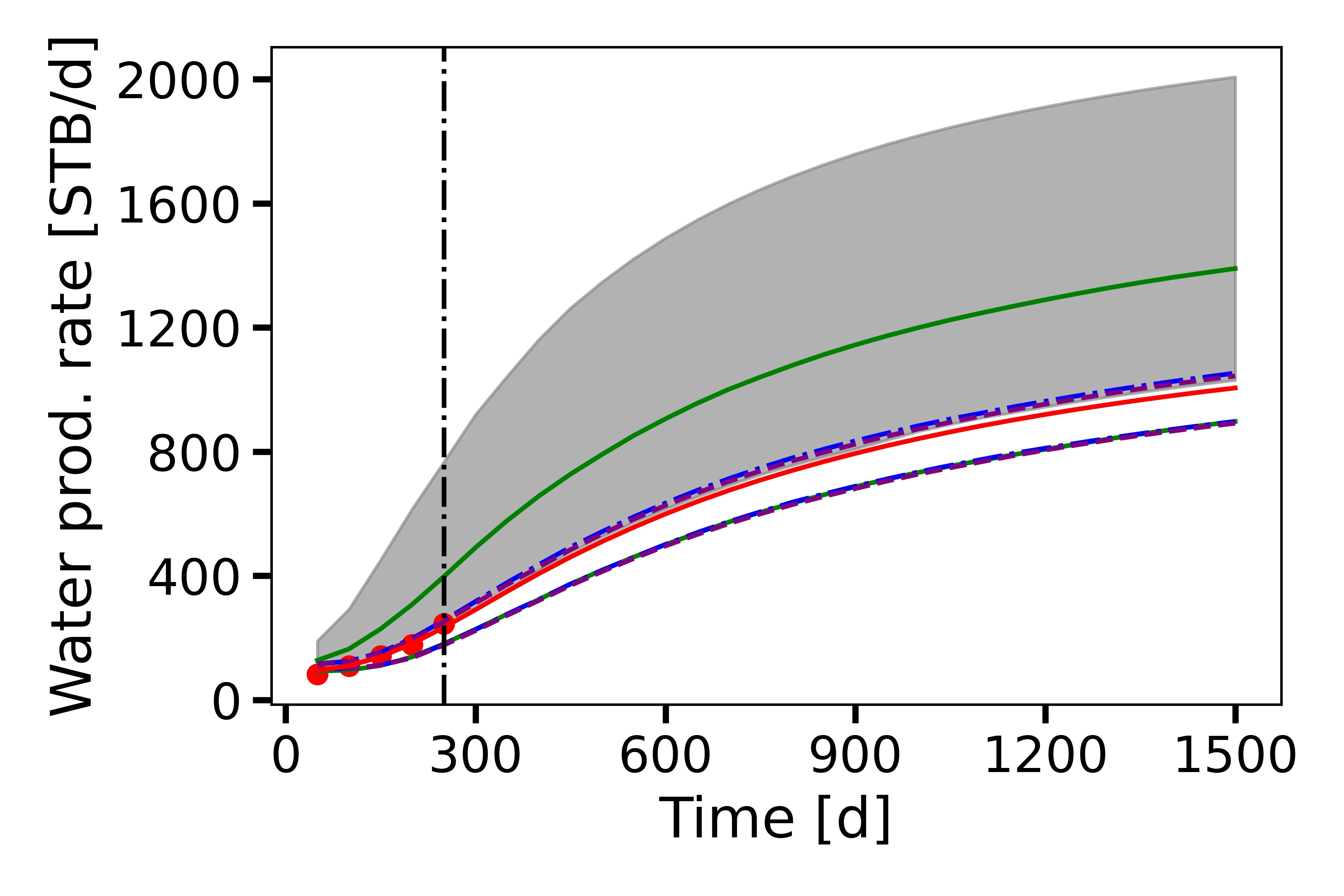}
        \caption{P1 water production rate}
    \end{subfigure}

    \vspace{1em}

    \begin{subfigure}[b]{0.48\textwidth}
        \centering
        \includegraphics[width=\textwidth]{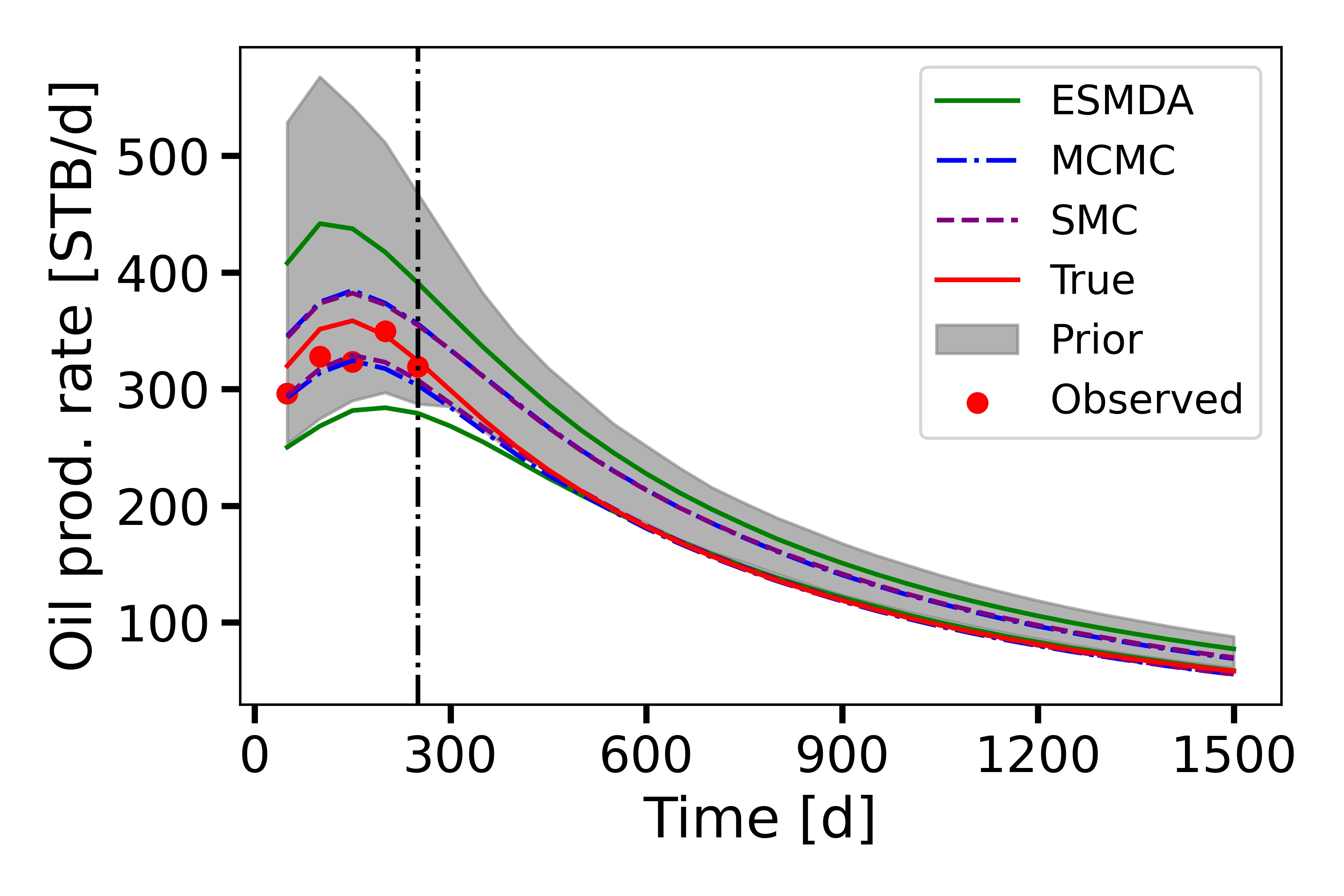}
        \caption{P6 oil production rate}
    \end{subfigure}
    \hfill
    \begin{subfigure}[b]{0.48\textwidth}
        \centering
        \includegraphics[width=\textwidth]{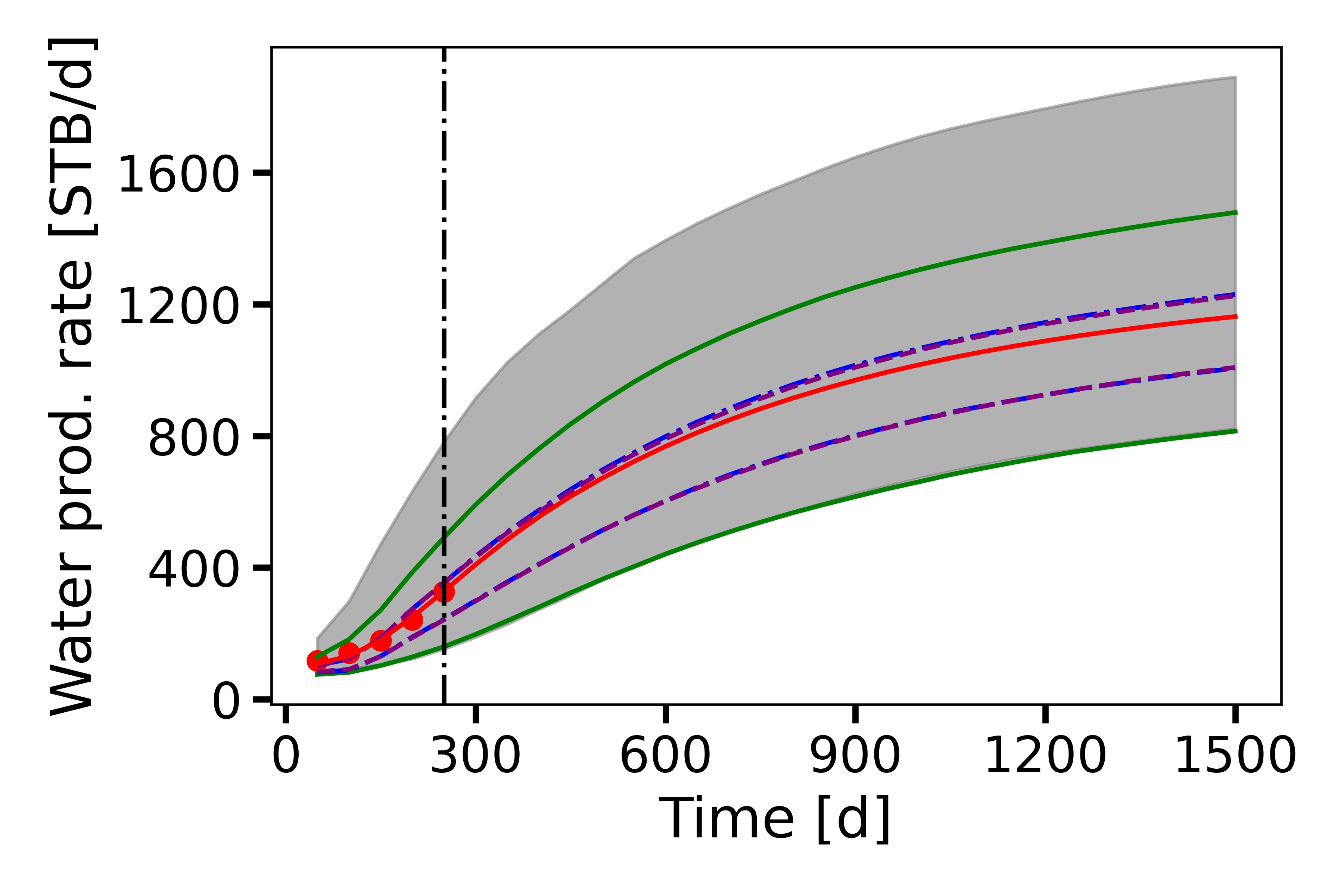}
        \caption{P6 water production rate}
    \end{subfigure}

    \vspace{1em}

    \begin{subfigure}[b]{0.48\textwidth}
        \centering
        \includegraphics[width=\textwidth]{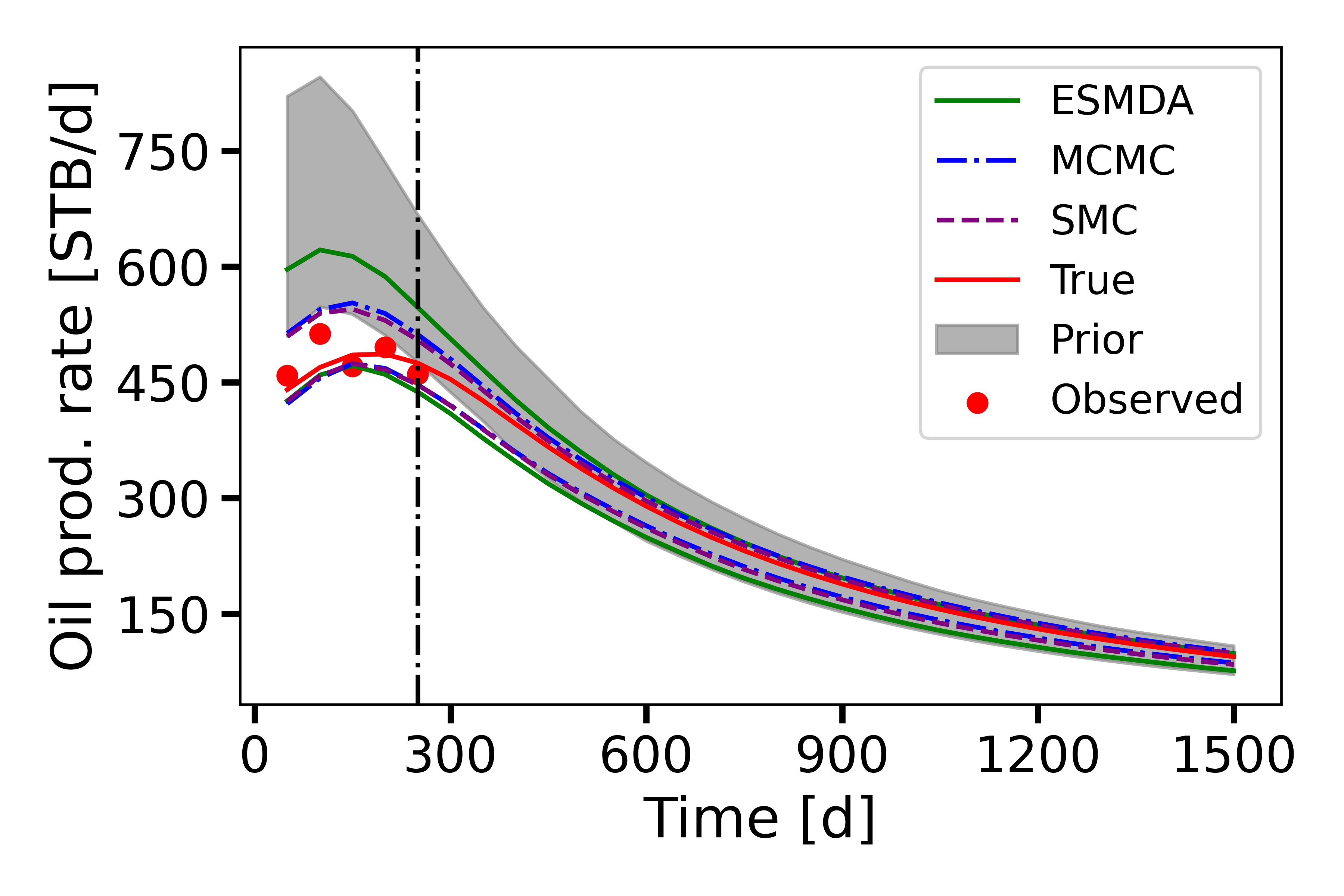}
        \caption{P3 oil production rate}
    \end{subfigure}
    \hfill
    \begin{subfigure}[b]{0.48\textwidth}
        \centering
        \includegraphics[width=\textwidth]{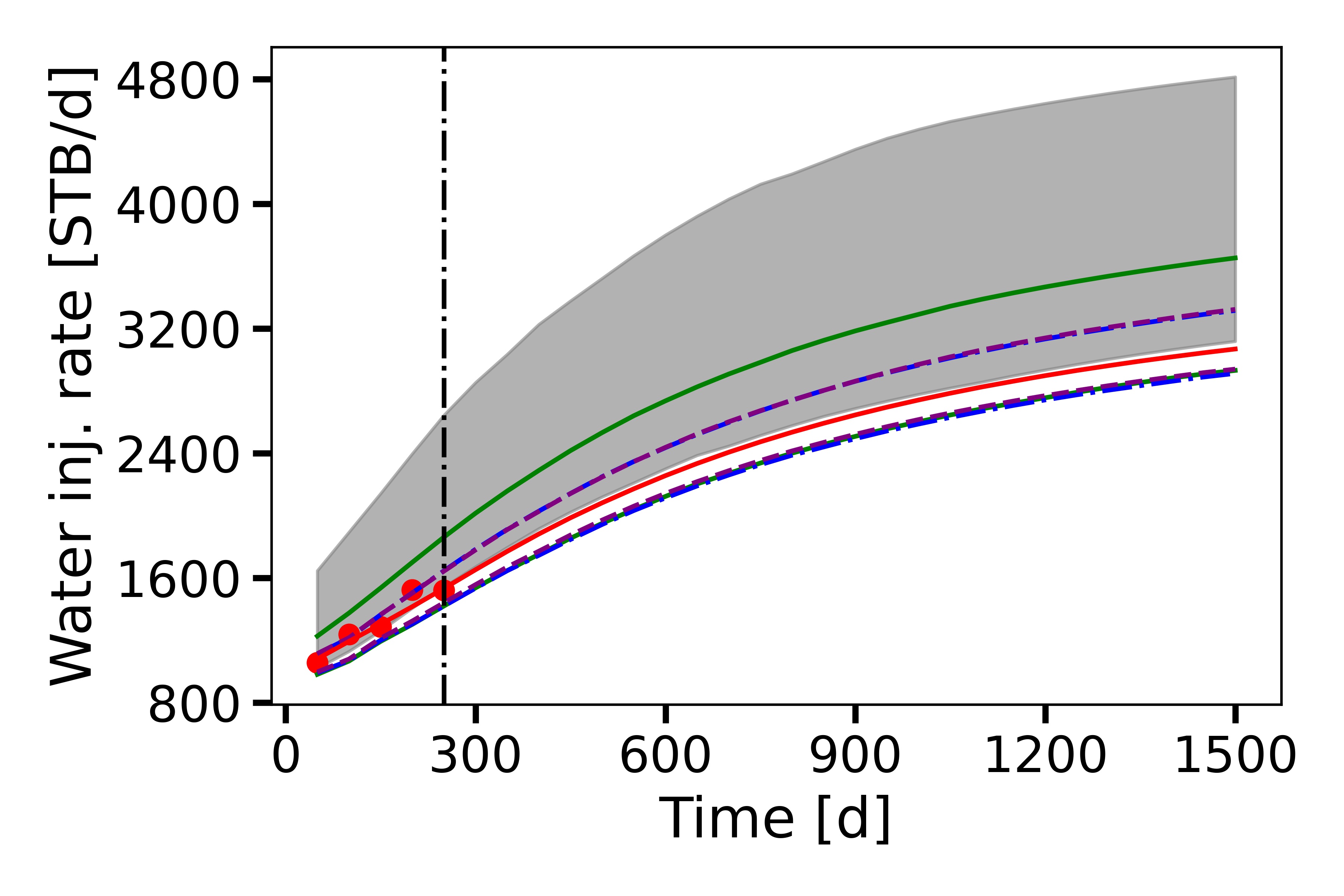}
        \caption{I3 water injection rate}
    \end{subfigure}

    \caption{DA results for selected wells (Case~1, DA for all methods performed in 3D-LDM latent space). Gray regions show the prior P$_{10}$--P$_{90}$ range. Green, blue, and purple lines denote the posterior P$_{10}$--P$_{90}$ ranges for ESMDA, MCMC, and SMC, respectively. Red points and red curves represent observed and true data. The vertical black dot-dash line indicates the end of the history matching period.}
    \label{fig:case1_post_rates}
\end{figure}

Posterior results for well flow rates for representative wells are shown in Figure~\ref{fig:case1_post_rates}. The prior P$_{10}$--P$_{90}$ region is shown in gray and the true and observed data are in red (these results are identical to those shown in Figure~\ref{fig:esmda_comparison_rates_latent}). The P$_{10}$--P$_{90}$ curves for latent-space ESMDA, MCMC and SMC are shown in solid green, dash-dotted blue, and dashed purple, respectively. Across different wells, and for both water and oil flow rates, the P$_{10}$--P$_{90}$ posterior curves for MCMC and SMC are in very close agreement. Conversely, there are large differences between ESMDA and the MCMC/SMC results. In particular, we consistently observe much less uncertainty reduction with ESMDA. This is particularly noticeable in the historical period (before 250~days) for oil production rates (e.g., Figure~\ref{fig:case1_post_rates}a,c) and in water production rates at later times (e.g., Figure~\ref{fig:case1_post_rates}b,d). 

Summary statistics, analogous to those in Table~\ref{table:esmda_metrics_da}, are reported in Table~\ref{table:case1_metrics_da}. The lower values of NV for MCMC and SMC (0.63), compared to that for ESMDA (0.74), indicate a higher degree of uncertainty reduction in the geomodels (recall NV is computed in model space). NV does not appear to be a very sensitive parameter, however, as these differences are relatively small. Nonetheless, this is consistent with the differences in the posterior ranges for the scenario parameters in Table~\ref{tab:case1_post_params} and the variogram and cluster function results in Figures~\ref{fig:realism_props_1} and~\ref{fig:realism_props_2}. Similarly, the values for RMSE ($\sim$56~STB/d versus $\sim$116~STB/d) and $\Delta$P$_{90-10}$ (0.18--0.19 versus 0.44) reflect differences in the accuracy of data matching and in uncertainty reduction in the well rates. These values are in alignment with the results in Figure~\ref{fig:case1_post_rates}. \textcolor{black}{The results in this section, which involve quantitative comparisons between latent-space ESMDA and rigorous Monte Carlo sampling methods, clearly confirm the phenomena observed for latent-space ESMDA in Section~\ref{sec:esmda_results}.}

Note that in the simulation-based latent-space ESMDA results presented in Section~\ref{sec:esmda_results}, we used $C_d$ in the ESMDA update (Eq.~\ref{eqn:esmda_update}). In the results here, in which the flow surrogate is applied, we use $C_{\text{tot}}$, given by Eq.~\ref{eq:total_model_err_cov}. Thus we expect some differences in the latent-space ESMDA results here relative to those in Table~\ref{table:esmda_metrics_da}. In particular, the inclusion of surrogate model error would be expected to increase RMSE, NV, and $\Delta$P$_{90-10}$, and this is indeed what we observe.

\begin{table}[htbp]
\centering
\caption{Performance metrics for DA algorithms (Case~1, DA performed in 3D-LDM latent space).}
\label{table:case1_metrics_da}
\begin{tabular}{@{}lcccc@{}}
\toprule
\textbf{Algorithm} 
& \textbf{RMSE} [STB/d]
& \textbf{NV} 
& \textbf{NV$_{\text{dum}}$} 
& $\mathbf{\Delta \text{P}_{90-10}}$ \\
\midrule
ESMDA & 116.19 & 0.74 & 0.99 & 0.44 \\
MCMC  & 56.02  & 0.63 & --   & 0.19 \\
SMC   & 55.89  & 0.63 & --   & 0.18 \\
\bottomrule
\end{tabular}
\end{table}

\subsection{Impact of algorithmic treatments}
\label{sec:impact}

\textcolor{black}{We now explore the impact of some key parameters on the results presented in Section~\ref{sec:case1_results}. In particular, we vary the latent space dimension of the 3D-LDM used with ESMDA, the $\hat{R}$ convergence criterion for MCMC, and the number of mutation steps for SMC. For brevity, we report limited sets of results for these cases.}

\textcolor{black}{It is useful to consider whether the low dimension of the 3D-LDM latent space, rather than the nonlinear character of the latent space, is the primary reason for the overestimation of posterior uncertainty with ESMDA. To assess this issue, we train two additional 3D-LDMs, which contain $2\times$ and $3\times$ the number of latent variables compared to the base case. For the base case we have $n_c = 1024$, so these new cases correspond to $n_c = 2048$ and $n_c=3072$. These 3D-LDM representations are constructed by increasing the number of latent variable channels from (1,16,16,4) with $n_c=1024$ to (2,16,16,4) and (3,16,16,4), respectively. We do not investigate smaller latent spaces, as $n_c=1024$ was found to provide an appropriate compromise between dimension reduction and generation quality for this setup~\citep{DiFederico2025}.}

\begin{figure}[htbp]
    \centering
    \begin{subfigure}[b]{0.48\textwidth}
        \centering
        \includegraphics[width=\textwidth]{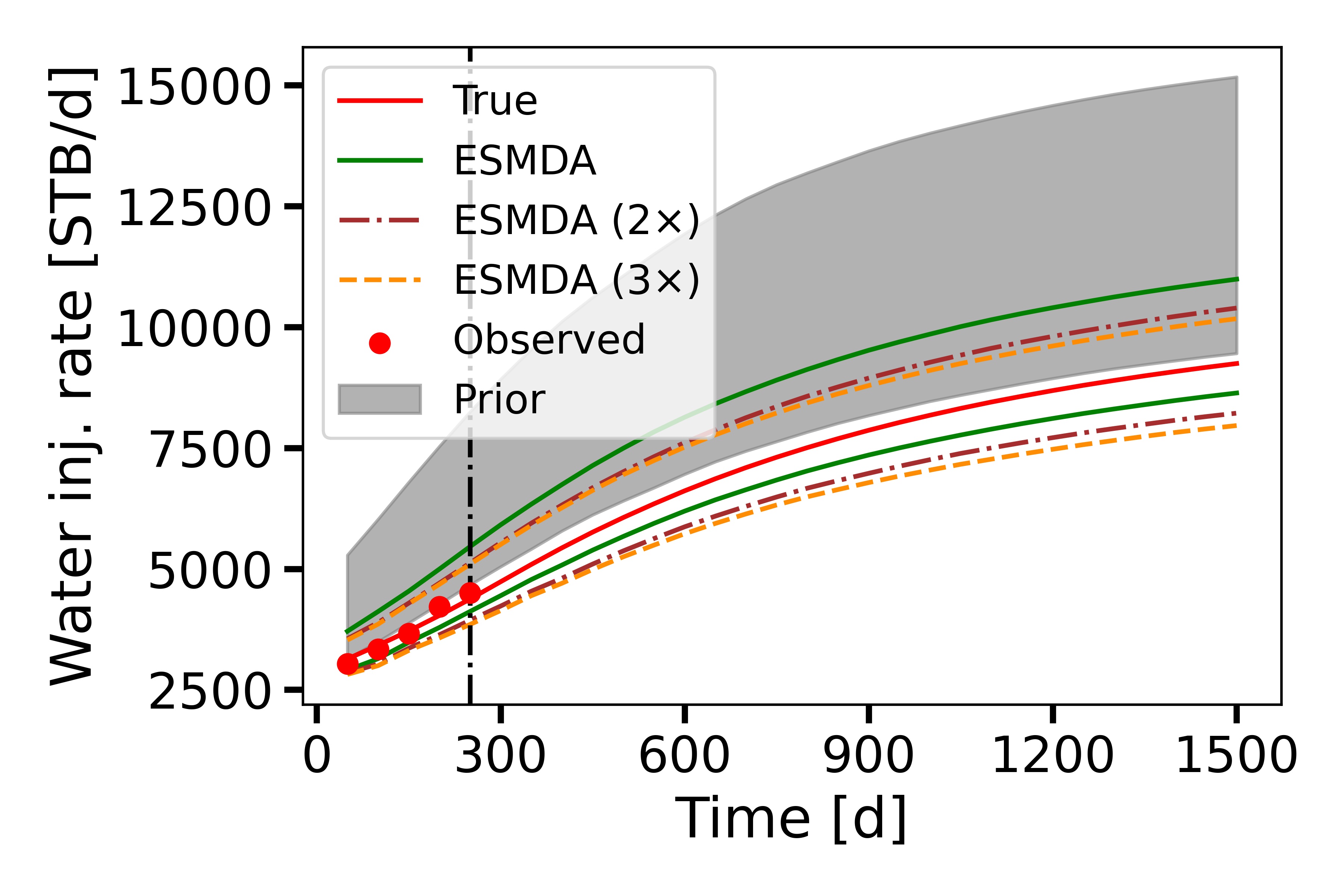}
        \caption{Field water injection rate}
        \label{fig:field_inj_lat}
    \end{subfigure}
    \hfill
    \begin{subfigure}[b]{0.48\textwidth}
        \centering
        \includegraphics[width=\textwidth]{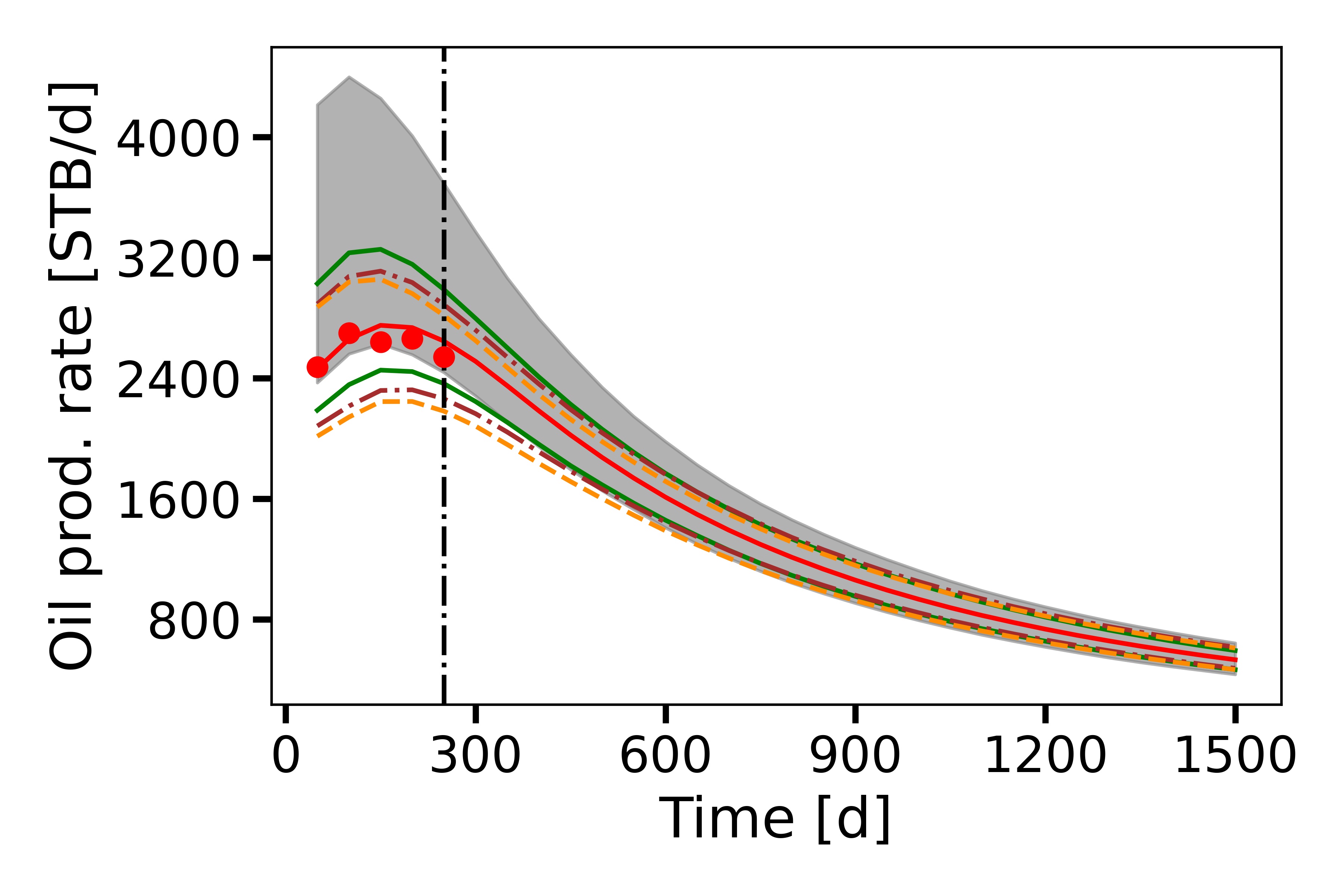}
        \caption{Field oil production rate}
        \label{fig:field_oil_lat}
    \end{subfigure}

    \vspace{1em}
    
    \begin{subfigure}[b]{0.48\textwidth}
        \centering
        \includegraphics[width=\textwidth]{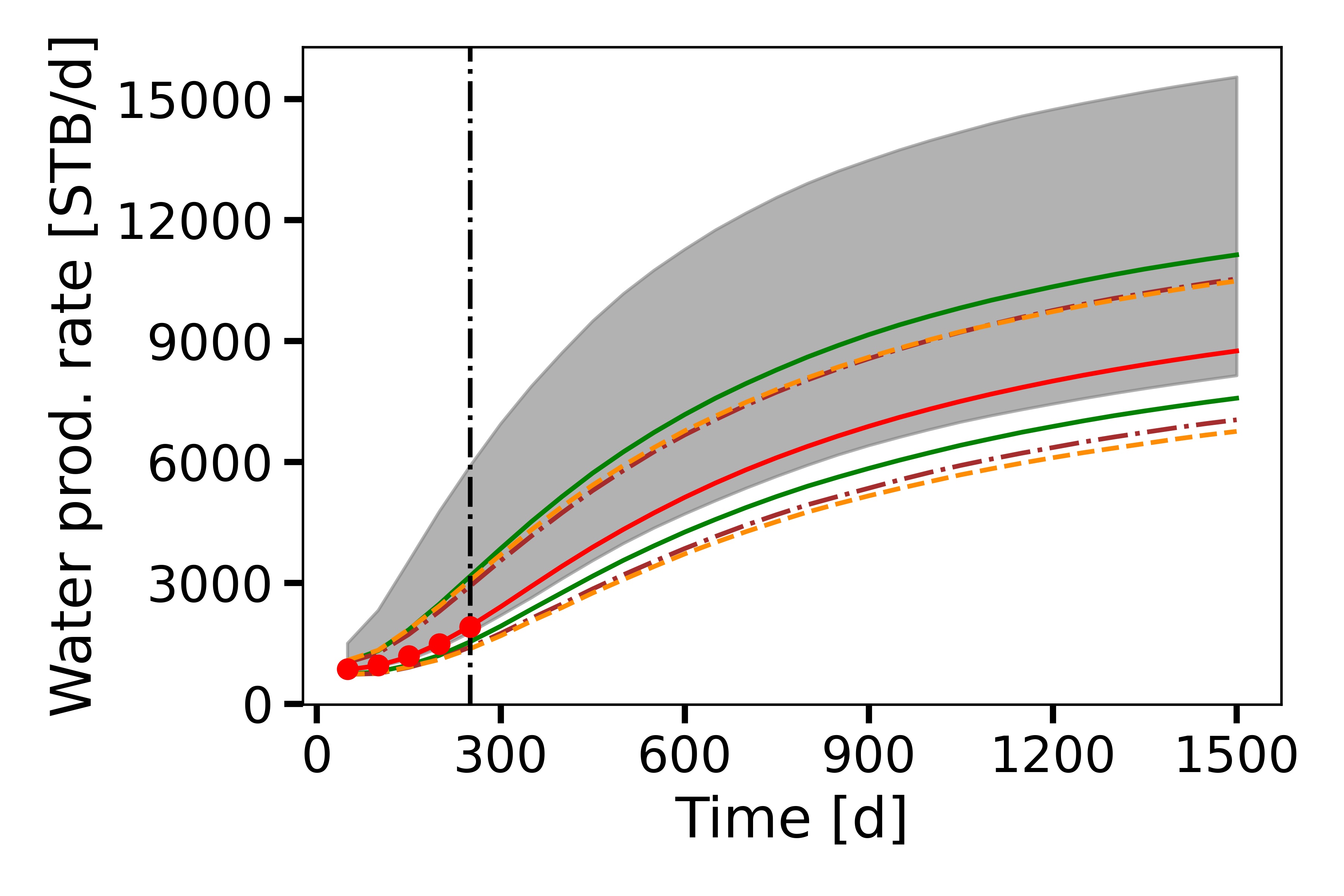}
        \caption{Field water production rate}
    \label{fig:rates_field_wat_lat}
    \end{subfigure}
    \hfill
    \begin{subfigure}[b]{0.48\textwidth}
        \centering
        \includegraphics[width=\textwidth]{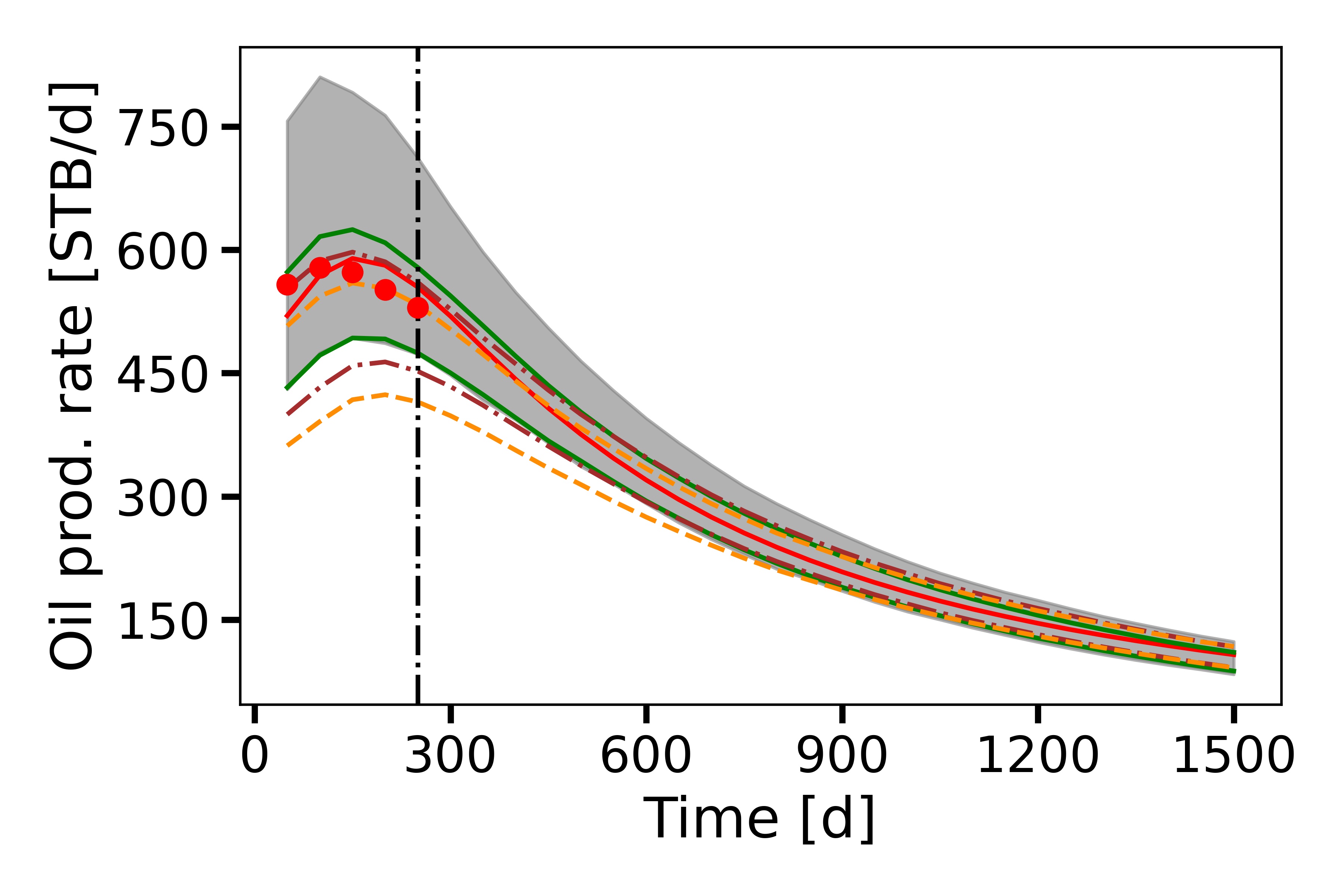}
        \caption{P2 oil production rate}
        \label{fig:rates_P2_OIL_lat}
    \end{subfigure}
    \caption{ESMDA results for field-level rates and oil rate for well P2 for different 3D-LDM latent space dimensions (Case~1). Gray regions show the prior P$_{10}$--P$_{90}$ range. Green, brown, and orange lines denote the posterior P$_{10}$--P$_{90}$ ESMDA ranges for the base case ($n_c=1024$) and with $2\times$ ($n_c=2048$) and $3\times$ ($n_c=3072$) the number of latent variables, respectively. Red points and red curves represent observed and true data. The vertical black dot-dash line indicates the end of the history matching period.}
    \label{fig:field_rates_lat}
\end{figure}

\textcolor{black}{The identical latent-space ESMDA workflow (using the same flow surrogate, ESMDA localization, etc., as in Section~\ref{sec:case1_results}) is then applied for Case~1. Results for field-level injection and production rates (field-level rates are computed by summing over all wells), and for oil production rate for the largest producer (P2), are shown in Figure~\ref{fig:field_rates_lat}. DA metrics for the various $n_c$ values are provided in Table~\ref{table:case1_metrics_da_latent_analysis}. The base case ESMDA corresponds to the case considered in Section~\ref{sec:case1_results} (shown in Figure~\ref{fig:case1_post_rates}). 
In this assessment, although NV decreases (from 0.74 to 0.65) as $n_c$ increases, RMSE and $\Delta$P$_{90-10}$ are affected very little, and they do not decrease monotonically with increasing $n_c$. These findings are reiterated in Figure~\ref{fig:field_rates_lat}, where we see the curves shift for the different $n_c$ values, but there is no clear trend showing decreased posterior uncertainty with increasing $n_c$. These results suggest that the overestimation of posterior uncertainty with ESMDA and 3D-LDM is not due solely to the low latent space dimension. We note, however, that there are many ways in which the 3D-LDM representation can be varied, and it is possible that other architectures will result in improved ESMDA results.} 


\begin{table}[htbp]
\centering
\caption{Performance metrics for ESMDA with different latent-space dimensions (Case~1, DA performed in 3D-LDM latent space).}
\label{table:case1_metrics_da_latent_analysis}
\begin{tabular}{@{}lcccc@{}}
\toprule
\textbf{Algorithm}
& \textbf{RMSE} [STB/d]
& \textbf{NV}
& \textbf{NV$_{\text{dum}}$}
& $\mathbf{\Delta \text{P}_{90-10}}$ \\
\midrule
ESMDA (base case, $n_c = 1024$)          & 116.19 & 0.74 & 0.99 & 0.44 \\
ESMDA ($n_c = 2048$)  & 105.52 & 0.68 & 0.98 & 0.40 \\
ESMDA ($n_c = 3072$)  & 106.74 & 0.65 & 0.98 & 0.44 \\
\bottomrule
\end{tabular}
\end{table}

\begin{figure}[htbp]
    \centering
    \begin{subfigure}[b]{0.48\textwidth}
        \centering
        \includegraphics[width=\textwidth]{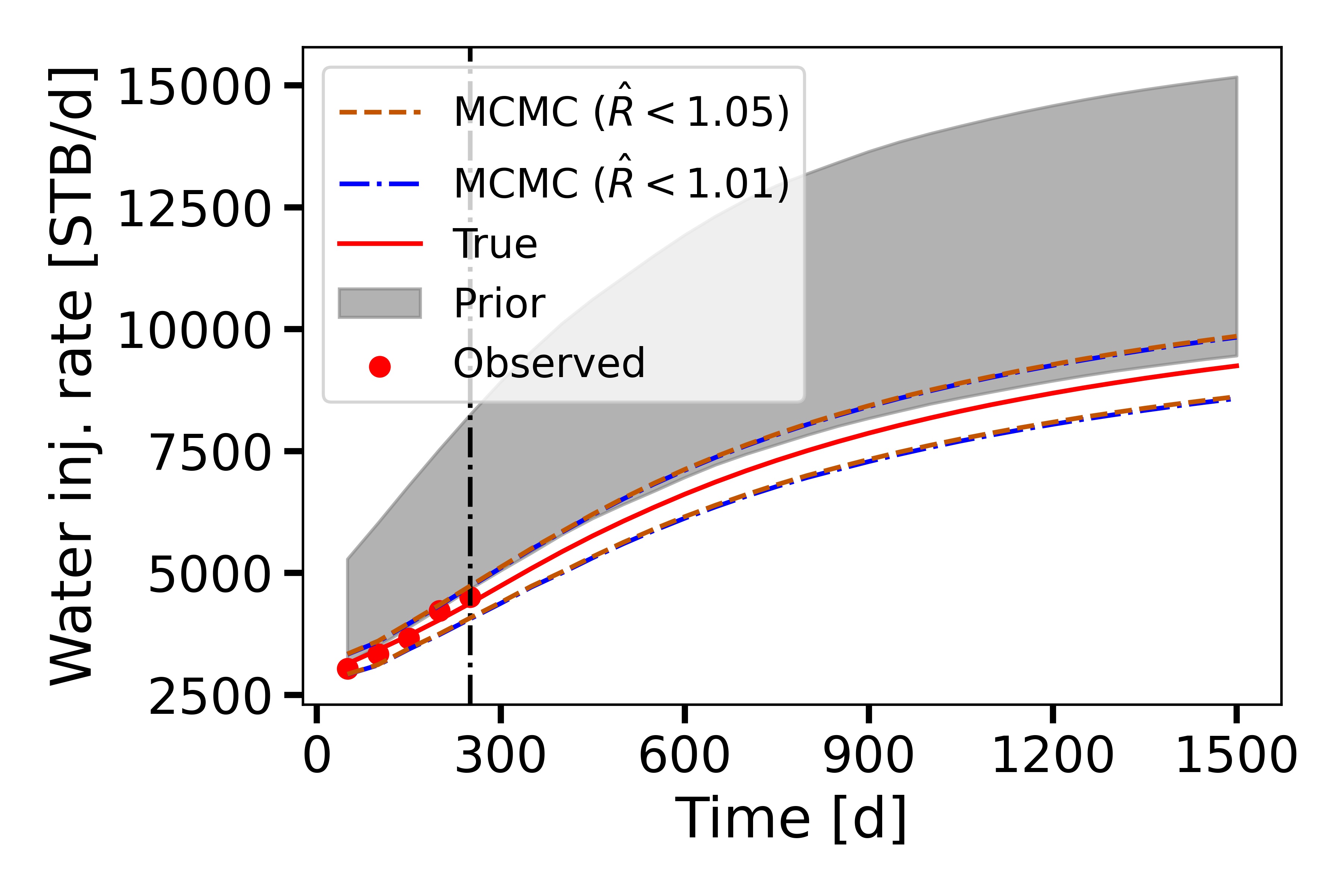}
    \caption{Field water injection rate}
    \label{fig:field_inj_lat_rhat}
    \end{subfigure}
    \hfill
    \begin{subfigure}[b]{0.48\textwidth}
        \centering
        \includegraphics[width=\textwidth]{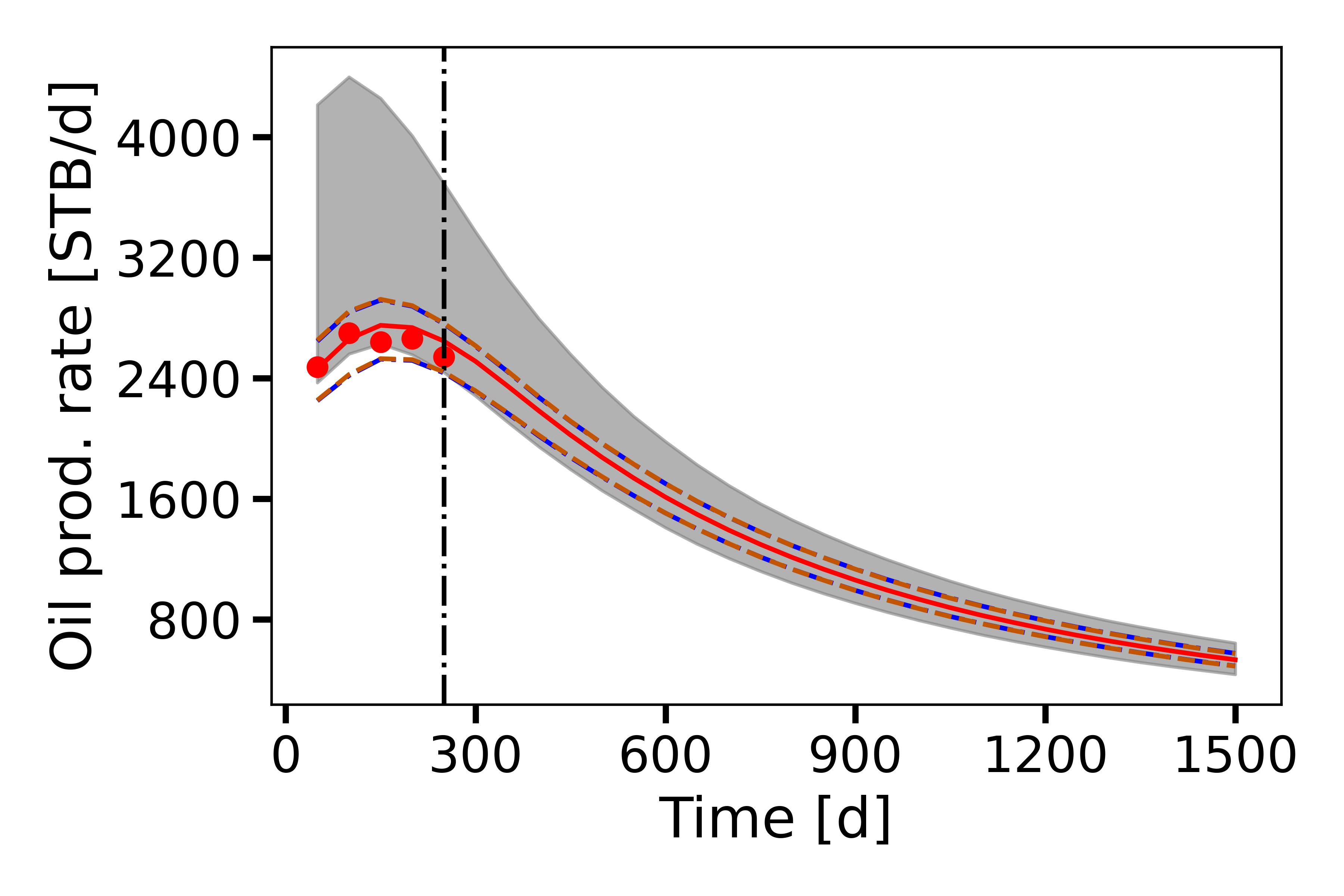}
        \caption{Field oil production rate}
        \label{fig:field_oil_lat_rhat}
    \end{subfigure}

    \vspace{01em}
    \begin{subfigure}[b]{0.48\textwidth}
        \centering
        \includegraphics[width=\textwidth]{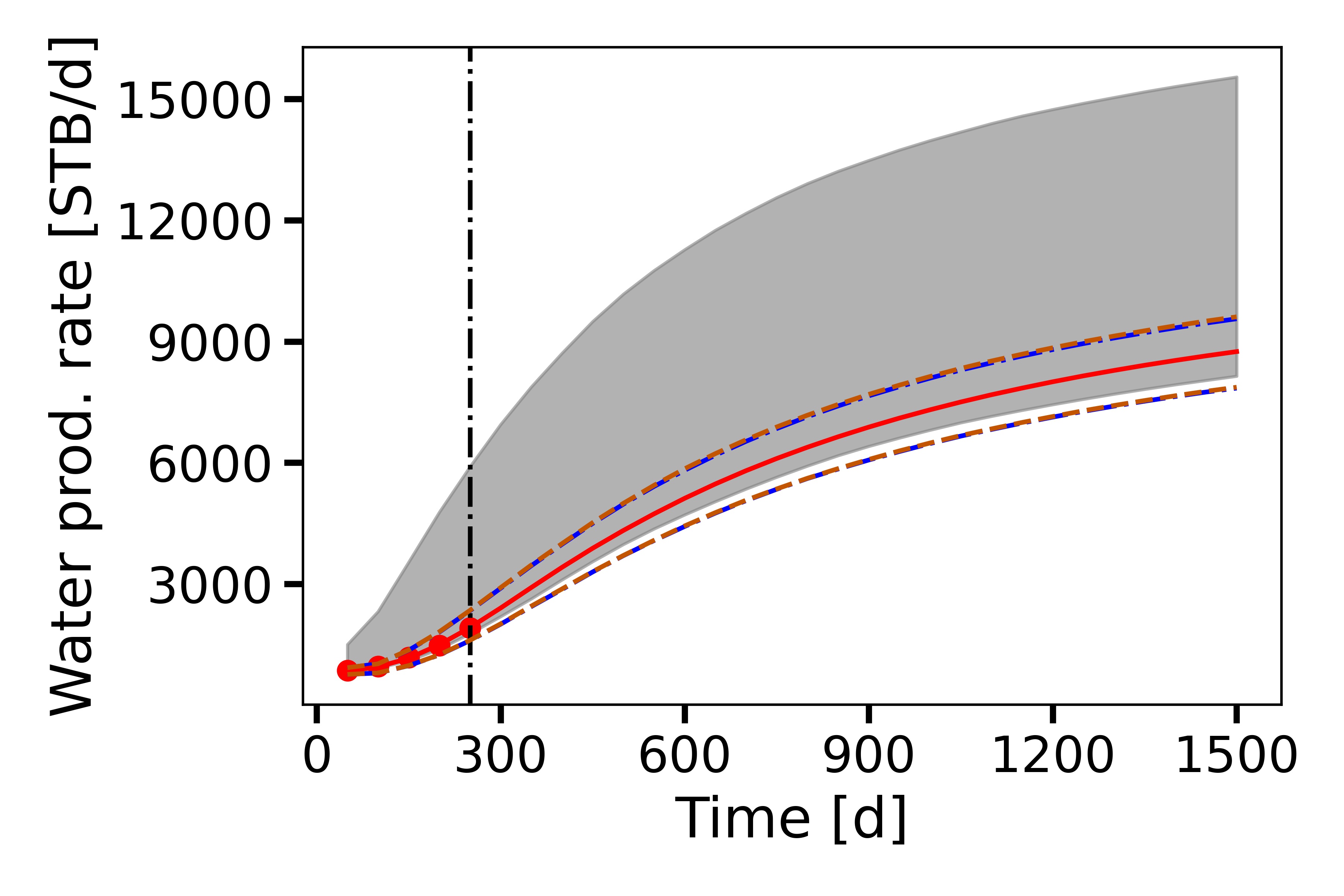}
        \caption{Field water production rate}
        \label{fig:field_wat_lat_rhat}
    \end{subfigure}
    \hfill
    \begin{subfigure}[b]{0.48\textwidth}
        \centering
        \includegraphics[width=\textwidth]{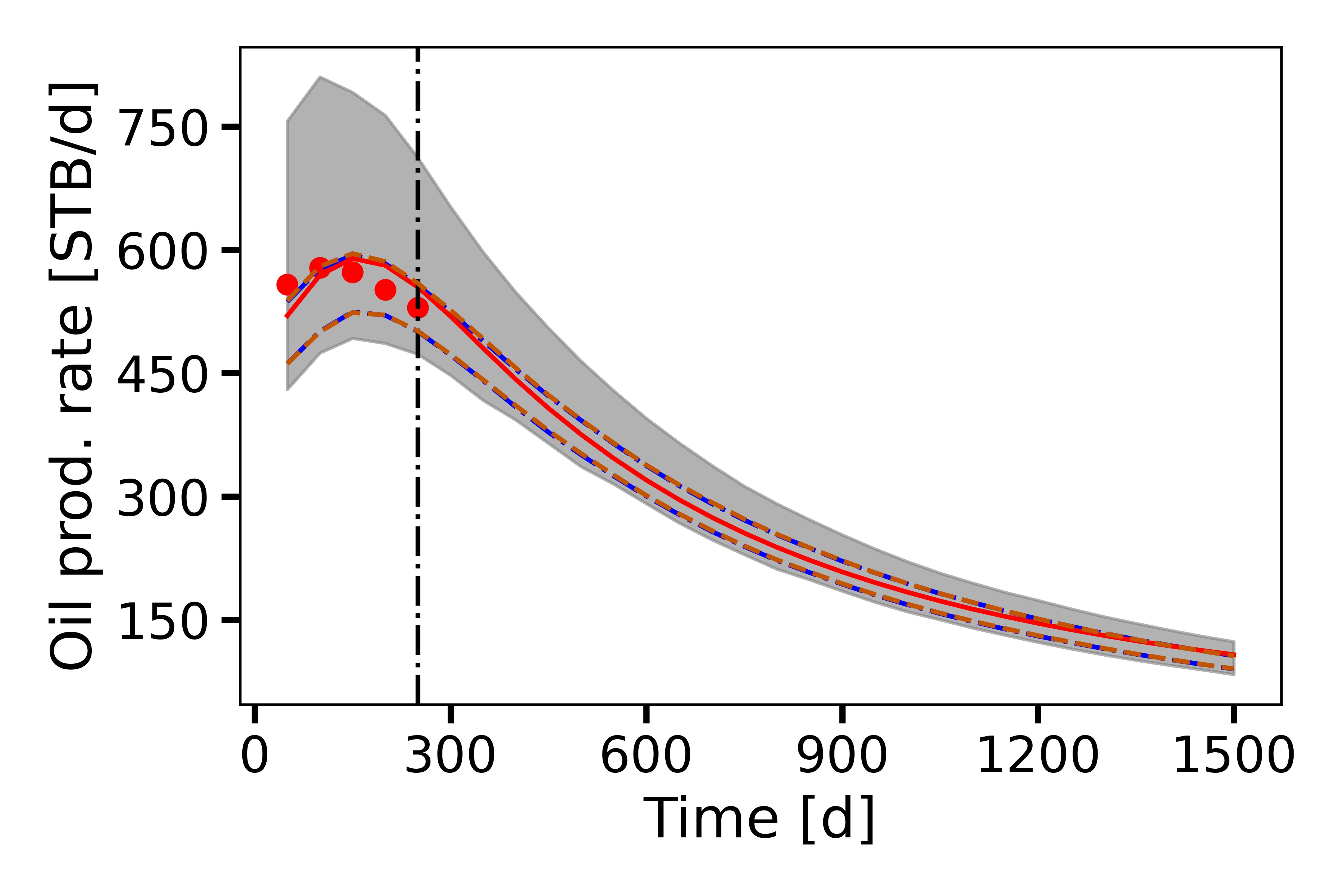}
        \caption{P2 oil production rate}
        \label{fig:rates_P2_OIL_mcmc_rhat_case88}
    \end{subfigure}
    \caption{MCMC results for field-level rates and oil rate for well P2 using two different convergence criteria  (Case~1, DA performed in 3D-LDM latent space). Gray regions show the prior P$_{10}$--P$_{90}$ range. Orange dashed and blue dash-dot lines denote the posterior P$_{10}$--P$_{90}$ ranges for MCMC stopped at $\hat{R} < 1.05$ and $\hat{R} < 1.01$, respectively. Red points and red curves represent observed and true data. The vertical black dot-dash line indicates the end of the history matching period.}
    \label{fig:case1_mcmc_rhat}
\end{figure}

\textcolor{black}{ We next consider the impact of the $\hat{R}$ value used to identify MCMC convergence. As noted earlier, we apply $\hat{R}<1.01$, but because this can be time-consuming to achieve in some cases, other studies have used $\hat{R}<1.05$. Results using these two values are shown in Figure~\ref{fig:case1_mcmc_rhat} and Table~\ref{table:case1_metrics_mcmc_rhat}. We observe a high level of agreement in the flow rate quantities between the two runs. This suggests that MCMC has indeed converged, and that the use of an even smaller $\hat{R}$ value would not be expected to have much effect on MCMC results. We note that the posterior distributions of the scenario parameters, though not shown here, are also essentially identical for the two convergence criteria.}


\textcolor{black}{Our final test involves SMC. In this case we assess sensitivity to the number of mutation steps by considering $N_{\text{mut}}=70$, in contrast to the $N_{\text{mut}}=35$ specification used in Section~\ref{sec:case1_results}. We provide only summary metrics (in Table~\ref{table:case1_metrics_mcmc_rhat}) for this assessment. These show negligible differences, indicating that the use of 35~mutation steps in SMC is sufficient. We note that posterior flow rates and scenario parameters with the two different $N_{\text{mut}}$ values, though not shown, are also essentially identical.}


\textcolor{black}{Taken in total, the results in this section demonstrate that the algorithmic parameters used in Section~\ref{sec:case1_results} are indeed appropriate. In subsequent results for Cases~2 and~3, we   therefore continue with these parameters, i.e., we specify $n_c=1024$ for 3D-LDM, apply $\hat{R}<1.01$ (or a maximum of 500,000 iterations) for the MCMC stop criteria, and use $N_{\text{mut}}=35$ within SMC.}

\begin{table}[htbp]
\centering
\caption{Performance metrics for MCMC with two convergence criteria and SMC with 35 and 70 mutation steps per iteration (Case~1, DA performed in 3D-LDM latent space).}
\label{table:case1_metrics_mcmc_rhat}
\begin{tabular}{@{}lccc@{}}
\toprule
\textbf{Algorithm}
& \textbf{RMSE} [STB/d]
& \textbf{NV}
& $\mathbf{\Delta \text{P}_{90-10}}$ \\
\midrule
MCMC ($\hat{R} < 1.05$) & 55.93 & 0.63 & 0.18 \\
MCMC ($\hat{R} < 1.01$) & 56.02 & 0.63 & 0.19 \\
SMC ($N_{\text{mut}}=35$) & 55.89 & 0.63 & 0.18 \\
SMC ($N_{\text{mut}}=70$) & 56.43 & 0.64 & 0.19 \\
\bottomrule
\end{tabular}
\end{table}

\subsection{DA results for Cases~2 and~3}
\textcolor{black}{Having characterized the behavior of the three algorithms for Case~1, we now assess whether these findings hold for true models with different geological scenario parameters.} The true models for Cases~2 and~3 are shown in Figure~\ref{fig:examples_petrel}e and~f. These correspond to quite different mud fractions, channel orientations and widths than in Case~1 (though all cases share the same prior distribution). For ESMDA, the algorithmic parameters are not modified from Case~1. For MCMC, for Case~2 we set $\beta_{\text{MCMC}}=0.05$ (this case terminated at 500,000~iterations at which $\hat{R}=1.02$, acceptance rate of 33\%) 
and for Case~3, $\beta_{\text{MCMC}}=0.1$ ($\sim$470,000 iterations, reached $\hat{R}<1.01$, acceptance rate of 27\%).
For SMC, 8~iterations using $\beta_{\text{mut}}=0.12$ are required for Case~2 (final acceptance rate of 16\%), and 6~iterations using $\beta_{\text{mut}}=0.20$ are needed for Case~3 (17\%). The choices for $\beta_{\text{MCMC}}$ and $\beta_{\text{mut}}$ are again based on limited numerical experimentation to achieve acceptance rates in the range 10--40\%. 

\begin{figure}[htbp]
    \centering

    \begin{subfigure}[b]{0.32\textwidth}
        \centering
        \includegraphics[width=\textwidth, trim=50 50 50 50, clip]{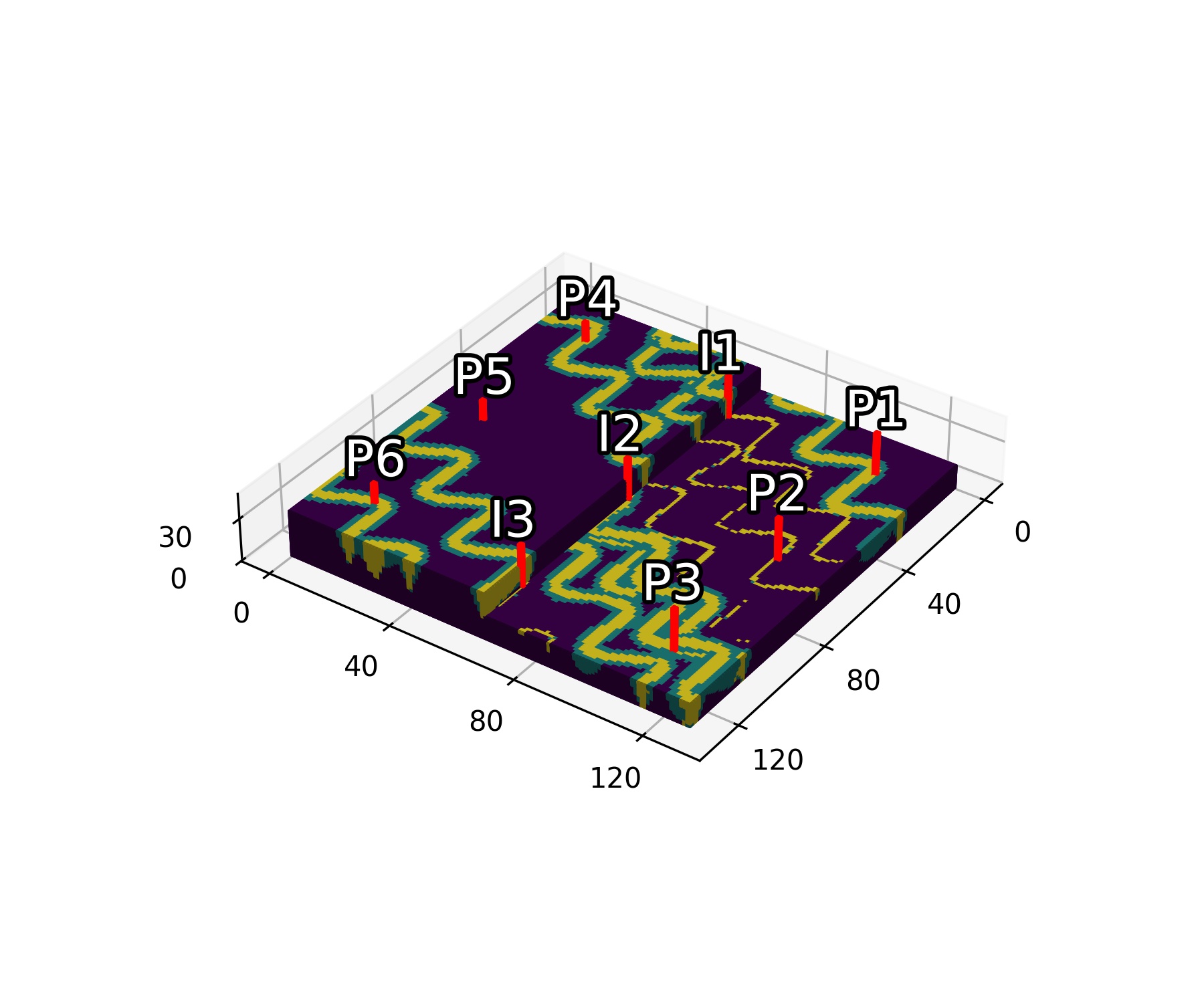}
        \caption{Case~2: ESMDA}
        \label{fig:case2_esmda}
    \end{subfigure}
    \hfill
    \begin{subfigure}[b]{0.32\textwidth}
        \centering
        \includegraphics[width=\textwidth, trim=50 50 50 50, clip]{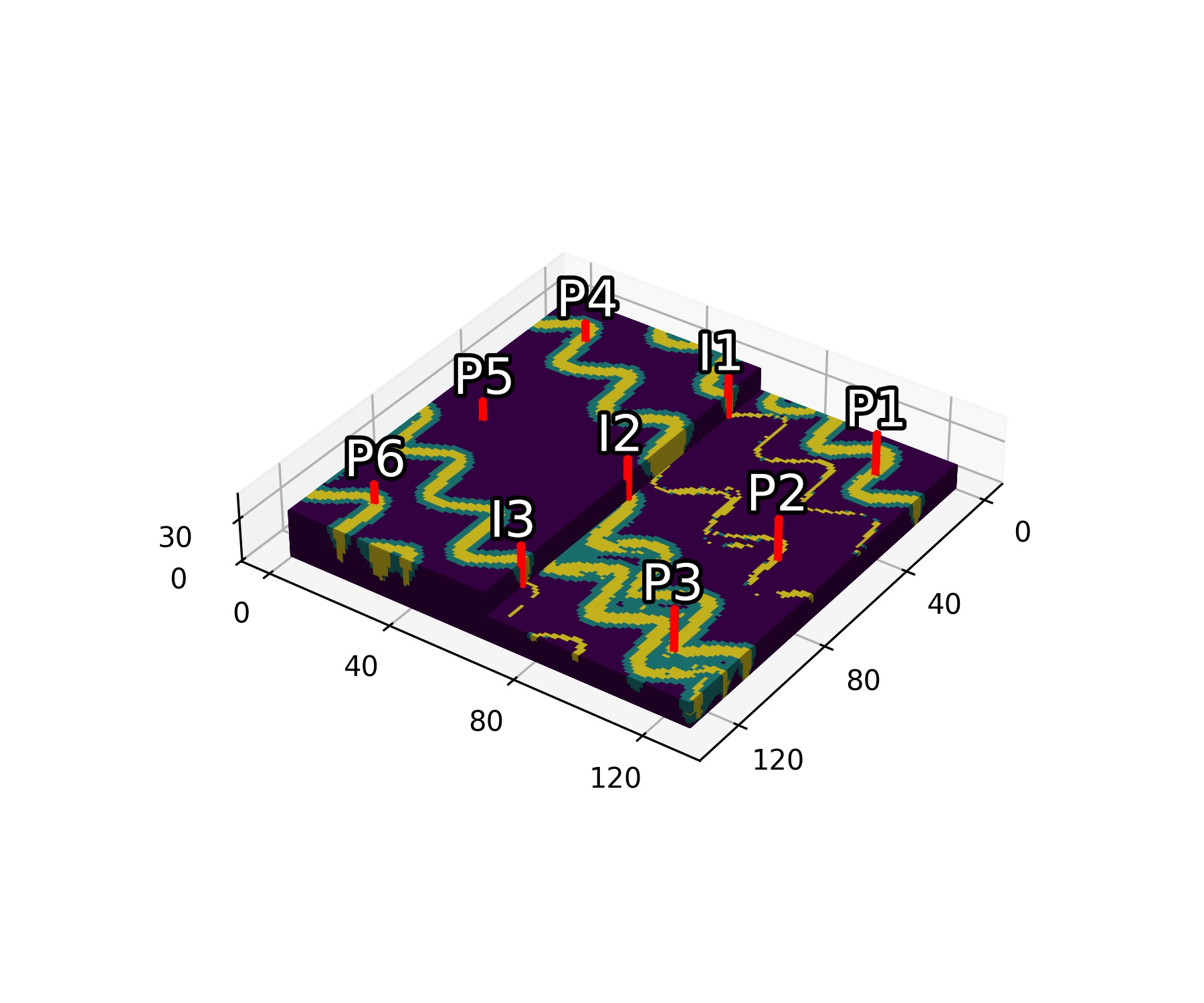}
        \caption{Case~2: MCMC}
        \label{fig:case2_mcmc}
    \end{subfigure}
    \hfill
    \begin{subfigure}[b]{0.32\textwidth}
        \centering
        \includegraphics[width=\textwidth, trim=50 50 50 50, clip]{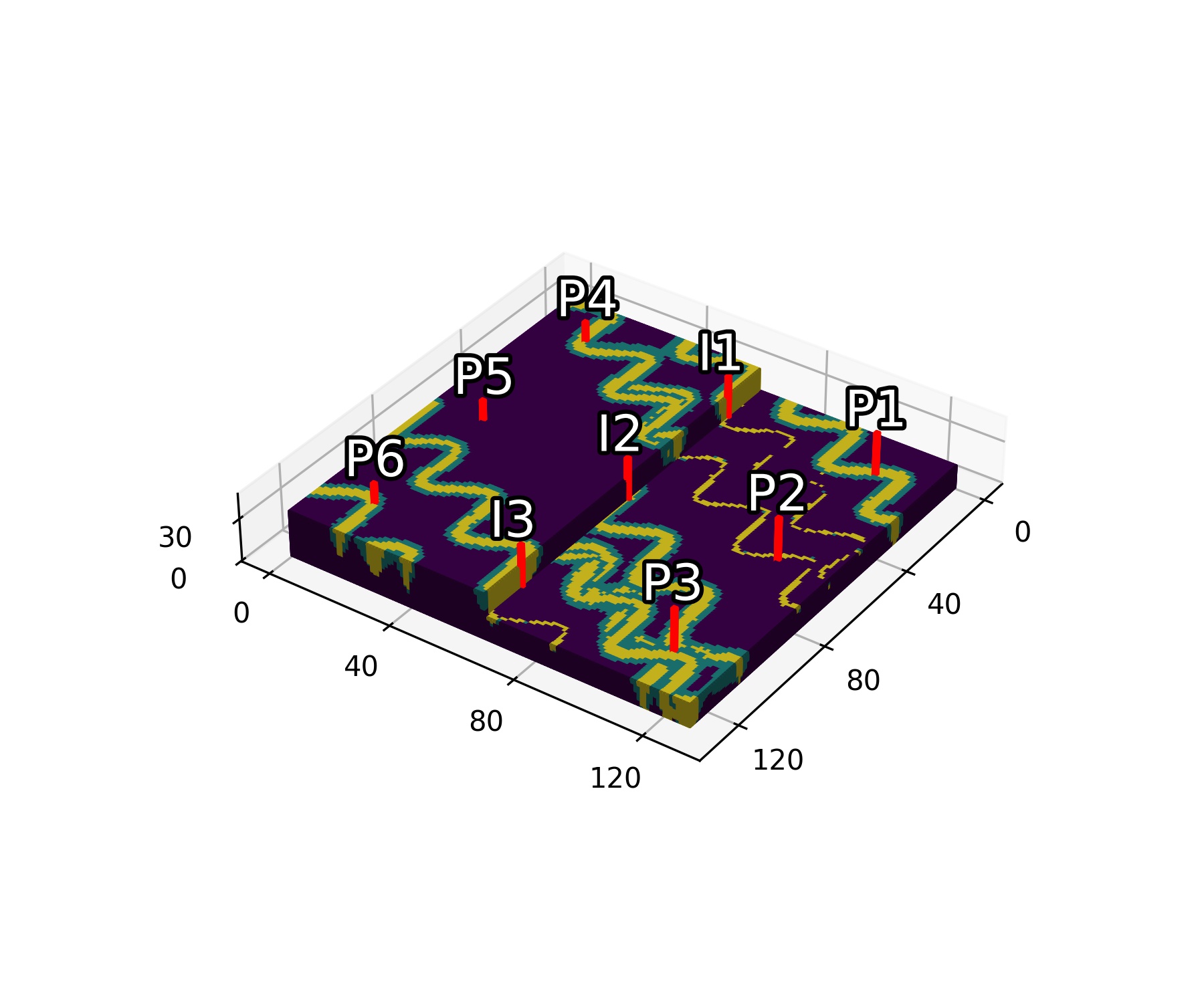}
        \caption{Case~2: SMC}
        \label{fig:case2_smc}
    \end{subfigure}

    \vspace{1em}

    \begin{subfigure}[b]{0.32\textwidth}
        \centering
        \includegraphics[width=\textwidth, trim=50 50 50 50, clip]{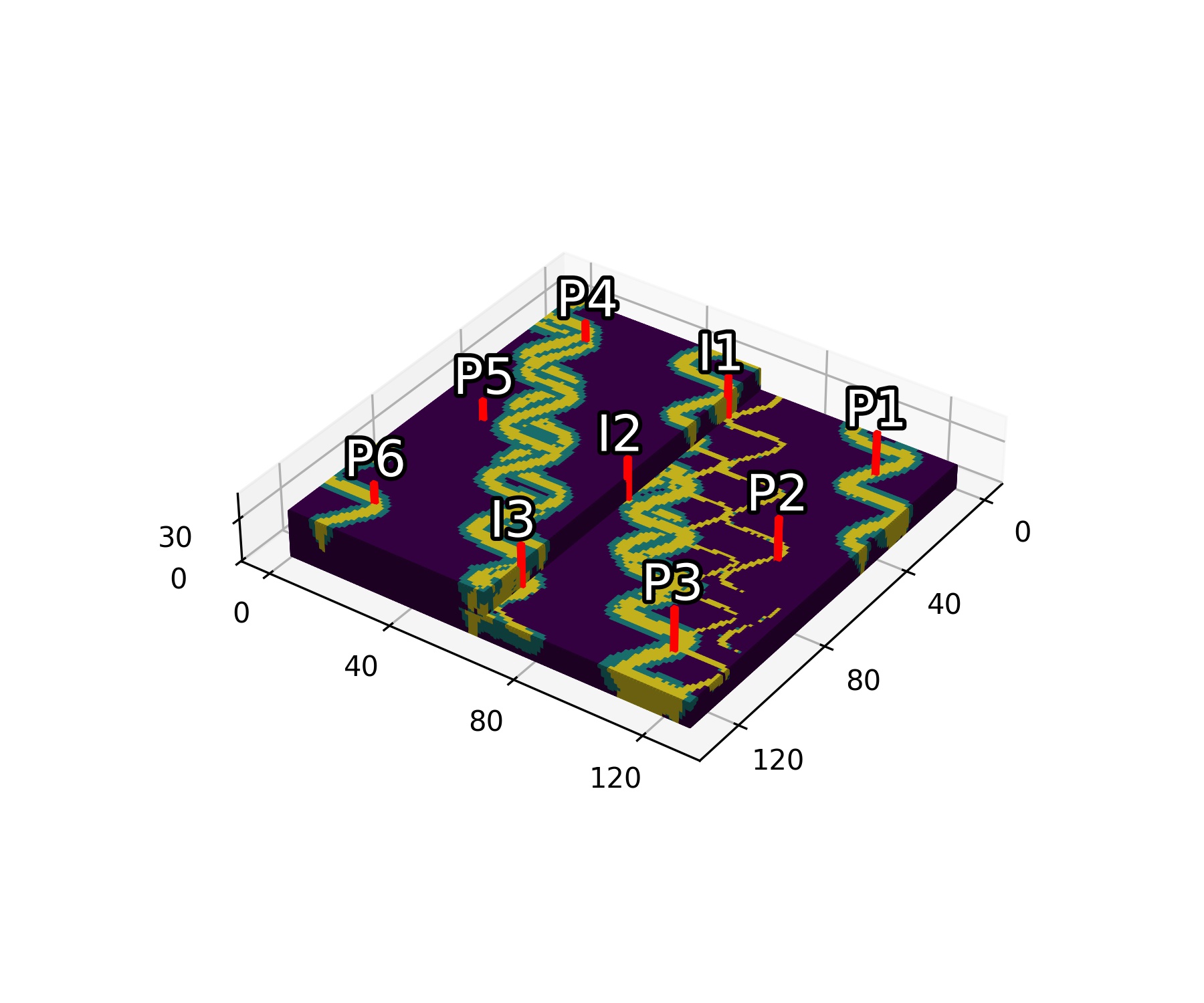}
        \caption{Case~3: ESMDA}
        \label{fig:case3_esmda}
    \end{subfigure}
    \hfill
    \begin{subfigure}[b]{0.32\textwidth}
        \centering
        \includegraphics[width=\textwidth, trim=50 50 50 50, clip]{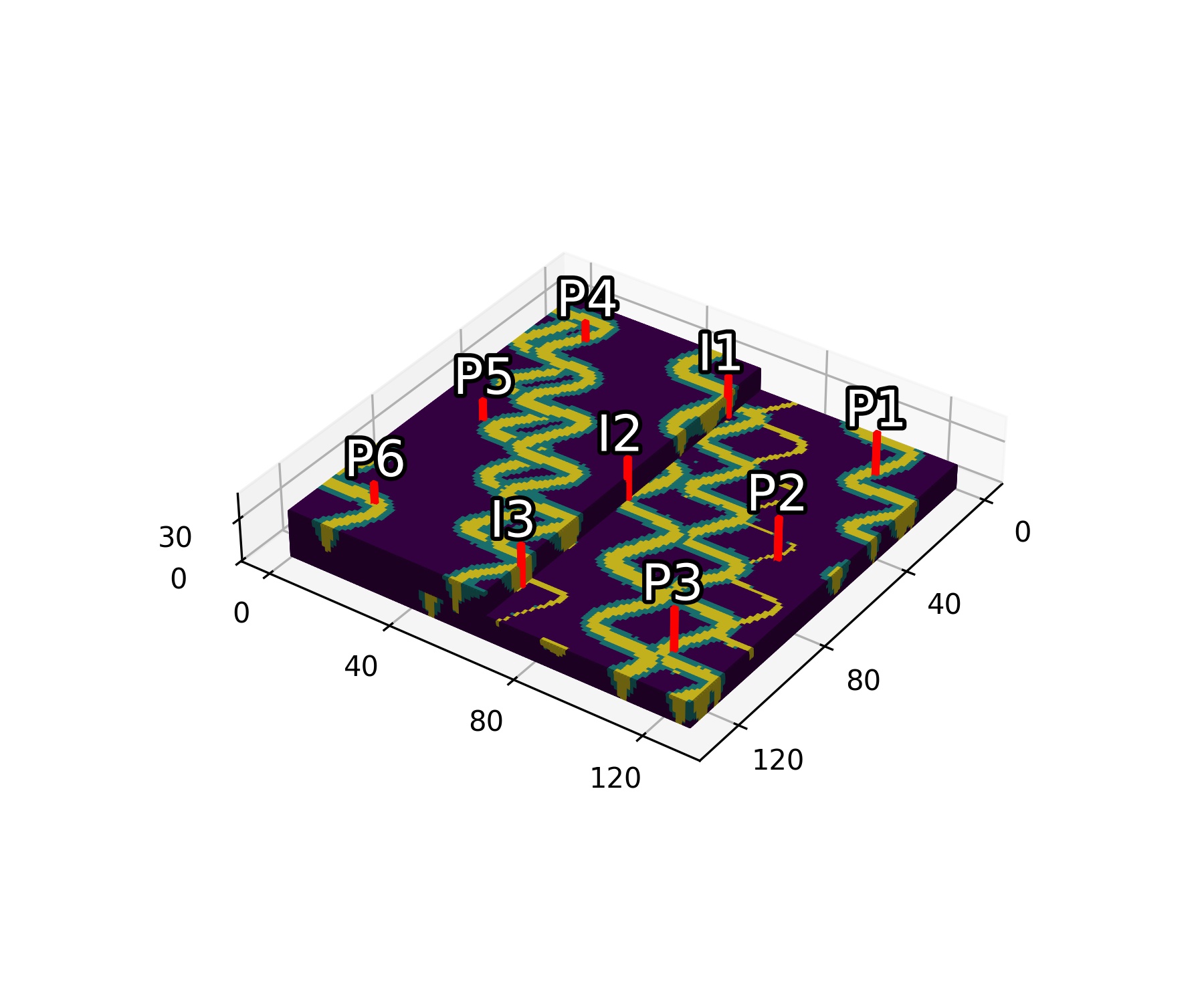}
        \caption{Case~3: MCMC}
        \label{fig:case3_mcmc}
    \end{subfigure}
    \hfill
    \begin{subfigure}[b]{0.32\textwidth}
        \centering
        \includegraphics[width=\textwidth, trim=50 50 50 50, clip]{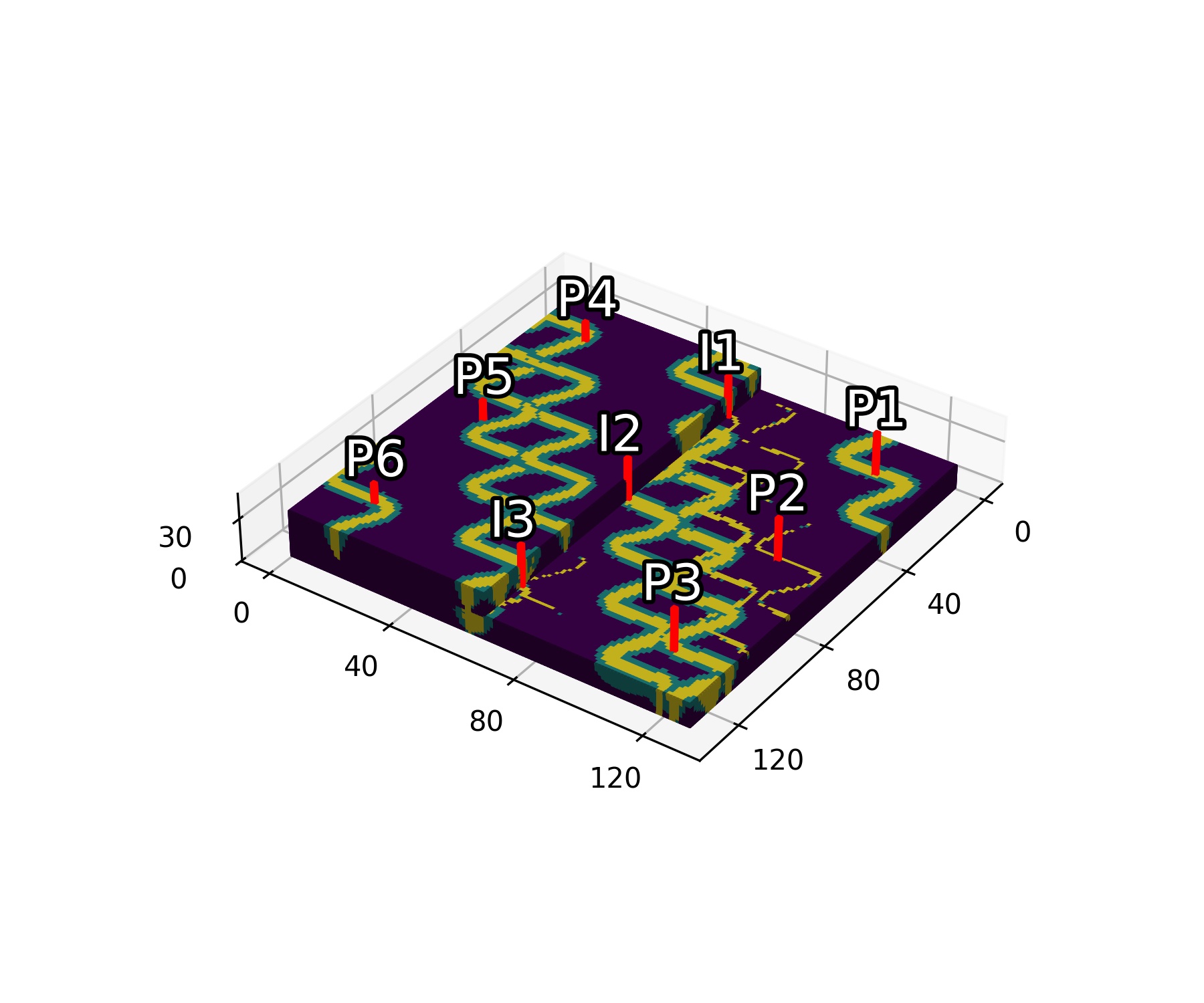}
        \caption{Case~3: SMC}
        \label{fig:case3_smc}
    \end{subfigure}

    \caption{Randomly selected posterior realizations obtained using ESMDA, MCMC, and SMC for Case~2 (top row) and Case~3 (bottom row). DA performed in the 3D-LDM latent space. All models maintain geological realism and display features consistent with the corresponding true models (shown in Figure~\ref{fig:examples_petrel}e and f).}

    \label{fig:case23_post_models}
\end{figure}

Figure~\ref{fig:case23_post_models} shows a randomly sampled realization from each of the posterior ensembles for ESMDA, MCMC, and SMC, for Case~2 (top row) and Case~3 (bottom row). The level of geological realism and visual similarity to the true models (shown in Figure~\ref{fig:examples_petrel}e and f) in the posterior samples are analogous to those in Case~1. In particular, posterior realizations for Case~2 all display channel orientations of approximately 30$^\circ$, while for Case~3, the channels are orientated at an angle of about 60$^\circ$. These orientations are in agreement with the true models. The channels in Cases~2 and~3 are also wider than those in Case~1, which is again consistent with the true models.

Summary statistics for the geological scenario parameters, which confirm these qualitative visual assessments, are reported in Table~\ref{tab:case2_case3_post_params}. All the algorithms give results of reasonable accuracy for the mean parameter values. As in Case~1, we observe more significant uncertainty reduction (smaller standard deviations) using the Monte Carlo-based methods than with ESMDA\textcolor{black}{, with the exception of $\theta_{\rm ch}$ for MCMC in Case~3}.  

\begin{table}[htbp]
\centering
\caption{DA algorithm performance for scenario parameters (Cases~2 and~3, DA performed in 3D-LDM latent space).}
\label{tab:case2_case3_post_params}
\begin{tabular}{@{}llcccccc@{}}
\toprule
& & \multicolumn{2}{c}{$\boldsymbol{f_m}$} 
& \multicolumn{2}{c}{$\boldsymbol{\theta_{\rm ch}}$} 
& \multicolumn{2}{c}{$\boldsymbol{w_{\rm ch}}$} \\
\cmidrule(lr){3-4} \cmidrule(lr){5-6} \cmidrule(lr){7-8}
& & Mean & Std & Mean & Std & Mean & Std \\
\midrule
\multirow{5}{*}{\textbf{Case 2}} 
& Prior & 0.795 & 0.043 & 45.0 & 8.66 & 5.00 & 0.577 \\
\cmidrule(lr){2-8}
& ESMDA & 0.784 & 0.016 & 37.1 & 4.90 & 5.51 & 0.360 \\
& MCMC  & 0.782 & 0.008 & 35.2 & 2.19 & 5.61 & 0.215 \\
& SMC   & 0.781 & 0.007 & 35.7 & 2.13 & 5.59 & 0.189 \\
\cmidrule(lr){2-8}
& True  & \multicolumn{2}{c}{{0.78}} & \multicolumn{2}{c}{33$^\circ$} & \multicolumn{2}{c}{5.5} \\
\midrule
\multirow{5}{*}{\textbf{Case 3}} 
& Prior & 0.795 & 0.043 & 45.0 & 8.66 & 5.00 & 0.577 \\
\cmidrule(lr){2-8}
& ESMDA & 0.785 & 0.021 & 54.7 & 4.70 & 5.44 & 0.423 \\
& MCMC  & 0.769 & 0.013 & 55.8 & 4.82 & 5.72 & 0.273 \\
& SMC   & 0.765 & 0.010 & 57.2 & 2.78 & 5.81 & 0.190 \\
\cmidrule(lr){2-8}
& True  & \multicolumn{2}{c}{0.75} & \multicolumn{2}{c}{58$^\circ$} & \multicolumn{2}{c}{5.9} \\
\bottomrule
\end{tabular}
\end{table}

\begin{figure}[htbp]
    \centering

    \begin{subfigure}[b]{0.48\textwidth}
        \centering
        \includegraphics[width=\textwidth, trim=0 0 0 0, clip]{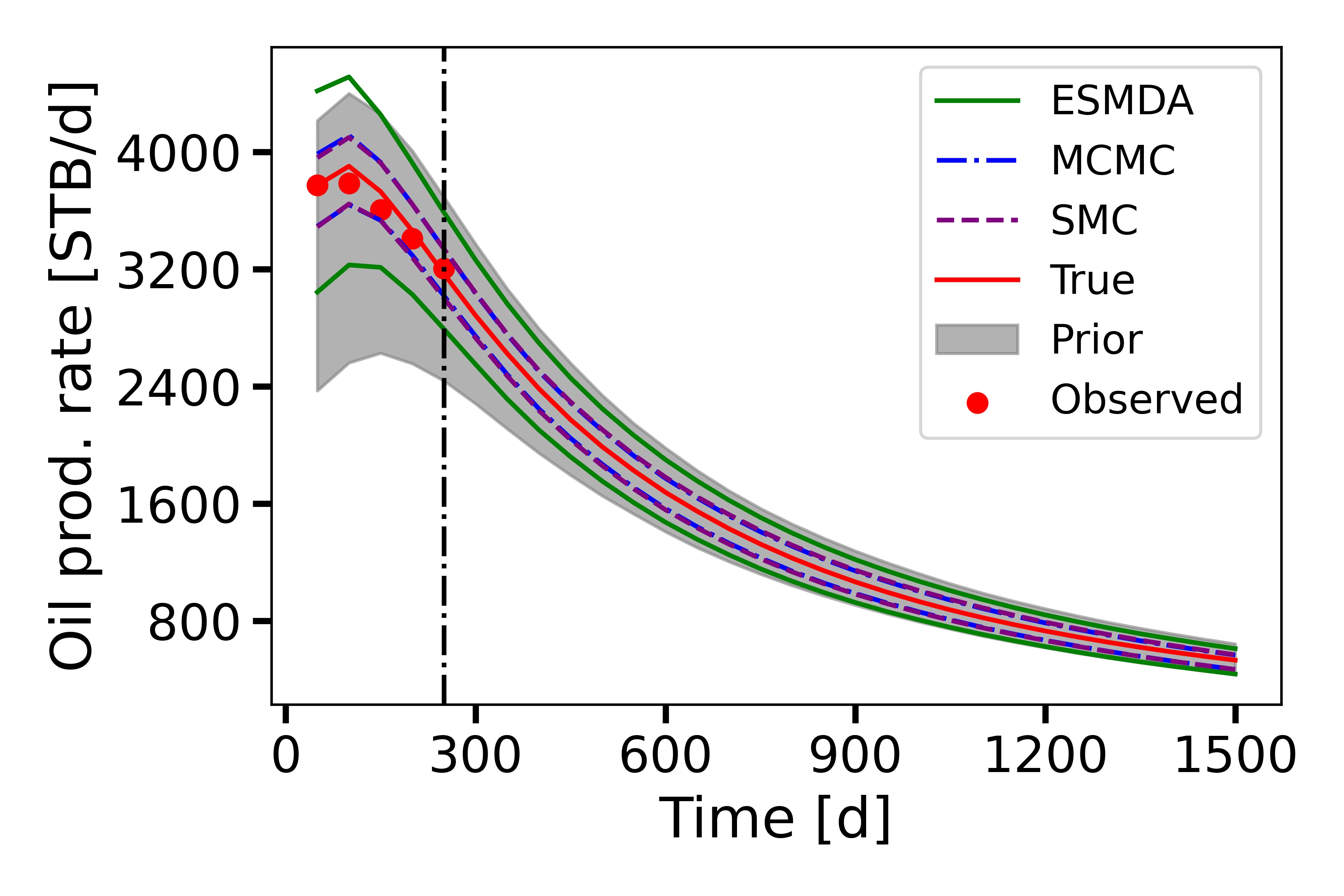}
        \caption{Case~2: field oil production rate}
    \end{subfigure}
    \hfill
    \begin{subfigure}[b]{0.48\textwidth}
        \centering
        \includegraphics[width=\textwidth, trim=0 0 0 0, clip]{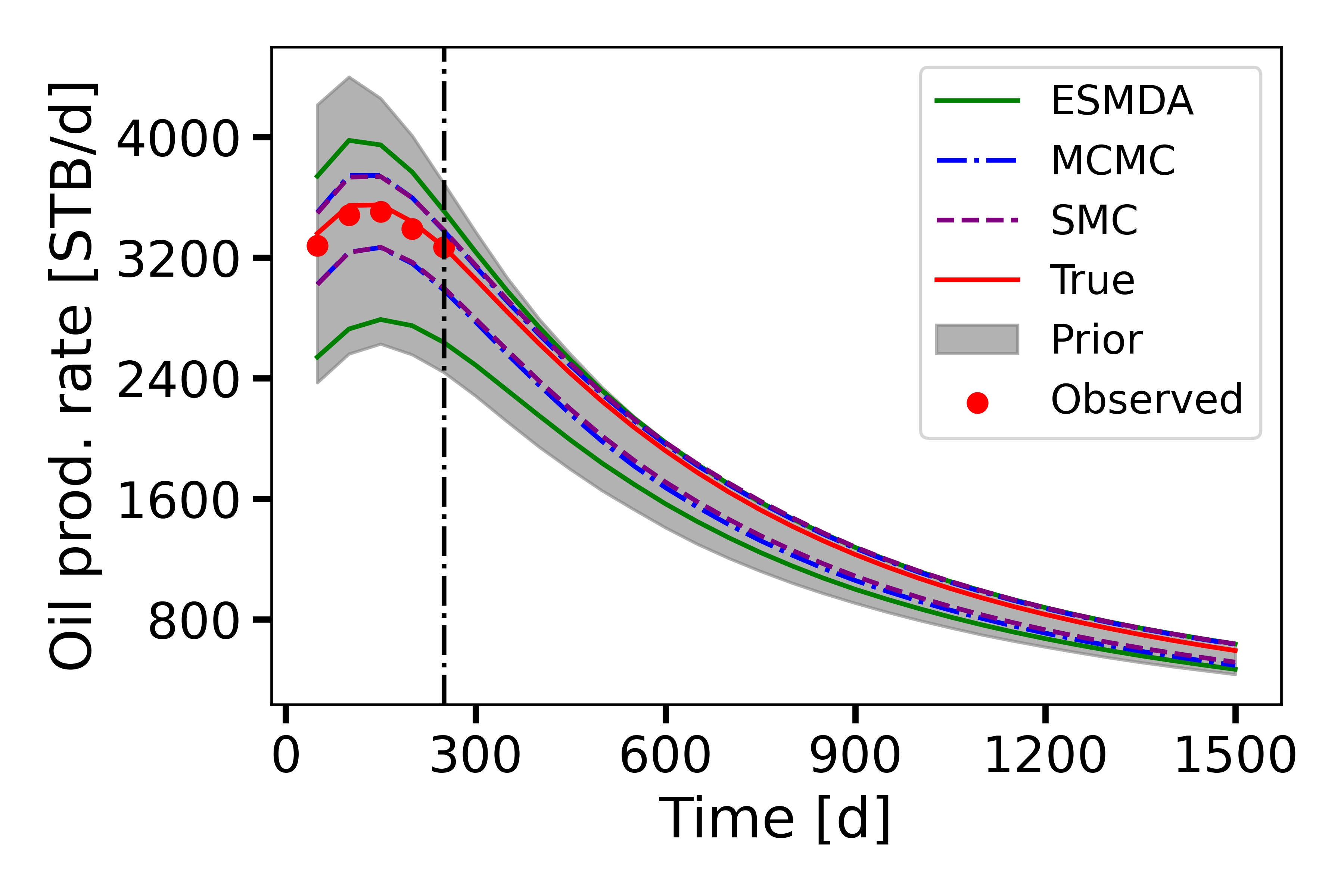}
        \caption{Case~3: field oil production rate}
    \end{subfigure}

    \vspace{1em}

    \begin{subfigure}[b]{0.48\textwidth}
        \centering
        \includegraphics[width=\textwidth, trim=0 0 0 0, clip]{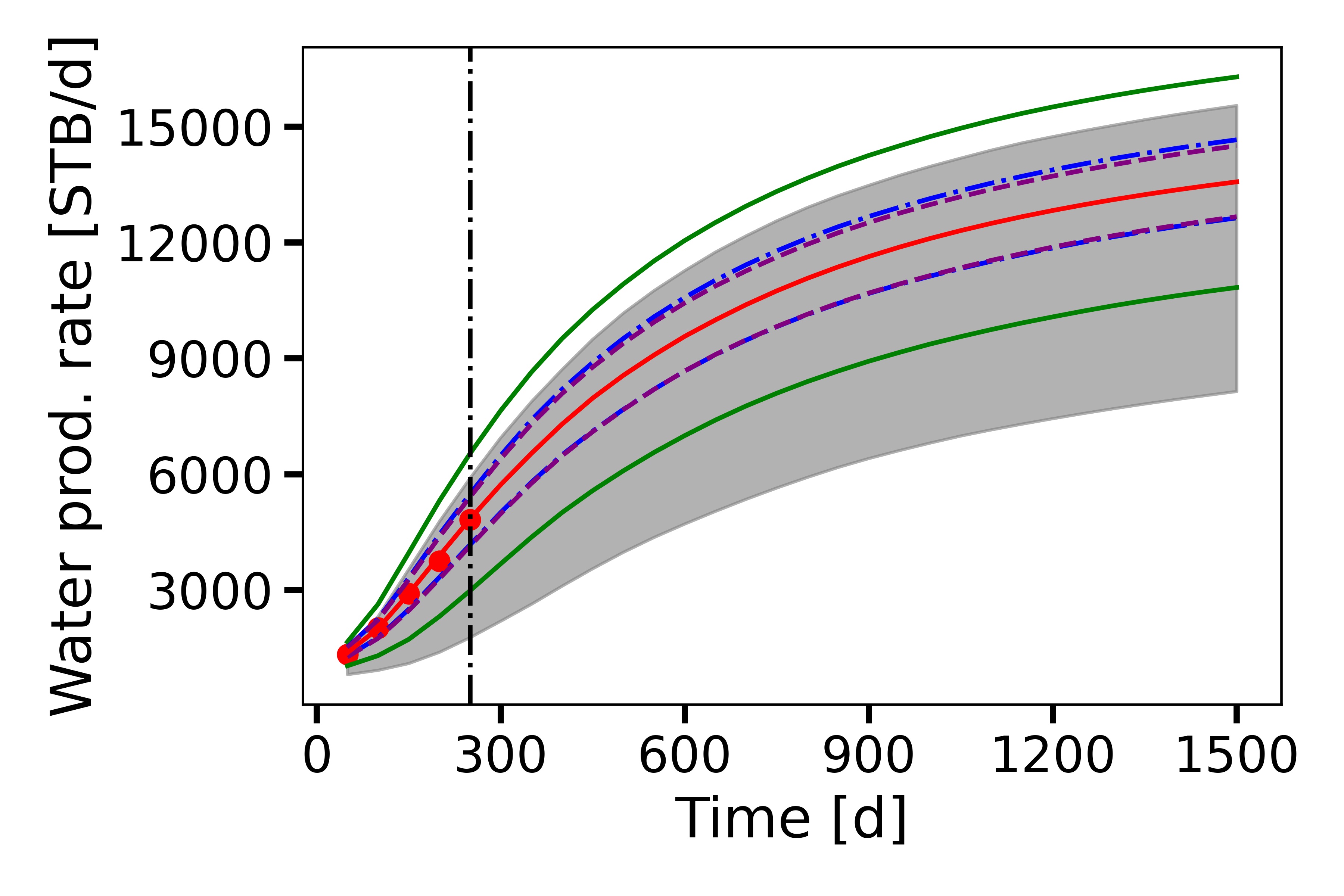}
        \caption{Case~2: field water production rate}
    \end{subfigure}
    \hfill
    \begin{subfigure}[b]{0.48\textwidth}
        \centering
        \includegraphics[width=\textwidth, trim=0 0 0 0, clip]{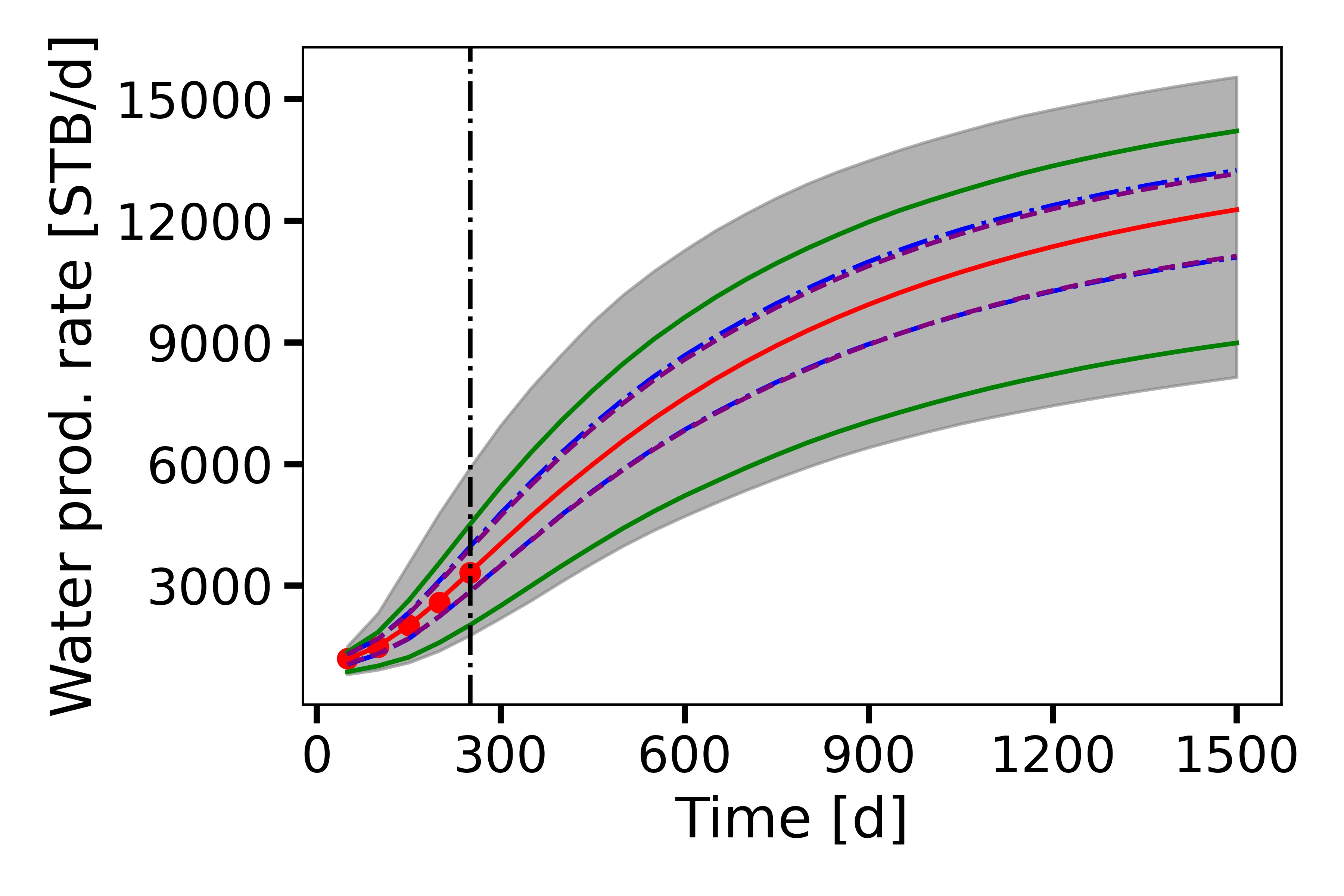}
        \caption{Case~3: field water production rate}
    \end{subfigure}

    \vspace{1em}

    \begin{subfigure}[b]{0.48\textwidth}
        \centering
        \includegraphics[width=\textwidth, trim=0 0 0 0, clip]{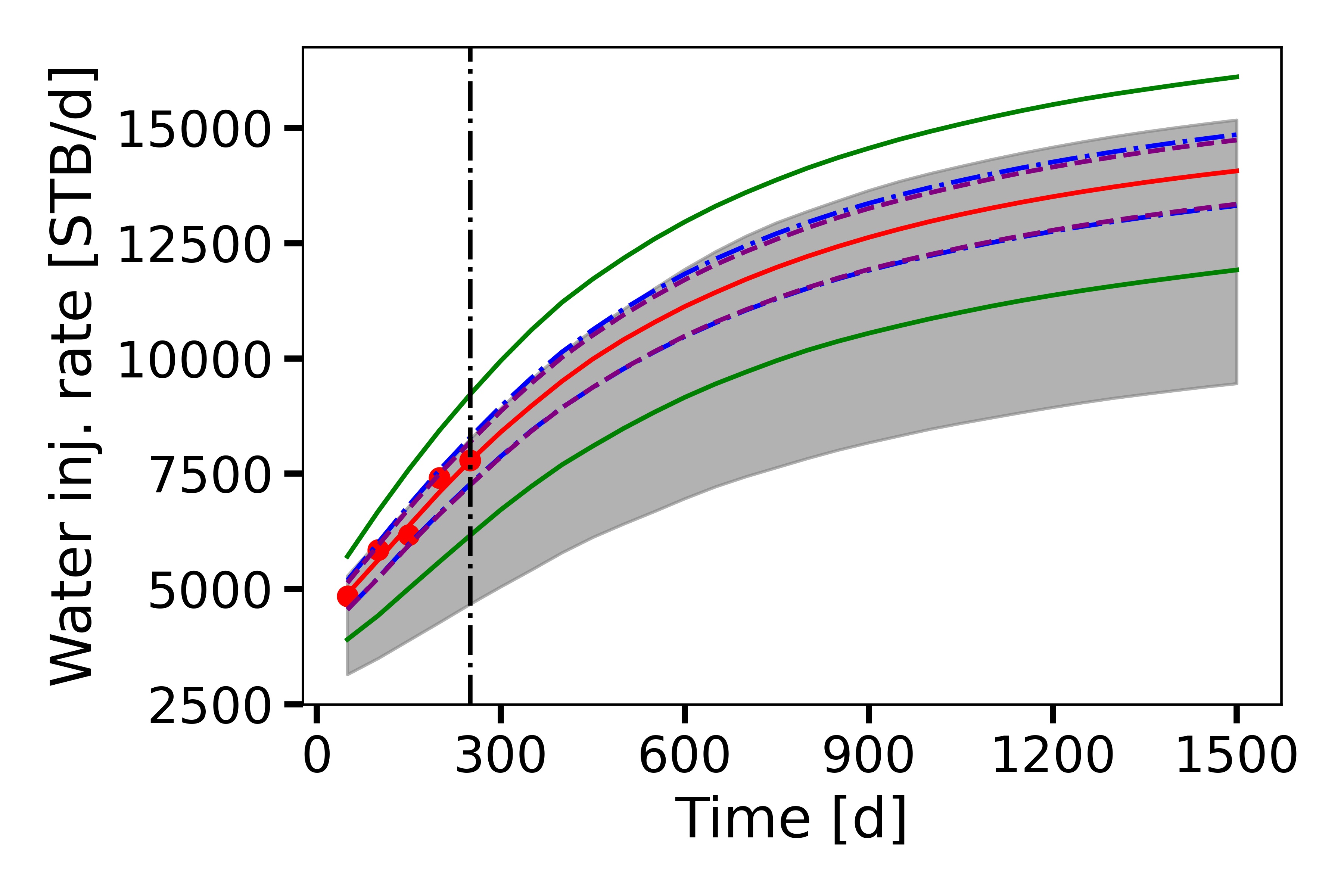}
        \caption{Case~2: field water injection rate}
    \end{subfigure}
    \hfill
    \begin{subfigure}[b]{0.48\textwidth}
        \centering
        \includegraphics[width=\textwidth, trim=0 0 0 0, clip]{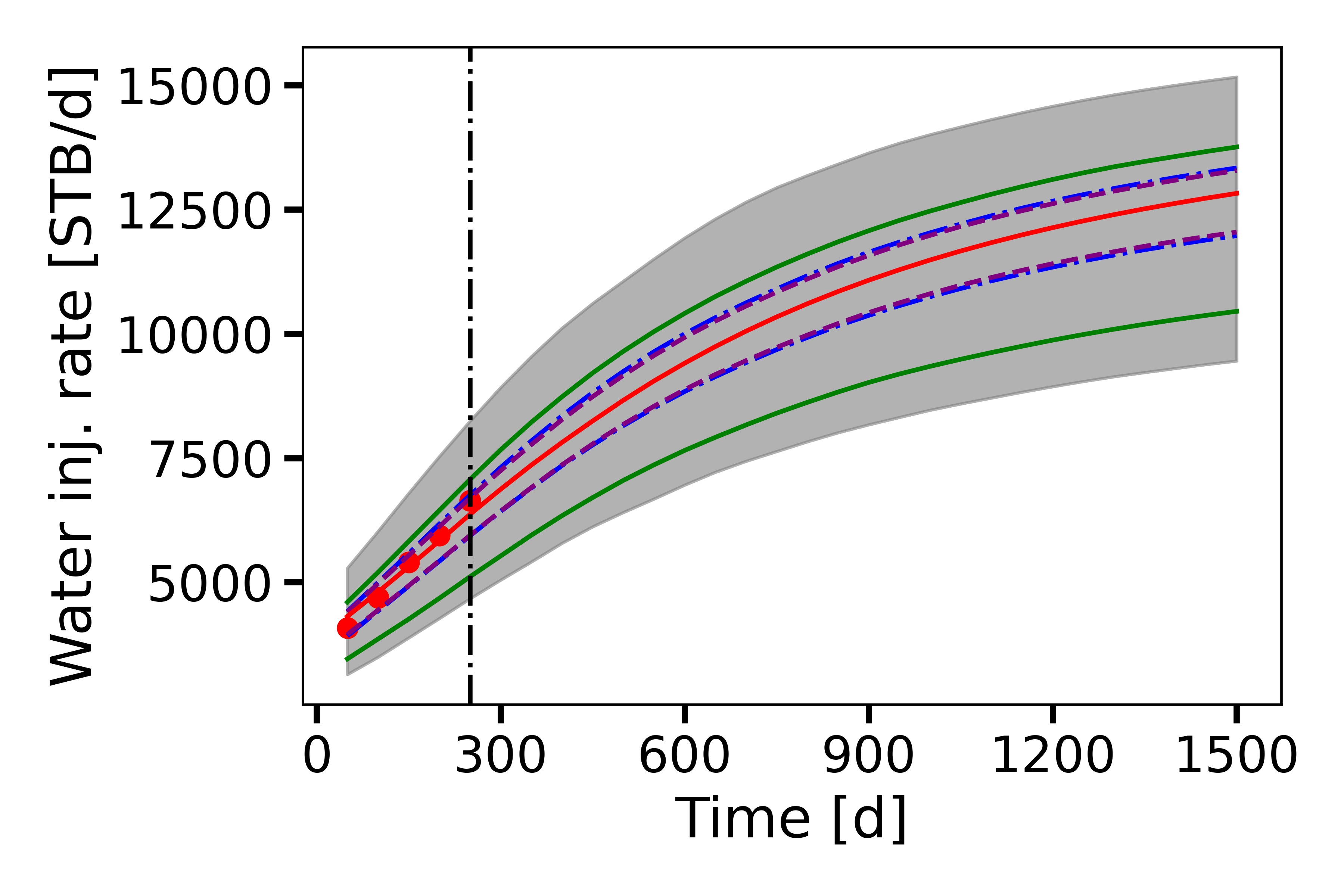}
        \caption{Case~3: field water injection rate}
    \end{subfigure}

    \caption{DA results for field rates for Case~2 (left column) and Case~3 (right column). DA for all methods performed in 3D-LDM latent space. Gray regions show the prior P$_{10}$--P$_{90}$ range, and green, blue, and purple lines denote the posterior P$_{10}$--P$_{90}$ ranges for ESMDA, MCMC, and SMC, respectively. Red points and red curves represent observed and true data. The vertical black dot-dash line indicates the end of the history matching period.}
    \label{fig:case23_post_rates}
\end{figure}

Rather than show well-by-well flow rates, for Cases~2 and~3 we present P$_{10}$--P$_{90}$ field-level responses for oil production, water production, and water injection. 
These results appear in Figure~\ref{fig:case23_post_rates}, with Case~2 results in the left column and Case~3 in the right column. Prior ranges, observed data points, and true curves are also shown. As in the Case~1 results, we again observe close agreement between MCMC and SMC, and much wider P$_{10}$--P$_{90}$ ranges for ESMDA. The discrepancies in the ESMDA posterior estimates are again quite noticeable at early times in the oil production curves and at late times in the water production curves. 

Finally, we report summary metrics for Cases~2 and~3 in Table~\ref{table:case23_metrics_da}. These results confirm our previous observations for Case~1. Namely, MCMC and SMC show a high degree of consistency, with more accurate data matching (RMSE) and more pronounced uncertainty reduction in model parameters (NV) and well rate responses ($\Delta$P$_{90-10}$) than ESMDA. 


\begin{table}[htbp]
\centering
\caption{Performance metrics for DA algorithms (Cases~2 and~3, DA performed in 3D-LDM latent space).}
\label{table:case23_metrics_da}
\setlength{\tabcolsep}{4pt}
\begin{tabular}{@{}lcccccccc@{}}
\toprule
& \multicolumn{4}{c}{\textbf{Case 2}}
& \multicolumn{4}{c}{\textbf{Case 3}} \\
\cmidrule(lr){2-5}
\cmidrule(lr){6-9}
\textbf{Alg.}
& \textbf{RMSE}
& \textbf{NV}
& \textbf{NV\textsubscript{dum}}
& $\mathbf{\Delta P_{90-10}}$
& \textbf{RMSE}
& \textbf{NV}
& \textbf{NV\textsubscript{dum}}
& $\mathbf{\Delta P_{90-10}}$ \\
\midrule
ESMDA & 204.44 & 0.80 & 0.97 & 0.77 & 136.86 & 0.88 & 0.98 & 0.64 \\
MCMC  & 63.44  & 0.67 & --   & 0.23 & 53.72  & 0.83 & --   & 0.23 \\
SMC   & 62.44  & 0.66 & --   & 0.22 & 51.08  & 0.78 & --   & 0.22 \\
\bottomrule
\end{tabular}
\end{table}

The findings reported in this section, which are consistent over the three cases considered, highlight important differences between computationally efficient, ensemble-Kalman-based algorithms (ESMDA) and computationally expensive, more rigorous Monte Carlo-based methods (MCMC, SMC). In particular, ensemble-Kalman methods, when combined with the highly nonlinear 3D-LDM parameterization, provide significantly overestimated posterior uncertainties. This is observed even using localization to limit spurious updates along with large ESMDA ensemble sizes and large numbers of assimilation steps. This behavior is the opposite of the ensemble-collapse problem that can occur with ensemble-Kalman methods operating in model space (with small ensemble sizes).

We attribute the problems observed with latent-space ESMDA to the use of Kalman-gain-based updates on the highly nonlinear, nonsmooth LDM latent space. The 3D-LDM mapping introduces a shift between the space in which updates are performed (latent space) and the space in which the data mismatch is evaluated (model space). As a result, Kalman-gain-based corrections may fail to identify optimal descent directions, causing ensemble members to oscillate around a residual level of uncertainty that cannot be further reduced with additional iterations. \textcolor{black}{Since this behavior stems from the nonlinearity of the parameterization, we expect the relative performance of ESMDA and Monte Carlo-type methods to extend beyond our specific test cases and setup. Consistent with this speculation, in a previous study~\citep{DIFEDERICO2025105755} with a 2D channelized case that included uncertain porosity and permeability within each facies, ESMDA with LDM appeared to similarly overestimate posterior uncertainty. The level of uncertainty reduction observed in that work was not analyzed in detail, however.} Our overall findings suggest that 3D-LDM (and other deep learning parameterization methods) may be more suitable for use with Monte Carlo-type algorithms, as their stochastic components allow for more reliable exploration of the latent space. SMC appears to be particularly appropriate, as it combines the large sample size and computational efficiency of ensemble methods with Monte Carlo sampling. Taken in total, our findings suggest that formal DA with deep learning parameterization is indeed achievable, provided that a fast and accurate flow surrogate is available.

\clearpage
\section{Concluding remarks}
\label{sec:conclusions}

In this study, we performed a systematic comparison of data assimilation (DA) algorithms for large-scale 3D systems parameterized with a latent diffusion model (3D-LDM). The geomodels considered represent three-facies fluvial channel systems characterized by hierarchical geological uncertainty. This means the geological scenario parameters (mud fraction, channel orientation, channel width) are themselves uncertain, as are the cell-by-cell facies indicators in each realization. The 
dataset for the 3D-LDM parameterization consisted of 3000 conditional realizations of dimensions $128 \times 128 \times 32$ (524,288 total grid blocks), while the latent space was of overall dimension 1024. 

In the first part of this study, we characterized a key trade-off that arises when an ensemble-Kalman algorithm is used for DA. Specifically, we showed that model-space updates achieved significant uncertainty reduction but produced geologically unrealistic posterior models, while latent-space updates preserved geological realism but exhibited limited uncertainty reduction. This finding motivated the application of rigorous posterior sampling methods -- Markov chain Monte Carlo (MCMC) and Sequential Monte Carlo (SMC) -- operating in the 3D-LDM latent space. To accommodate the large number of forward evaluations required by MCMC and SMC, we developed a fast and accurate deep learning surrogate model that predicts well rate time series directly from the input geomodel. The model was trained using flow results for the same 3000 realizations as were used for the construction of the 3D-LDM representation. Surrogate model error was characterized and incorporated into the DA workflow through a full covariance treatment.

We then applied ESMDA, MCMC, and SMC (all operating in the 3D-LDM latent space) to three synthetic test cases characterized by different geological scenario parameters. For all cases, posterior geomodels preserved geological realism through the 3D-LDM parameterization. MCMC and SMC were found to be in essential agreement with one another, providing more accurate data matching and more pronounced uncertainty reduction than latent-space ESMDA. 
ESMDA, even using 1000-member ensembles, consistently overestimated posterior P$_{10}$--P$_{90}$ intervals for well rates by a factor of \textcolor{black}{2.3--3.3 relative to MCMC, and 2.4--3.5 relative to SMC}. This suggests that standard ensemble-Kalman methods, though computationally efficient, may not be ideally suited for use with parameterizations such as 3D-LDM. SMC was found to be a particularly attractive option, as it combines the parallel efficiency of ensemble methods with rigorous Monte Carlo sampling. We therefore recommend using SMC in combination with deep learning parameterizations, though a fast surrogate model for flow is essential. \textcolor{black}{To our knowledge, this is the first study to assess the posterior sampling accuracy of ensemble-Kalman methods, used in conjunction with a deep learning geological parameterization (in this case LDM), through detailed comparisons against formal Monte Carlo sampling procedures.}

Future work could involve the development of treatments that enable ensemble-Kalman methods to provide proper uncertainty reduction when combined with 3D-LDM or other nonlinear parameterizations. 
On the DA algorithm side, this may entail adjusting the data-parameter covariance matrix to account for the nonlinearity of the mapping or performing some type of ``equivalent'' model-space update. On the parameterization side, it could involve modifying the 3D-LDM algorithm to obtain a more physically interpretable latent space via the use of regularized loss functions or scenario parameter conditioning. The development of a ``one-shot'' inversion capability, in which posterior 3D-LDM-based geomodels are generated directly from observed data, could also be pursued. These solutions may, however, require larger training datasets or frequent retraining when the simulation setup is modified. We note that other investigators~\citep{
laloy_gradient, lopez_nonlinear_generator, multi_disentangled, fully_conv_vae} have observed limitations in DA when using VAE and GAN-based parameterizations (with potentially more entangled latent spaces than LDM), and there has been some work in these directions with other parameterizations. A better quantification of the smoothness (or lack of smoothness) of the latent space may help facilitate the choice of DA algorithm, or provide an estimate of the degree of uncertainty reduction that is achievable. The development of specific metrics for this purpose is thus of interest. 

The application of the overall methodology to real models should also be pursued. \textcolor{black}{This would require retraining (and possibly enhancing) the 3D-LDM for the target geology and hard data. Additional uncertain geological parameters might also enter the framework. The flow surrogate in this case would need to be extended to handle the actual well settings, which could involve switching from rate to BHP control. The modular design of our workflow, which involves separate geological parameterization, flow surrogate, and DA components, should act to facilitate extensions to real systems.}


\section*{Declaration of competing interest}
The authors declare that they have no known competing financial interests or personal relationships that could have appeared to influence the work reported in this paper. The last author is on the editorial board of this journal.

\section*{Acknowledgments}
We are grateful to the Stanford Smart Fields Consortium and the Stanford Graduate Fellowship in Science and Engineering for financial support. We also thank the SDSS Center for Computation for providing the computational resources used in this work.\\

\section*{Data and code availability}
\label{sec:code}
The datasets and codes for the flow surrogate model and latent-space data assimilation are available at \url{https://github.com/guidodf09/ldm_data_assimilation}.
These codes are based on the implementations provided in the \url{https://github.com/huggingface/diffusers} and \url{https://github.com/Project-MONAI/GenerativeModels} repositories. The Python libraries \texttt{diffusers} and \texttt{monai} are used in our implementation.

\newpage
\vfill\eject

\newpage
\bibliographystyle{cas-model2-names.bst}
\bibliography{sn-bibliography}

\newpage

\end{document}